\documentclass[12pt]{article}

\usepackage[utf8]{inputenc}

\usepackage{epsfig,pgf,wrapfig,hyperref}
\usepackage{tabu,booktabs,longtable} 
\usepackage{amssymb,latexsym,amsmath,xcolor,url}
\usepackage{graphicx}

\newcommand{\Nens}{{N_{ens}}}

\newcommand{\mut}{\hat{\mu}_t}
\newcommand{\sigmat}{\hat{\sigma}^2(t)}

\title{EpiCovDA: a mechanistic COVID-19 forecasting model with data assimilation}

\author{Hannah R. Biegel and Joceline Lega \\
\\
\small Department of Mathematics, University of Arizona\\
\small 617 N. Santa Rita Avenue, Tucson, AZ 85721}

\begin{document}

\maketitle

\begin{abstract}
We introduce a minimalist outbreak forecasting model that combines data-driven parameter estimation with variational data assimilation. By focusing on the fundamental components of nonlinear disease transmission and representing data in a domain where model stochasticity simplifies into a process with independent increments, we design an approach that only requires four core parameters to be estimated. We illustrate this novel methodology on COVID-19 forecasts. Results include case count and deaths predictions for the US and all of its 50 states, the District of Columbia, and Puerto Rico. The method is computationally efficient and is not disease- or location-specific. It may therefore be applied to other outbreaks or other countries, provided case counts and/or deaths data are available.
\end{abstract}

\section{Introduction}
An increasingly common application of epidemiological modeling is outbreak forecasting, as exemplified by a variety of recent ``challenges'' aiming to predict the burden caused by the flu on the US healthcare system \cite{Biggerstaff18,McGowan19,Reich19}, or case counts of dengue \cite{Johansson19}, chikungunya \cite{DelValle18}, and neuroinvasive West Nile virus disease \cite{WNC}. It is no longer rare to see government officials relying on model predictions to guide public health decisions \cite{Eubank20} and a future in which the general public is knowledgeable about and routinely refers to epidemiological forecasts may not be too distant. Improving the reliability and the speed at which such forecasts are created is therefore an important aspect of mathematical modeling.

The COVID-19 pandemic \cite{Sun20,CCDC20} has brought epidemiological modeling to the forefront of scientific research. Compartmental models of different levels of complexity (see for instance \cite{Hethcote00,Keeling11} for a discussion of the fundamental principles of epidemiological modeling) applied to the general population or to different age groups have been used to explore disease risk and assess the effectiveness of a variety of mitigation scenarios \cite{Davies20,Gosce20,Salje20,Tian20,Zhang20}. Metapopulation approaches, combined with mobility data, have informed the spread of contagion between different regions or countries and documented the effectiveness of travel restriction measures \cite{Chinazzi20,Kucharski20,Li20,Tian20,Wu20}. Other efforts have emphasized statistical analyses \cite{Kraemer20,Lonergan20} and, at a more local level, agent-based modeling \cite{Koo20,Xue20}. For mechanistic models, an important trade-off in the case of new, emerging diseases, is to balance model complexity with limited information on parameter values: on the one hand, too simple a model is likely to miss essential aspects of disease dynamics; on the other hand, lack of knowledge about sensitive parameters may lead to forecasts with so much uncertainty that they become uninformative \cite{Edeling20}. Because different methodologies lead to forecasts that perform optimally under different conditions, it is now common to develop ensemble models that combine predictions from different approaches into a single forecast \cite{Ray18,Reich19}. Such ensembles have consistently been shown to be overall more reliable than any individual model used to create them \cite{Solazzo13,Kioutsioukis14,Ray18,McGowan19,Reich19,Ray20,Cramer21}.

In this article, we introduce a novel and computationally efficient forecasting methodology that relies on a small number of parameters. Our approach combines two key elements: ICC curves and variational data assimilation (VDA). ICC curves \cite{Lega16,Lega20} are representations of outbreak dynamics in the incidence vs. cumulative-cases (ICC) plane. Remarkably, empirical observation reveals that when represented in this fashion, incidence data fluctuate about a mean ICC curve associated with the deterministic SIR (susceptible, infected, removed; \cite{Kermack27}) compartmental model. Such a curve has only 4 parameters and encompasses the entire deterministic SIR dynamics in a single equation \cite{Lega20}. Although there is currently no mathematical proof that this behavior is universal, it has been observed for a variety of diseases, spreading under different circumstances \cite{Lega16,Lega20,Sahneh21}. Moreover, a first theoretical justification was provided in \cite{Sahneh21}: in the limit of large populations, the trajectory in the ICC plane of a stochastic, network-based SIR model results from a Gaussian process with independent increments, of mean given by the deterministic ICC curve. Consequently, the first element of our modeling approach is the assumption that the time dynamics of a generic outbreak follows an iterative process dictated by a local SIR ICC curve, with additive noise.

The VDA \cite{Law15,Reich15} step uses incidence data to estimate the 4 parameters of the local ICC curve by balancing two constraints: the parameters should (i) define an ICC curve that is as close as possible to the observed incidence data and (ii) be compatible with pre-established prior distributions. Each parameter estimation obtained in this manner leads to one forecasted trajectory for future case counts, which is converted into an incidence forecast for the next 1 through 4 weeks ahead. Probabilistic forecasts are obtained by repeating the VDA step after perturbing the reported epidemiological data and priors with suitably chosen noise, and by assimilating data on windows of pre-specified lengths (3, 5, and 14 days in the case of COVID-19). Fitting parameters to very recent (e.g. the last few days) and more distant (e.g. the last two weeks) incidence reports adds modeling flexibility to capture the effects of  ongoing trends, such as changes in social distancing attitudes. 
This procedure of combining ICC curves with VDA leads to a primary case counts forecaster, which involves a minimal number of core parameters (four) and has minimal computational burden (since parameters are estimated by a minimization procedure rather than Markov chain Monte Carlo (MCMC) sampling \cite{ENDO19}). Deaths forecasts are obtained by adding a linear regression layer to the model, which provides an estimate of future deaths as a fraction of delayed case counts. As detailed below,
the linear coefficient and the delay are estimated from data and are time- and location-dependent.

Although our methodology can be transported to other diseases or locations, the present model, EpiCovDA, was created to forecast COVID-19 case counts and deaths in the US, its 50 states, the District of Columbia, and Puerto Rico. Its predictions have been regularly submitted to the University of Massachusetts Amherst COVID-19 repository \cite{COVID-19FCR} and are displayed, together with forecasts from other groups and an ensemble model, on the COVID-19 Forecast Hub \cite{COVID-19FC} and on the CDC COVID-19 forecasting page \cite{CDC-F}.

The rest of this article is organized as follows. Section \ref{Sec:Methods}
provides details on ICC curves, their use to find prior distributions, the VDA implementation, and the obtention of probabilistic case counts and deaths forecasts. Section \ref{Sec:Scoring} presents the scoring methodology. Section \ref{Sec:US_Forecasts} describes the performance of EpiCovDA at forecasting cases and deaths in the US. Section \ref{Sec:Discussion} summarizes our results and considers when a simple outbreak forecaster like EpiCovDA may be most able to contribute to public health efforts.

\section{Forecasting Methodology} \label{Sec:Methods}
In this section, we review data sources, explain how outbreak dynamics are captured by ICC curves, describe the data assimilation procedure, and provide details on how probabilistic forecasts are obtained.

\subsection{Data sources}
\label{sec:data_sources}

Minimal requirements for data sources include daily or weekly recordings of cumulative confirmed cases. For forecasts of disease-related deaths, corresponding cumulative data are also required. It should be noted that public health data are inherently variable due to irregular reporting patterns (for instance case counts go down over the weekend), backfill (revised counts for past reports), and revised numbers without specified dates (which therefore cannot be retroactively backfilled). Although different repositories have different ways of handling such corrections, the overall trends are the same.

Many data sources of COVID-19 cases are available online, including the well publicized Johns Hopkins University (JHU) dashboard \cite{JHUCD}. When we started this work, the COVID Tracking Project at {\it The Atlantic} \cite{CVDT} (CTP) included early case and death counts in all of the US States which at the time were not available from JHU. Since then, the two datasets have become more comparable and consistent, although the CTP stopped collecting data after 03/06/2021. 
Here, we use CTP data covering the period from early March to mid-October 2020.
For each state, the historical data of the cumulative number of confirmed (either clinical or laboratory diagnosis) cases, the daily incidence of cases, the cumulative number of COVID-19-attributed deaths, and the daily incidence of deaths were downloaded through the publicly available API \cite{CVDT}.

\subsection{ICC curves}
\begin{figure}[hbtp]
\centerline{\includegraphics[width=0.9\linewidth]{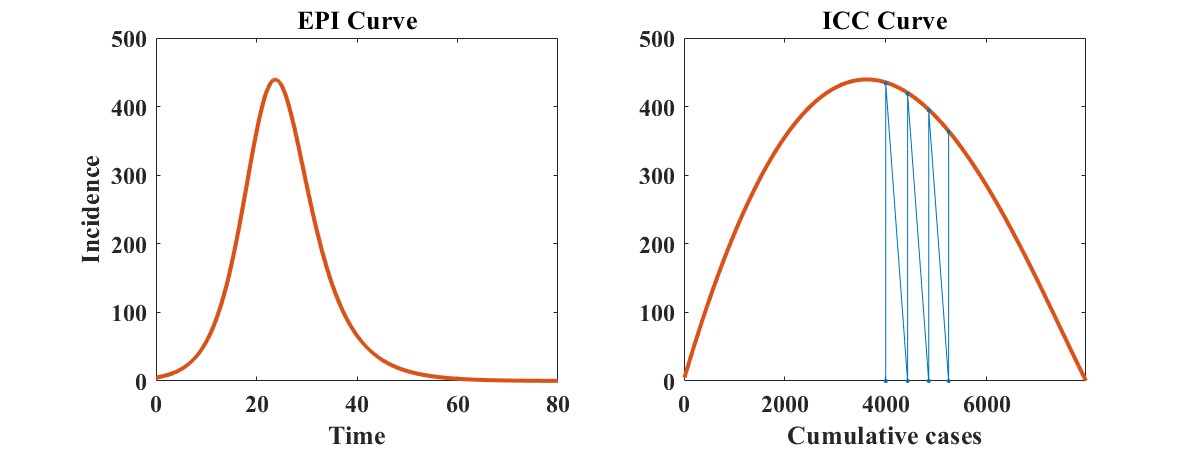}} 
\caption{\label{fig:EPI_ICC} Left: Epidemiological curve (incidence as a function of time) for a trajectory of the SIR model with $\beta = 0.5$ and $R_0 = 2.$ Right: The ICC curve corresponds to the same trajectory, but in the $({\mathcal I}, C)$ plane.}
\end{figure}

ICC (incidence vs. cumulative-cases) curves \cite{Lega20} provide a novel description of disease dynamics. They
differ from traditional epidemiological (EPI) curves via {\it a nonlinear transformation of the horizontal axis}, in which the time variable is replaced by a monotonic function thereof, specifically the cumulative number of cases. Figure \ref{fig:EPI_ICC} shows the effect of such a transformation in the case of the SIR (susceptible, infected, removed; \cite{Kermack27}) compartmental model. 
In the left panel, the EPI curve represents incidence $\mathcal I$ (the derivative of the cumulative number of cases $C$) as a function of time; in the right panel, incidence is plotted as a function of $C$.

Advantages of the ICC representation over the EPI curve include:
(i) the concavity has constant sign before the outbreak peaks, (ii) the time variable is no longer explicitly present, and (iii) in the case of the deterministic SIR model for a disease spreading in a population of known size, there is a {\it unique set of parameters} that minimizes the root mean square error between epidemiological data points in the $({\mathcal I}, C)$ plane and the ICC curve \cite{Lega20}. Moreover, the time course of a simulated outbreak may be directly obtained from the ICC curve by successive iterations, as illustrated by the saw-tooth curve in the right panel of Figure \ref{fig:EPI_ICC}: given a value of $C$, the corresponding incidence may be read off the ICC curve and added to $C$ in order to estimate the cumulative number of cases after one additional unit of time. Repeating this process leads to a time series of cumulative cases that simulates an outbreak. 

Reporting noise may be included for instance by replacing each estimate of ${\mathcal I}(C)$ by a Poisson random variable of mean ${\mathcal I}(C)$. Noise due to the stochasticity of disease spread should also be taken into account. In the case of the SIR model, it was shown in \cite{Sahneh21} that in the limit of large population size $N$, the scaled incidence $\beta S I / N$ observed when $C$ cumulative cases have been reported is normally distributed with mean ${\mathcal I}(C)/N$ given by  \eqref{eq:ICC_Curve} below with $\kappa = 1$, and variance equal to 
\[
\frac{1}{N} \beta^2 \left( -\frac{1}{R_0} \ln\left(1-\frac{C}{N}\right) + \frac{1}{R_0^2}\frac{C/N}{1-C/N}\right) \left(1-\frac{C}{N}\right)^2,
\]
where $\beta$ is the contact rate of the disease, $N$ is the size of the population involved in the outbreak ($N > C$), and $R_0$ is the basic reproductive number.

\subsection{EpiGro}
\label{sec:EpiGro}
\begin{figure}[hbtp]
\centerline{\includegraphics[width=\linewidth]{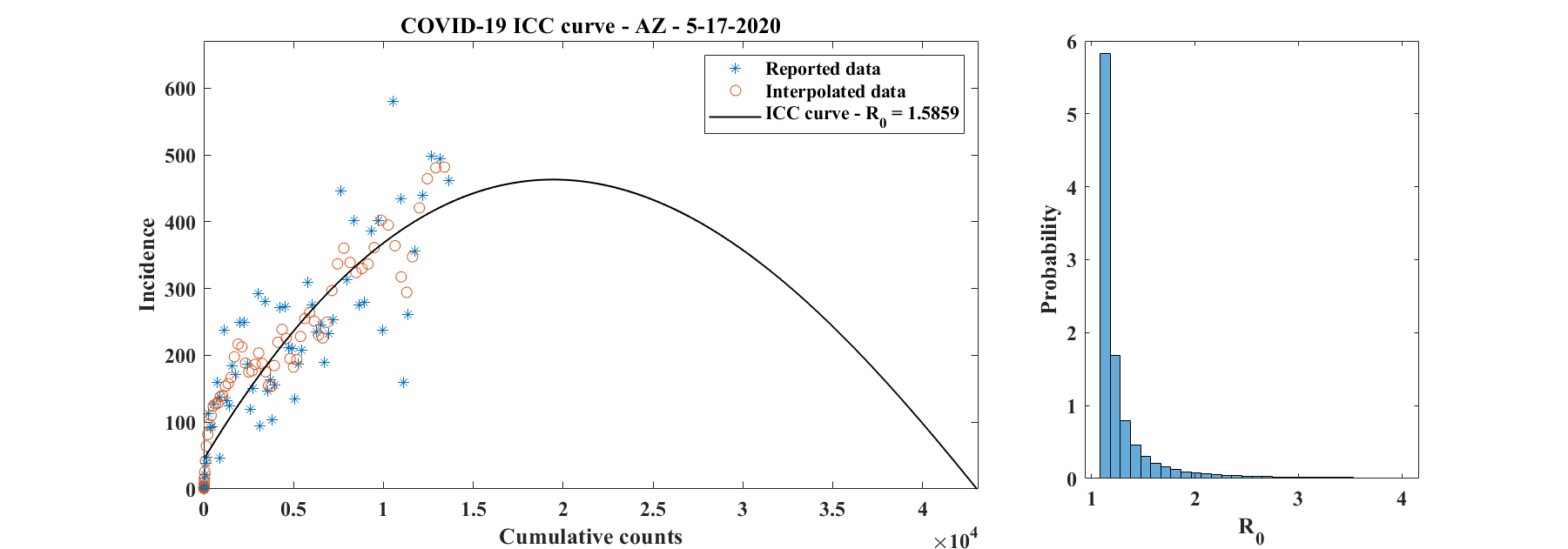}} 
\caption{\label{fig:EpiGro_v2} Left: Epidemiological data and optimal ICC curve for COVID-19 in Arizona through May 17, 2020. Right: Estimated distribution of the basic reproductive number $R_0$. COVID-19 case data provided by The COVID Tracking Project at {\sl The Atlantic} under a CC BY 4.0 license \cite{CVDT}.}
\end{figure}

A parabolic approximation of the ICC curve led to the point value (i.e. non-probabilistic) forecasting model {\it EpiGro} \cite{Lega16}, which won the 2014-15 DARPA Chikungunya Challenge \cite{DelValle18}. In this case, since ${\mathcal I} = d C / d t$ is a quadratic function of $C$, the cumulative number of cases $C$ follows logistic dynamics, an approach that had been independently identified as a useful forecasting tool \cite{Chowell14,Pell16}. 
Version 2.0 of EpiGro uses the exact formulation of the ICC curve for the SIR model given in \cite{Lega20}: 
\begin{equation}
{\mathcal I}(C) = \beta \left(C+ \frac{N}{R_0} \ln\left(1-\frac{C}{N}\right) - \frac{N}{R_0} \ln(\kappa)\right)
\left(1 - \frac{C}{N}\right),
\label{eq:ICC_Curve}
\end{equation}
where $\beta$, $N$, and $R_0$ are defined above, and $\kappa$ represents initial conditions. There is a complete equivalence between trajectories of the SIR model and of the differential equation $d C / d t = {\mathcal I}(C)$, in the sense that knowledge of one implies knowledge of the other, and vice versa \cite{Lega20}. Moreover, as previously stated, a unique vector of parameters $(\beta, \gamma = \beta / R_0, \kappa)$ minimizes the $\ell_2$ norm between the ICC curve of the SIR model and given epidemiological data points, for $N$ known. If $N$ is unknown, for instance due to the existence of transmission clusters, or because of under-reporting, a range of values of $N$ is considered, leading to a range of possible parameter values. Figure \ref{fig:EpiGro_v2} illustrates the output of EpiGro v.2.0. The left panel shows the ICC curve that best fits the May 17, 2020 COVID-19 epidemiological data for the state of Arizona. The right panel displays the distribution of $R_0$ values associated with values of $N$ near optimum. As explained in Section \ref{sec:Priors}, the parameters that provide the optimal ICC curve are used to build a prior distribution for the variational data assimilation step of EpiCovDA.

\begin{figure}[ht]
\centerline{\includegraphics[width=0.9\linewidth]{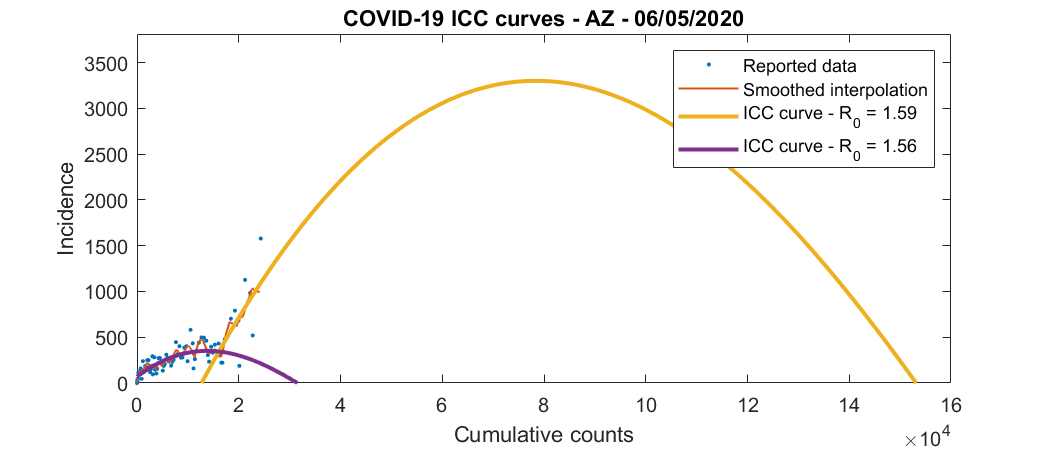}} 
\caption{\label{fig:ICC_AZ} COVID-19 incidence data (dots) as of June 5, 2020, plotted as a function of cumulative cases for the state of Arizona, together with two ICC curves. The smaller ICC curve fits the reported data for 50 days preceding May 5, 2020. The larger ICC curve fits the reported data for 14 days preceding June 5, 2020. COVID-19 case data provided by The COVID Tracking Project at {\sl The Atlantic} under a CC BY 4.0 license \cite{CVDT}.}
\end{figure}

In the case of an epidemic with more than one wave, or in the presence of social distancing or other mitigation efforts, the resulting ICC curve typically no longer resembles the simple shape shown in the right panel of Figure \ref{fig:EPI_ICC}. However, even in such a situation, different ICC curves can still be locally fitted to the data: for a specified set of consecutive data points, the optimal parameters $(\beta, \gamma, N, \kappa)$ are found by minimizing the $\ell_2$ norm between the ICC curve and the selected data points, while keeping $R_0$ bounded (for COVID-19, we set $\max(R_0)=4$). The result of such a procedure is illustrated in Figure \ref{fig:ICC_AZ} for the state of Arizona. The final sizes of the two ICC curves plotted on this figure differ by an order of magnitude, consistent with the significant increase in the number of cases after social-distancing measures were relaxed.
Because the larger ICC curve in Figure \ref{fig:ICC_AZ} is shifted along the $C$ axis, it crosses the $C = 0$ axis at a negative value of $\mathcal I$, which corresponds to a value of $\kappa$ larger than 1 in \eqref{eq:ICC_Curve}. 

An important advantage of ICC curves is that since they can be fitted to incidence data locally, 
they can also be used to produce short-term forecasts of the course of an outbreak: barring significant changes in mitigation efforts, future incidence is expected to oscillate about the ICC curve that best fits recent data. In what follows, we explain how to use variational data assimilation to identify parameters and quantify forecast uncertainty.

\subsection{Estimation of prior distributions}
\label{sec:Priors}
\begin{figure}[hbtp]
\centerline{\includegraphics[width=1.1\linewidth]{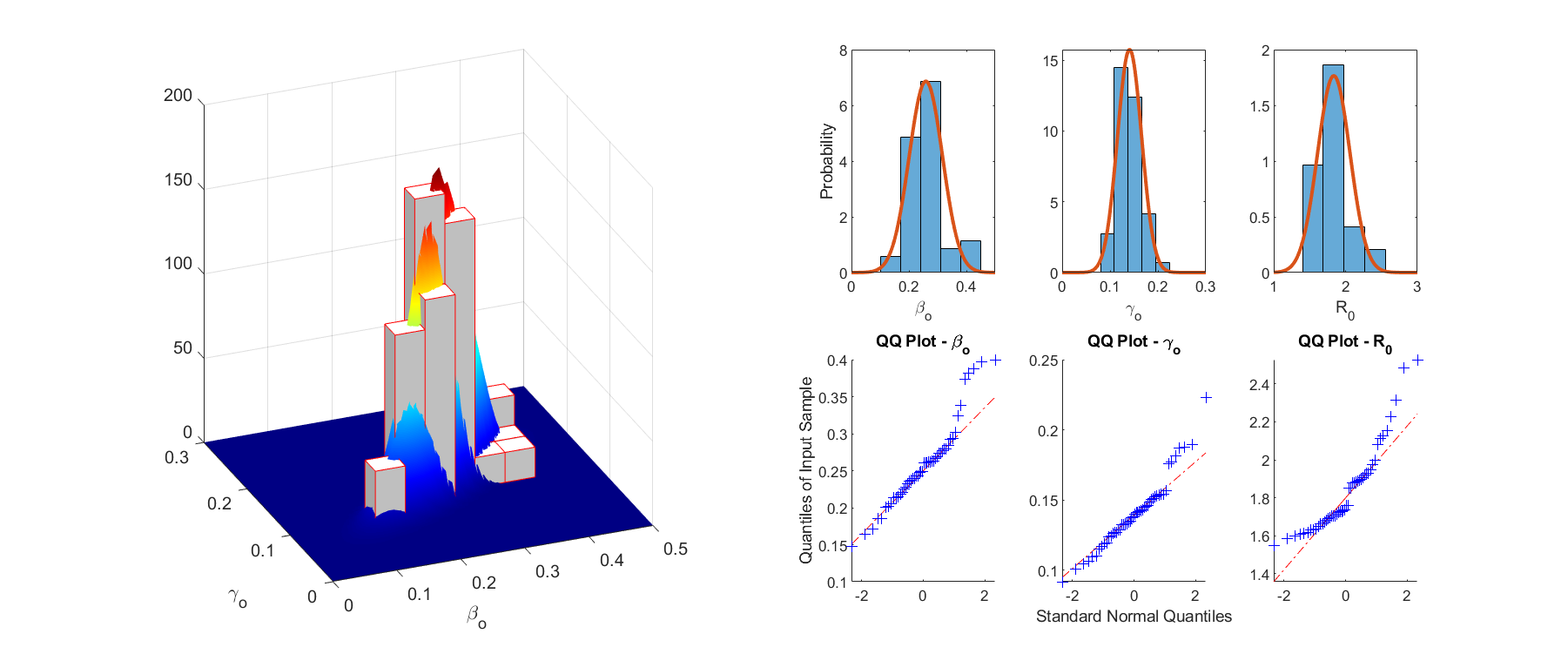}} 
\caption{\label{fig:Priors} Prior distribution for $\beta$ and $\gamma$ used in the variational data assimilation step (using 4/27/2020 data). Left: joint histogram of the optimal parameters $\beta_o$ and $\gamma_o$ for the US and all of the US states with more than 1000 cases on April 1, 2020, except VT and ND. The bivariate normal distribution of mean vector $\mu_0 = \big(\langle \beta_o \rangle, \langle \gamma_o \rangle \big)$ and covariance matrix $B_0 = \text{Cov}(\beta_o,\gamma_o)$ is shown for comparison. Right, top row: corresponding histograms of $\beta_o$, $\gamma_o$, and $R_0 = \beta_o / \gamma_o$. The solid curve shown in each panel is a normal distribution of mean and variance equal to the sample mean and sample variance respectively. Right, bottom row: quantile-quantile plots for $\beta_o$, $\gamma_o$, and $R_0$.}
\end{figure}

Priors on parameters used in the variational data assimilation step of EpiCovDA are identified with EpiGro v.2.0 as follows. For any US state that had more than 1000 cases on April 1, 2020, we compute an optimal set of parameters $(\beta, \gamma, \kappa)$ for a range of values of $N$, according to the formulas provided in \cite{Lega20}. We then select the value of $N$ that minimizes the $\ell_2$ error between the ICC curve and the corresponding data points. This defines a set $\mathcal S_o$ of optimal values $\{\beta_o, \gamma_o, \kappa_o, N_o\}$. The prior on the parameters $\beta$ and $\gamma$ is chosen to be a bivariate normal distribution of mean vector $\mu_0 = \big(\langle \beta_o \rangle, \langle \gamma_o \rangle \big)^T$ and covariance matrix $B_0 = \text{Cov}(\beta_o,\gamma_o)$, where $\langle \beta_0\rangle$ and $\langle \gamma_0 \rangle$ are the means of $\{\beta_0\}$ and $\{\gamma_0\}$, respectively. This distribution is shown in the left panel of Figure \ref{fig:Priors}, together with the normalized two-dimensional histogram of the points $(\beta_o, \gamma_o) \in {\mathcal S}_o$. Two outliers, VT and ND, are not included in this plot. The corresponding histograms, estimated marginal distributions of $\beta_o$ and $\gamma_o$, and quantile-quantile (QQ) plots are also 
displayed, together with the histogram, QQ plot, and estimated Gaussian distribution of $R_0$. Linear regression between $\beta_o$ and $\gamma_o$ gives an overall estimation of $R_0$ for the initial phase of the COVID-19 outbreak in the US equal to $1.841$. The data used to estimate the prior distribution were downloaded on 4/27/20, which is before the evaluation period presented here.

\subsection{Variational data assimilation} \label{sec:VDA}

Often called ``4D-Var'' from its origins in numerical weather prediction, variational data assimilation (VDA)\cite{Rhodes09,Kalnay07} uses a Kalman Filter-like loss function \cite{Kalman60,Law15,Reich15}
that includes consecutive time observations and penalizes differences between model and observations, as well as parameter departure from prior estimations.
In this section, we adapt the general methodology of Bayesian data assimilation 
\cite{Law15,Reich15} to the context of disease outbreaks. The ICC perspective introduced above makes it possible to describe the dynamics in terms of a discrete, deterministic dynamical system with additive noise. We keep the discussion as general as possible by including two sources of noise, model noise and observation noise. Specific assumptions related to our model are introduced at the end of this section. 

For a given location, we consider the discrete dynamics of the cumulative number of cases $C_k$, where $k \in {\mathcal K}$ is an index that measures time in days and $\mathcal K$ is a window of fixed length. We assume that the indices in $\mathcal K$ correspond to consecutive days and write for $k, k+1 \in {\mathcal K}$,
\begin{equation}
C_{k+1}
= C_k + F(C_k,\theta) + \xi_k,\label{eq:C_map}
\end{equation}
where $\theta=(\beta,\gamma,N,\kappa)$, and $\{\xi_k, k \in {\mathcal K}\}$ are independent identically distributed realizations of a mean zero and variance $\sigma^2_\xi$ normal random variable that accounts for model errors. The map from $C_k$ to $C_{k+1}$ is defined by the ICC curve introduced in Section \ref{sec:EpiGro}. We thus write $F(C,\theta) = {\mathcal I}(C)$, where ${\mathcal I}(C)$ is given by \eqref{eq:ICC_Curve} with parameters $\beta$, $R_0 = \beta / \gamma$, $N$, and $\kappa$. If the disease followed the SIR model exactly, then each $\xi_k$ would be zero and the dynamics of $C$ would be fully deterministic. Later on we make this assumption to simplify the data assimilation step.

Let $X_k = (C_k, \theta)$ be a multidimensional state variable that includes the quantity to be modeled and parameter values. We assume that for $k \in {\mathcal K}$, the parameters $\theta$ are unknown but constant, i.e. that the period of time over which the data is assimilated is short enough for any changes in mitigation efforts to be negligible. We may therefore define a map $\mathcal F$ between $X_k$ and $X_{k+1}$ as
\begin{align*}
X_{k+1} & = {\mathcal F}(X_k) + \Xi_k, \quad \Xi_k = (\xi_k, 0),\\
 {\mathcal F}(X_k) & = (C_k+F(C_k,\theta), \theta).\\
\end{align*}
Because $X_{k+1}$ only depends on $X_k$ and $\Xi_k$, where $\Xi_k$ is independent of the dynamics of $\{X_j\}_{j=1}^k$, the process $\{X_j\}_{j\ge1}$ has Markovian structure. As such, the probability density function for the collection
$$X = (X_{k_m-1},X_{k_m}, \cdots, X_{k_M}),$$
where $\displaystyle k_m = \min\mathcal K$ and $\displaystyle k_M = \max\mathcal K$, is given at $X=x$ by
$$p(x)=p(x_{k_m-1},x_{k_m},...,x_{k_M})=\left(\prod_{k=k_m}^{k_M}p(x_k|x_{k-1})\right)p(x_{k_m-1}),$$
where we assume
$$p(x_{k_m-1})=\pi(\theta|c_{k_m-1})\,p(c_{k_m-1})=\pi(\theta)\,p(c_{k_m-1}).$$
The final assertion that $\theta$ is independent of $C_{k_m-1}$ reflects the assumption that data from sufficiently far in the past may be governed by a different parameter vector $\theta'$ which may not actually provide information on the value of $\theta$.
In the last equation, $\pi(\theta)$ represents a prior on model parameters $\theta=(\beta,\gamma,N,\kappa)$. We assert a prior of the form
$$\pi(\beta,\gamma,N,\kappa)=\pi_1(\beta,\gamma)\,\pi_2(N,\kappa)$$
where $\pi_1$ is a multivariate $\mathcal N(\mu_0,B_0)$ density. This assumes independence between $(\beta,\gamma)$ and $(N,\kappa)$. The particular choice of $\pi_1$ is discussed in the previous section and the choice of $\pi_2$ will be discussed later.

Moreover, because $X_{k+1}$ is the sum of two independent random variables ${\mathcal F}(X_k)$ and $\Xi_k$, we may write
\begin{align*}
p\left(x_{k+1} \vert x_k \right) & \propto \exp\left(-\frac{1}{2 \sigma_\xi^2}\, (c_{k+1} - c_k - {\mathcal I}(c_k))^2 \right).
\end{align*}
The goal of the variational data assimilation is to estimate the posterior mode of $\theta$ for use in prediction.

Epidemiological reports typically provide consecutive observations of $C_j$ and/or equivalently of $I_j=C_j-C_{j-1}$ for $j = 1, 2, \cdots$. A reported measurement $G_k$ of $I_k$ results from adding observation noise $\eta_k$ to the first coordinate of $X_k-X_{k-1}$, 
$$G_k = I_k + \eta_k.$$
For simplicity, we assume that the $\eta_k$ are independent and normally distributed with mean zero and variance $\sigma^2_k$, $\eta_k \sim \mathcal N(0,\sigma_k^2)$. We therefore write the conditional density of $G|X$ at $g|x$
\begin{align*}
p(g \vert x) &= \prod_{k=k_m}^{k_M} p(g_k \vert x) = \prod_{k=k_m}^{k_M} p(g_k \vert c_k-c_{k-1})\\
& \propto \prod_{k=k_m}^{k_M} \exp\left(-\frac{1}{2 \sigma^2_k}\, (g_k - c_k+c_{k-1})^2 \right),
\end{align*}
where $G = (G_{k_m}, \cdots, G_{k_M})$. Although we allow the variance $\sigma^2_k, \ k \in {\mathcal K}$ to vary, we still assume independence of $\eta$ from $X$ at any point in time. With Bayes' theorem, we may now compute the posterior of $\theta$ given the observations $G=g$:
\begin{align*}
	p(\theta|g) & =\int p(\theta,c|g)dc=\int p(x|g)dc \propto\int p(g|x)p(x)dc\\
	 & \propto\int \exp\big(- \frac{1}{2} \mathcal L(\theta|g,c)\big) \pi_2(N,\kappa) \, p(c_{k_m-1})dc,
\end{align*}
where $c=(c_{k_m-1},c_{k_m},...,c_{k_M})$,
\begin{align*}
\mathcal L(\theta|g,c) =&  \sum_{k=k_m}^{k_M} (g_k - c_k+c_{k-1})^2 / \sigma^2_k + \frac{1}{\sigma_\xi^2} \sum_{k = k_m}^{k_M} (c_k-c_{k-1} - {\mathcal I}(c_{k-1}))^2\\
& + \big(\beta - \langle \beta_o \rangle, \gamma - \langle \gamma_o \rangle \big) B_0^{-1}  \big(\beta - \langle \beta_o \rangle, \gamma - \langle \gamma_o \rangle \big)^T \ge 0,
\end{align*}
and $B_0$ is the covariance matrix of the parameters $\beta$ and $\gamma$ estimated in the previous section. When $\pi_2$ is chosen to be uniform over a region $\mathbf{a} \times \mathbf{b}$, the posterior mode of $\theta$ is given by
$$\hat\theta(G)=\arg\max_\theta p(\theta|G)=\arg\max_\theta \int\exp\big(- \frac{1}{2} \mathcal L (\theta|G,c)\big)p(c_{k_m-1})dc,$$
assuming uniqueness of the maximizer. Additionally, if we neglect model noise ($\xi_k = 0$ in \eqref{eq:C_map}), then $c_k-c_{k-1} = {\mathcal I}(c_{k-1})$, making $c$ a function of $\theta$ and of the initial condition $c_{k_{m-1}}$. As a consequence, the posterior mode reduces to 
\[
\hat\theta(G) =\arg\max_\theta p(\theta|G)
=\arg\min_\theta \big(\mathcal L_{\mathcal K} (\theta|G)\big),
\]
where we have assumed that $p(c_{k_{m-1}}) = \delta(c_{k_{m-1}} - \textbf{C}_{k_{m-1}})$, i.e. $c_{k_{m-1}}$ is known, and
\begin{align*}
\mathcal L_{\mathcal K}(\theta|g) =&  \sum_{k=k_m}^{k_M} (g_k - {\mathcal I}(c_{k-1}))^2 / \sigma^2_k\\
& + \big(\beta - \langle \beta_o \rangle, \gamma - \langle \gamma_o \rangle \big) B_0^{-1}  \big(\beta - \langle \beta_o \rangle, \gamma - \langle \gamma_o \rangle \big)^T \ge 0.
\end{align*}
In the above expression for $\mathcal L_{\mathcal K}(\theta | g)$, the $c_{k-1}$ should be computed from $\theta$ and $C_{k_{m-1}}$ by iterating the map $\mathcal F$, i.e. 
\[
c_{k-1}= C_{k_{m-1}} + \sum_{j = k_m}^{k-1} {\mathcal I}(c_{j-1}).
\]
However, approximating the value of $c_{k-1}$ with
\[
C_{k-1}= \mathbf{C}_{k_{m-1}} + \sum_{j = k_m}^{k-1} G_j 
\]
 was seen to be more efficient and yield comparable or improved forecasts. Because the ICC map $\mathcal I$ is nonlinear, the landscape defined by $\mathcal L_{\mathcal K}(\theta|G)$ is likely to be intricate. In practice, we compute a local minimizer of
\begin{align*}
\mathcal L_{\mathcal K}(\theta|G) =&  \sum_{k=k_m}^{k_M} (G_k - {\mathcal I}(C_{k-1}))^2 / \sigma^2_k\\
& + \big(\beta - \langle \beta_o \rangle, \gamma - \langle \gamma_o \rangle \big) B_0^{-1}  \big(\beta - \langle \beta_o \rangle, \gamma - \langle \gamma_o \rangle \big)^T \ge 0
\end{align*}
found with the MATLAB function {\tt fminsearch} initialized at the parameter values $(\beta_0,\gamma_0) = (\langle\beta_o\rangle,\langle\gamma_o\rangle)$, $N_0$ as 1/3 of the state population, and $\kappa_0 = 1 + 100/N_0$. We do not impose any bounds on the range $\mathbf{a} \times \mathbf{b}$ where the distribution $\pi_2$ is supported, although we enforce $N \ge C_{k_M}$, $\beta > 0, \ \gamma > 0$, and $R_0 \le 20$.

The resulting assimilated vector of parameters, $\hat{\theta}$ is then used to create a single prediction for the trajectory of the outbreak through numerical integration of \eqref{eq:ICC_Curve} with a pre-specified initial condition, for example $C_{k_M}$. This numerical integration yields $C(t)$ for $t\ge k_M$. To obtain the forecasted incidence $I(t)$ we use
$$I(t) = \mathcal I(C(t-1)), $$
where if we want the daily forecasted incidence for approximately the next month, we would use $t = k_M+1, k_M+2, k_M+3,\hdots, k_M+31$.

\subsection{Pseudo-Observations} \label{sec:pseudo-observations}

The Bayesian approach described in the previous section provides a construction of the pdf of $\theta | G$, namely 
$$ \pi(\theta | G) \propto \exp(- \mathcal L_{\mathcal K}(\theta | G)/2).$$
However, due to nonlinearities in the expression of $\mathcal L_{\mathcal K}(\theta | G)$, specifically because of terms of the form $\mathcal{I}(C_{k-1})$ where $\mathcal{I}$ applies \eqref{eq:ICC_Curve} with parameters given by $\theta$, sampling this distribution would require a computationally expensive procedure. Instead, we generate pseudo-observations, $\{G^i\}_i$, $G^i=(\tilde{G^i}_k)_{k\in\mathcal K}$, and repeat the VDA steps with perturbed values of $\langle \beta_o \rangle$ and $\langle \gamma_o \rangle$ \cite{Chen12,Bardsley12,Bardsley14} to obtain an ensemble of assimilated vectors of parameters $\{\hat \theta_i\}_i$.

To generate the pseudo-observations, we first smooth the reported incidence data by twice averaging over a 7-day moving window, as described in the Supplementary text. 
The resulting incidence $S_k$ is assumed to be close to the true state of the system on day $k$, and thus close to the true ICC curve. As a consequence, the smoothed incidence values may be used to estimate the initial condition $\mathbf{C}_{k_m-1} = \sum_{j = 1}^{k_m -1} S_j$. Then, for each $k \in \mathcal K$, we generate a pseudo-observation of $G_k$ as 
$$\tilde{G}_k = S_k + \eta_k,$$
where $\eta_k$ is sampled from $\mathcal{N}(0,S_k)$, so that $\{\tilde{G}_k\}_{k\in\mathcal K}$ is comparable to the set $\{G_k\}_{k\in\mathcal K}$ from the VDA section. We thus obtain a new set of ``observations,'' $G^1 = (\tilde{G^1}_{k_m},\tilde{G^1}_{k_m+1},\hdots, \tilde{G^1}_{k_M})$ which, when combined with the VDA methodology, leads to a new assimilated vector of parameters $\hat{\theta}_1$. Repeating this process many times yields a collection of pseudo-observations $\{G^i\}_i$ and assimilated parameter vectors $\{\hat \theta_i\}_i$. It is assumed that the distribution of the $\{\hat \theta_i\}_i$ provides an approximation of the true parameter distribution $\pi(\theta | G)$.

\subsection{Ensemble Generation} \label{sec:EnsembleGeneration}

Each assimilated state $\hat{\theta}_i$ creates an incidence forecast as described at the end of Section \ref{sec:VDA}, using the specific initial condition
\[
C_{k_M} = \sum_{j = 1}^{k_M} S_j.
\]
We repeat this sampling and forecasting process for three choices of $\mathcal K$: last 3 days, last 5 days, and last 14 days. The use of recent data points allows adjustment for changes in mitigation efforts and reporting to be quickly reflected in the estimate of $\theta$.

The resulting ensemble $\{I_i(t)\}_{i = 1}^{\Nens}$ of predicted incidence values is used to create a probabilistic forecast described, for example, by a histogram or quantiles for each future $t$.

\subsection{COVID-19 case counts forecasts} \label{sec:case_forecasts}
For each discrete time $t$, we augment the ensemble of incidence forecasts $\{I_i(t)\}_{i = 1}^{N_{ens}}$
with an equal number of samples from a normal distribution centered at $\mut$ with variance $\sigmat = \zeta \cdot \max\{\mut,v_t\} $, where
\[
\mut = \frac{1}{\Nens} \sum_{i=1}^\Nens I_i(t), \qquad v_t = \frac{1}{\Nens-1}\sum_{i=1}^\Nens (I_i(t) - \mut)^2,
\]
and $\zeta$ is an inflation parameter that can be tuned for calibration. Here, $\zeta$ is forecast-specific and defined as
\begin{equation} 
\zeta = \max \left\{
q_{t_0}/v_{t_0}, 1
\right\}
\label{eq:zeta_1}
\end{equation}
where $t_0$ refers to the first day of the forecast and
\begin{equation}
    q_{t_0} = \text{Var}(\{S_k - G_k : k = t_0-10, t_0-9,...,t_0-1\}).
    \label{eq:zeta_2}
\end{equation}
Our choice to augment the day-ahead forecast ensemble was motivated by 1) a desire to add support in the histogram of $\{I_i(t)\}_{i = 1}^\Nens$ around the ensemble mean, and 2) a desire to augment with a normal distribution with the same or greater variance as the original ensemble. For $\zeta=1$, the expression for $\sigmat$ reflects the belief that the observed incidence is likely to be $\textit{Poisson}(\mut)$, and when $\mut$ is large, $\mathcal N(\mut,\mut)$ is a good continuous approximation for the  $\textit{Poisson}(\mut)$ distribution. We allow for $\zeta>1$ when the variance of the recent reported data is large compared to the variance of the forecast ensemble. This reduces over-confidence in forecasts when the data are highly variable. Sampling from the $\mathcal N(\mut,v_t)$ distribution may result in negative ``observations'' of incidence, so we adjust for this by setting the negative sample to $0$. In most cases, this adjustment is unnecessary due to the size of $\mut$ and $v_t$.

After augmentation, we have an ensemble of $2 \Nens$ day-ahead point forecasts for the entire duration of the forecasting period. These values are combined into probabilistic day-ahead case forecasts, described by 23 quantiles, $q_\alpha$ for $\alpha \in \{0.01, 0.025, 0.05, 0.1, 0.15, \hdots,$ 0.85, 0.9,0.95, 0.975, 0.99$\}$. Each quantile, $q_\alpha$, is smoothed using a moving average across a 5 day window and rounded to the nearest integer. As a final check to correct for the (rare) possibility that smoothing might remove the monotonicity of $\{q_\alpha\}$ at a given day, we reorder the quantiles so they are monotonically increasing.

\subsection{COVID-19 deaths forecasts} \label{sec:death_forecasts}

\begin{figure}[hbtp]
\centerline{\includegraphics[width=1.0\linewidth]{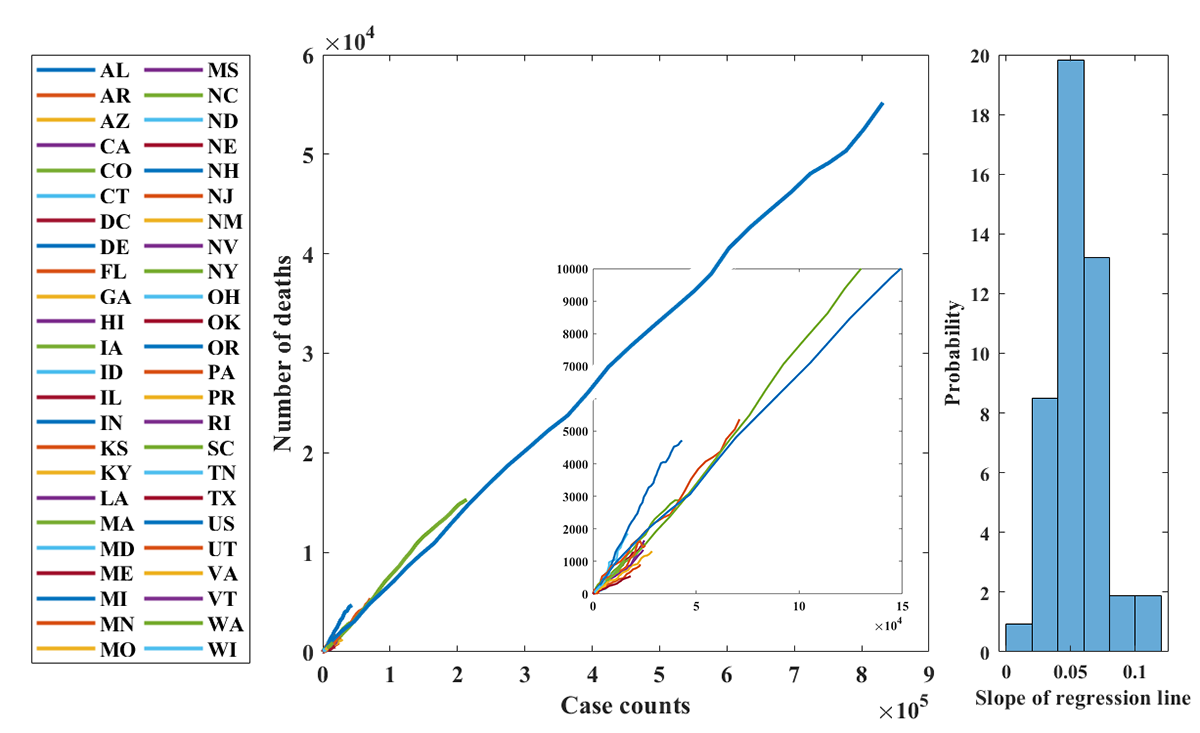}} 
\caption{\label{fig:Case_Cts_Deaths} Left: Number of cumulative COVID-19 deaths reported on day $t+\tau$ as a function of the cumulative case counts on day $t$ for the entire US and all of the states that had registered more than 500 cases and at least 10 deaths by 5/17/2020, based on data from the COVID Tracking Project. The inset is an enlargement of the region near the origin. Right: Histogram of the slopes of the linear regressions between $D(t + \tau)$ and $C(t)$ for the data shown in the left panel. COVID-19 case data provided by The COVID Tracking Project at {\sl The Atlantic} under a CC BY 4.0 license \cite{CVDT}.}
\end{figure}

In the case of COVID-19 in the US, a striking relationship in early outbreak data is observed between daily case counts and reported deaths. Specifically, for each state, we are able find a value of $\tau$ in days such that the relationship between $D(t + \tau)$ and $C(t)$  is almost linear, where $D(t)$ and $C(t)$ are the cumulative number of deaths and the cumulative number of cases on day $t$, respectively. Figure \ref{fig:Case_Cts_Deaths} shows the resulting plots for the entire US, as well as for states that had more than 500 cases and 10 deaths by May 17th, 2020. In each case, $\tau$ is chosen to optimize the correlation (minimum RMSE) between $D(t+\tau)$ and $C(t)$. The value of $\tau$ varies from state to state, between 3 and 12 days. The right panel of Figure \ref{fig:Case_Cts_Deaths} shows a normalized histogram of the slopes $a$ of the linear regressions $D(t+\tau) = a \, C(t)$, and suggests an initial case-fatality ratio of about 5\%.

As a consequence of the strong correlations observed in these data, forecasts for deaths are made as a proportion of delayed case counts forecasts, $D(t) = a\, C(t - \tau)$. Location and date specific delays and regression slopes $a$ are calculated for each forecast to account for differences in reporting and testing over time and region. Specifically, a linear regression is performed at each location between the sum of delayed and smoothed case incidence values and the sum of smoothed death incidence values over the most recent period of $N_c$ days for which data are available. The default $N_c$ is set at 10 days. Exceptions are made in the cases of AK ($N_c = 20$), HI ($N_c = 20$), VT ($N_c = 50$), and if more than five of the last 10 days had 0 deaths reported ($N_c = 20$). We optimize these regressions on the delay $\tau$, which takes values between 0 and 21 days. Larger values of $\tau$ relate to lengthier illness before death, potentially due to improvement in treatment of hospitalized patients. For death predictions that occur within $\tau$ days of the forecast date, for which the value of $C(t - \tau)$ can be calculated from the data, we use a normal distribution $\mathcal N(D(t),D(t))$, centered at the proportion of the appropriate smoothed delayed cases $D(t) = aC(t-\tau)$.

\section{Scoring Methodology} \label{Sec:Scoring}

We evaluate EpiCovDA forecasts across $N_w = 20$ weeks. Every week, forecasts are made with data released for Sunday, to predict 1-,2-,3-, and 4-week ahead case and death cumulative numbers, where the target week day is always Saturday. This was chosen to make forecasts comparable to those displayed and submitted to the COVID-19 Forecast Hub. We provide two different scoring metrics, defined in the sections below: absolute error for point forecasts, and interval scoring at the $\alpha = 0.05$ level for probabilistic forecasts.

\subsection{Point Forecast Scoring} \label{sec:point_scoring}

We define the point forecast for a given target to be the median of the corresponding probabilistic forecast described in the case and death forecasting sections. Consequently, we use the absolute error to evaluate these forecasts, since such a scoring function is consistent for the median   \cite{Gneiting11}. Moreover, this also guarantees that the resulting scoring rule is proper. The absolute error for a location-specific target $T$ of a forecast made on day $M$ is 
$$ \text{Err}(M,T) = | m(M,T) - y(T) |,$$
where $m(M,T)$ is the median of the forecast made on day $M$ and $y(T)$ is the truth value of the target $T$ according to The COVID Tracking Project \cite{CVDT}. We report $\text{Err}(M,T)$ per 100,000 people. Absolute errors (per 100,000 people) are summarized by calculating their mean (MAE) and median (MedAE) over $N_w$ weeks and the 53 forecasted locations.

\subsection{Probabilistic Forecast Scoring} \label{sec:prob_scoring}
We use the interval scoring method described in \cite{Bracher20,Gneiting07}.  Specifically, the interval score of the $(1-\alpha)\times 100\%$ prediction interval is defined to be
$$\text{IS}_\alpha(M,T) = (u -  l) + \frac{2}{\alpha} \times (l - y)\times \mathbf{1}(y<l) + \frac{2}{\alpha} \times (y-u) \times \mathbf{1}(y>u),$$
where $l$ and $u$ are the lower and upper bounds, respectively of the central $(1-\alpha)\times 100\%$ prediction interval for the forecast made on day $M$ for target $T$ and $y$ is the corresponding truth for target $T$. Interval scores are also reported per 100,000 people.

\subsection{Calibration}

We furthermore report the forecast calibration as measured by interval coverage. Specifically, for the 10\%, 20\%,$ \hdots,$  90\%, 95\%, 98\% central prediction intervals as given by the forecast quantiles, we calculate the proportion of times the corresponding interval captured the truth. A forecast can be considered well-calibrated when the coverage rate is close to the interval size, e.g., when the 95\% prediction interval captures the truth about 95\% of the time. A perfectly accurate forecast will always have 100\% coverage; an over-confident forecast will have lower than nominal coverage; and an under-confident forecast will have above nominal coverage.

\section{Forecasting performance for COVID-19 in the US}
\label{Sec:US_Forecasts}
\begin{figure}[htbp]
\begin{center}
	\includegraphics[width=.7\linewidth]{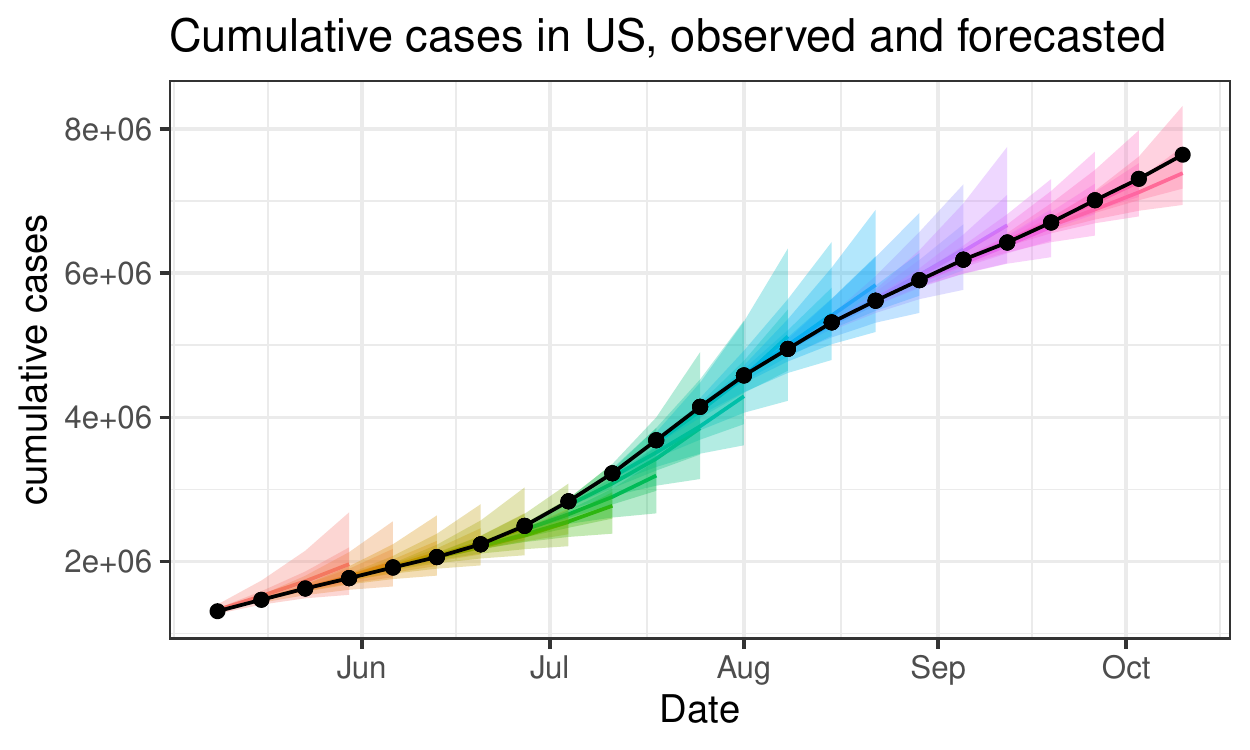}\\
	\includegraphics[width=.7\linewidth]{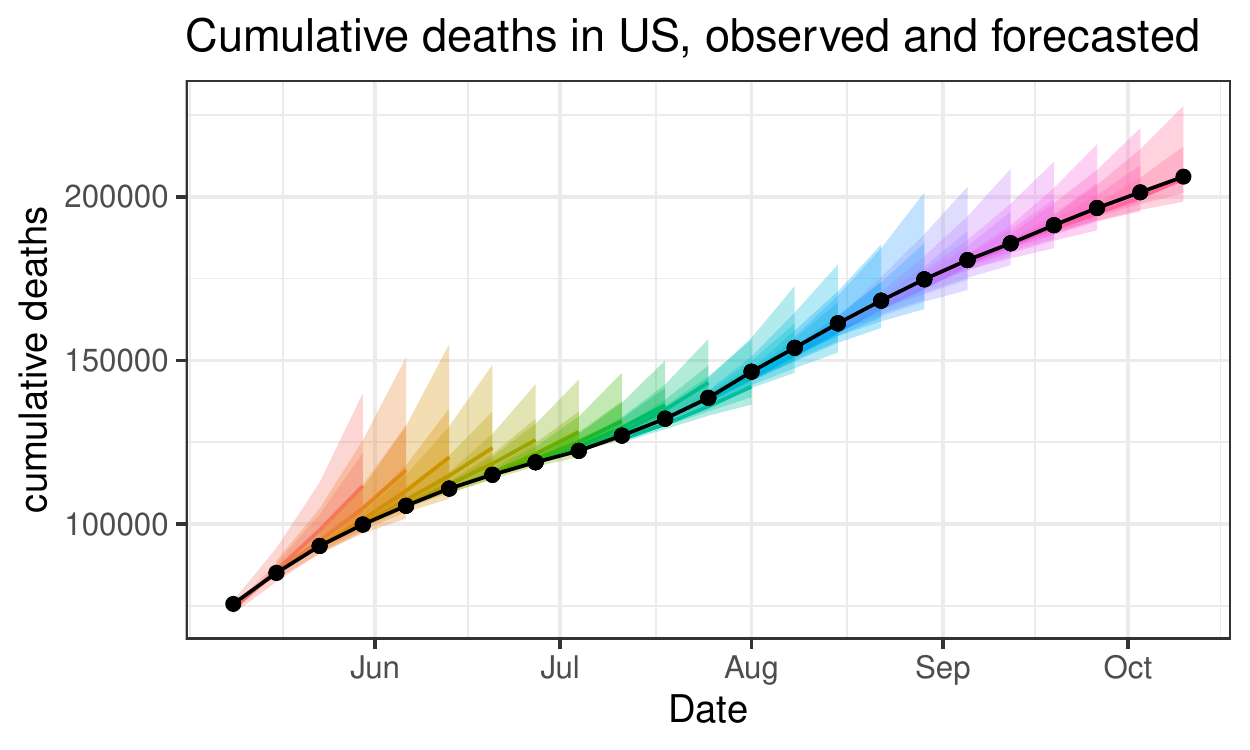}
\end{center}
\caption{\label{fig:fans_US} EpiCovDA weekly US forecasts. Probabilistic forecasts are shown in the form of the median (solid colored line) and the 50\% and 95\% central prediction intervals. The truth is the black solid line in each figure. Top: cumulative case count forecast. Bottom: cumulative deaths forecast. COVID-19 case data provided by The COVID Tracking Project at {\sl The Atlantic} under a CC BY 4.0 license \cite{CVDT}.}
\end{figure}

\begin{figure}[htbp]
\centering 
	\includegraphics[width=0.95\linewidth]{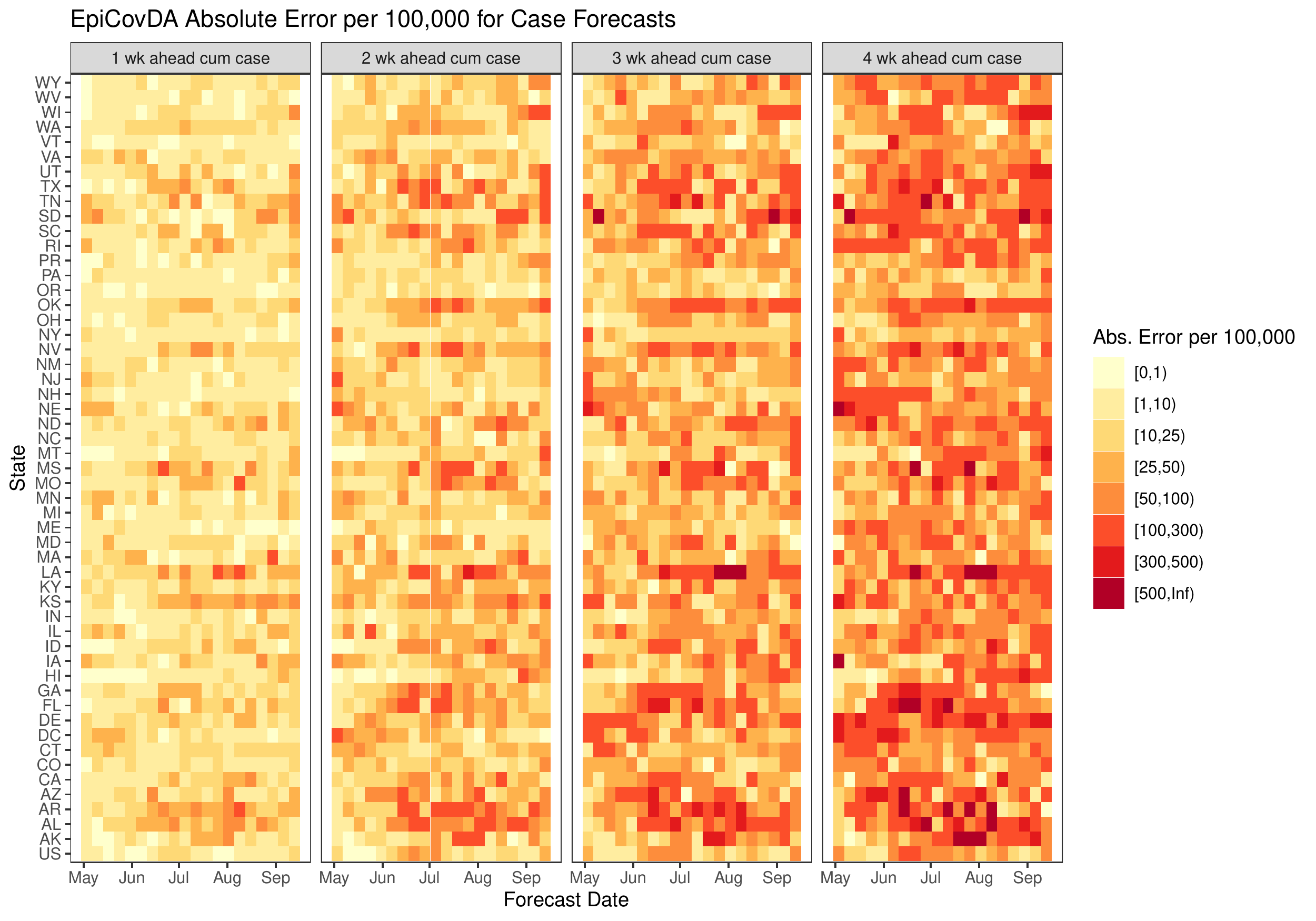}
\caption{\label{fig:HM_CC} Absolute error for case count forecasts, one through four weeks ahead of the forecast date. Each state corresponds to a row and each rectangle is a forecast week. The color scale ranges from less than 1 to more than 500 cases per 100,000 population.}
\end{figure}
For the analysis presented here, and for each US state, D.C., and Puerto Rico, a single data stream, downloaded from the COVID Tracking Project \cite{CVDT} on 11/16/2020, provides both the input data used by EpiCovDA to make its forecasts (only data prior to each forecast date are used), and the truth to which forecasts are compared weekly, for a period of 4 weeks after the forecast date. Because they were released after the evaluation period (mid-May to mid-October 2020), these data include ``backfill'' (retroactive) corrections. They are therefore more accurate and provide a more stable environment to evaluate the performance of the model. They also include the summer of 2020, which witnessed a wave of cases in the US. The priors however were obtained from ``live'' data (as of 4/27/20) as the COVID-19 pandemic developed in the US. The reader is referred to Section \ref{sec:data_sources} for further description of data sources, and to \cite{Cramer21} for an evaluation of forecasts made by the present and 26 other models, as well as a baseline model, on ``live'' weekly reports over an extended period of time.  In what follows, the 52 locations where forecasts are made are referred to as ``states.''

Figure \ref{fig:fans_US} shows EpiCovDA forecasts for cumulative case counts (top) and cumulative deaths (bottom) in the US, over 20 weeks from mid-May to mid-October 2020. These probabilistic forecasts, obtained by combining state-level predictions, are displayed in the form of 50\% and 95\% central prediction intervals (colored ``fans''); the point forecast corresponds to the median of the forecasted sample, shown as a solid colored curve for each 4-week forecasting period. Similar plots for the state-level forecasts as well as incident case and death national-level forecasts are provided in the Supplementary Materials. 
As detailed in Section \ref{sec:case_forecasts}, cumulative forecasts are obtained by adding incidence forecasts to an estimate of the current cumulative number of cases.  Predictions capture the truth with good accuracy, although steep increases in cumulative numbers are often associated with under-predictions (for case counts; see top panel of Figure \ref{fig:fans_US}) or over-predictions (in the case of deaths; see bottom panel).

\begin{figure}[ht]
\centering 
	\includegraphics[width=0.95\linewidth]{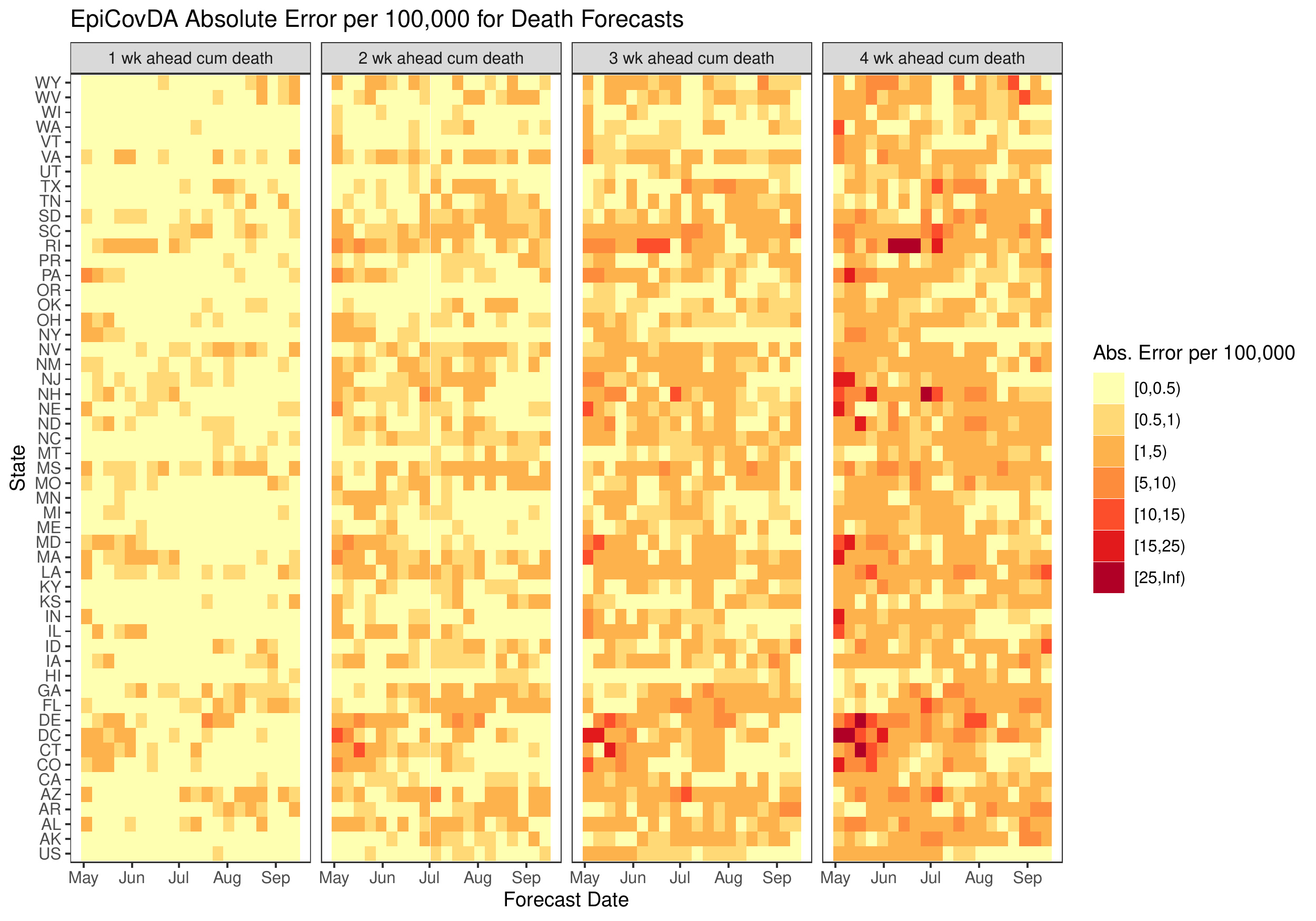}
\caption{\label{fig:HM_D} Absolute error for deaths forecasts, one through four weeks ahead of the forecast date. Each state corresponds to a row and each rectangle is a forecast week. The color scale ranges from less than 0.5 to more than 25 deaths per 100,000 population.}
\end{figure}

For each state and target type (case counts or deaths, forecasted 1 to 4 weeks ahead of time), we report the absolute error (AE) as a measure of point forecast performance, which is a consistent scoring function for the median \cite{Gneiting11}. Figure \ref{fig:HM_CC} displays the AE on case counts per 100,000 population for each of the state-level forecasts and for the US
(bottom row), for each week of the 4-week forecasting period. The color range shows a typical error of less than 25 cases per 100,000 population in the first week, increasing to a few hundred per 100,000 population after 4 weeks. The AE at the US level (bottom row) is much lower due to the averaging effect of combining state results, and does not exceed 300 cases per 100,000 population even 4 weeks after the forecast date. Similar results for death forecasts (Figure \ref{fig:HM_D}) show typical AE values of less than 5 deaths per 100,000 population after one week, and no more than 25 deaths per 100,000 population after 4 weeks, with few exceptions. At the US level, the AE does not exceed 5 deaths per 100,000 population over the 4-week forecasting period. As previously indicated, the model predicts incidence over the next 4 weeks and cumulative forecasts are obtained by adding these predictions to the cumulative number estimated on the day of the forecast; as a consequence, the 1-week (respectively $p$-week) ahead AE is an estimate of the error on the number of cases (or deaths) that will be reported over a period of 1 week (respectively $p$ weeks) after the forecast date.

To evaluate the forecasted probability distribution function, we use the interval scoring method of \cite{Bracher20,Gneiting07}, which penalizes central prediction intervals that are too wide or fail to capture the truth (see details in Section \ref{sec:prob_scoring}). A perfect score of zero would correspond to a highly confident forecast (with zero variance) exactly on target. Heat maps showing the 95\% interval scores for case counts and deaths forecasts are displayed in the Supplementary Materials. The scores per 100,000 population increase as forecast targets go further into the future, and their values are higher than the corresponding AE, as expected. Scores for the entire US are significantly lower (and thus better) than for individual states, as was the case for the AE (Figures \ref{fig:HM_CC} and \ref{fig:HM_D}). 

\begin{figure}[ht]
\centering 
	\includegraphics[width=.85\linewidth]{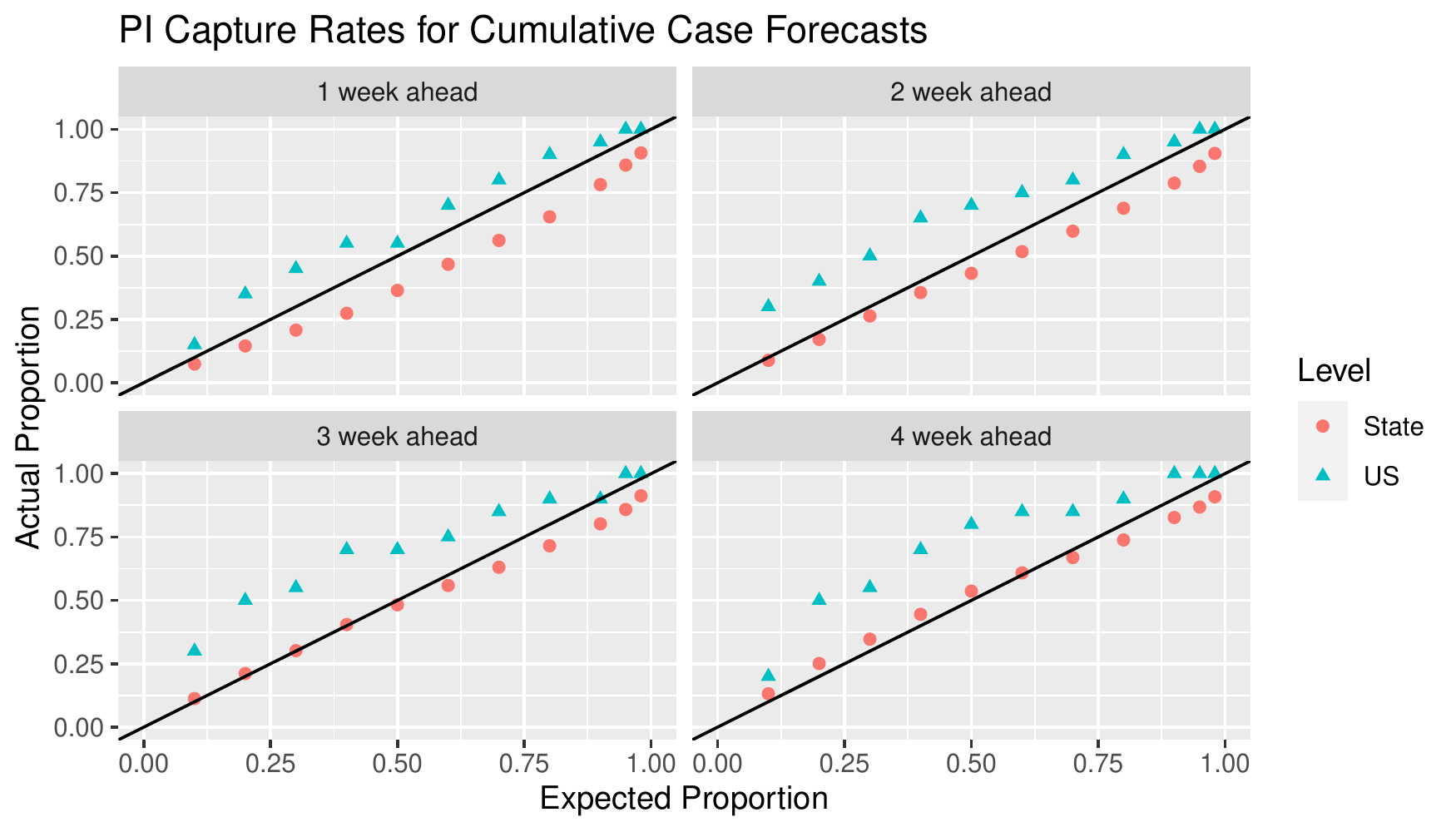}

\caption{\label{fig:CR_CC} Case count forecasts calibration. Each panel shows prediction interval capture rates for each forecast type, evaluated over all state forecasts (dots) and for the US forecast (triangles).}
\end{figure}

\begin{figure}[ht]
\centering 
	\includegraphics[width=.85\linewidth]{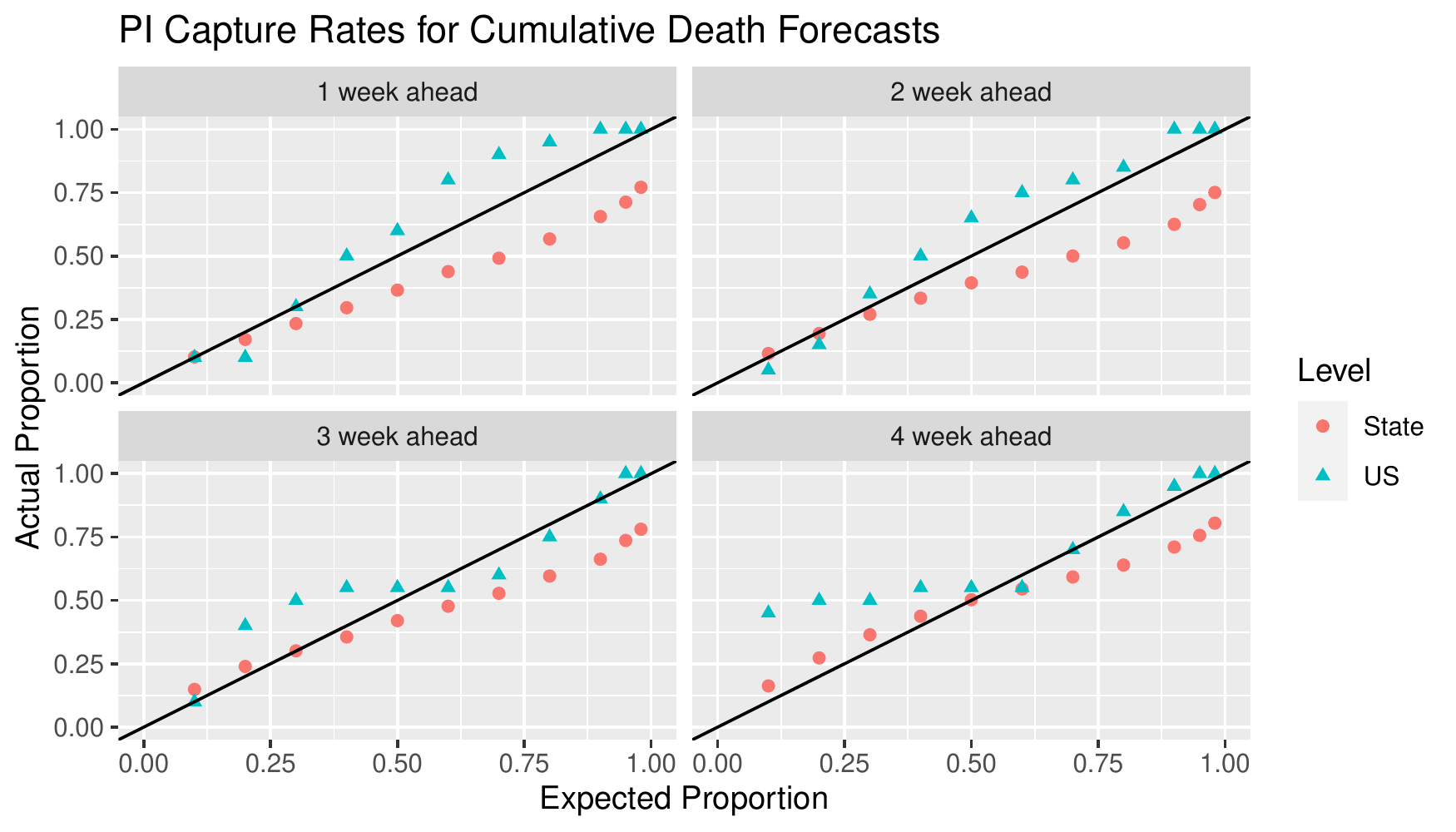}

\caption{\label{fig:CR_D} Death
forecasts calibration. Each panel shows prediction interval capture rates for each forecast type, evaluated over all state forecasts (dots) and for the US forecast (triangles).}
\end{figure}

Perhaps more intuitive than interval scores, capture rates of central prediction intervals are displayed in Figures \ref{fig:CR_CC} and \ref{fig:CR_D} for both case counts and deaths. For each value $x$ on the horizontal axis of each panel, the $y$ coordinate measures the proportion of times the truth falls within the $x$\% central prediction interval. The expectation is that $y$ should be close to $x$ since on average a random number drawn according to a given probability distribution function should fall 10\% of the times in the associated 10\% central prediction interval, 50\% of the time in the 50\% central prediction interval, etc. An over-confident forecast would typically result in $y < x$, and an under-confident forecast would correspond to $y > x$, although the latter condition would also be satisfied by a forecast that is always on target since, in such a case, all central prediction intervals would capture the truth 100\% of the time. Both figures show that EpiCovDA case counts and deaths forecasts are well calibrated.

As a final benchmark, we compare EpiCovDA point forecasts for cumulative deaths to those of the COVIDhub Ensemble \cite{Ray20,Cramer21}. The ensemble model uses the Johns Hopkins University (JHU) data \cite{JHUCD} as truth, whereas EpiCovDA is based on The COVID Tracking Project (CTP) data \cite{CVDT}. Although the two data streams are similar, small differences can nevertheless significantly affect absolute error estimates, as illustrated in Table \ref{tab:point-score-comp}. The first two rows display the mean absolute error per 100,000 (MAE) and median absolute error per 100,000 (MedAE) of a specific model, calculated over all forecasts (20 weeks and all locations); the next 8 rows show similar results for each target type (1 through 4 weeks ahead). Column 1 summarizes the performance of the version of EpiCovDA presented in this article, with CTP data used to run and score the model. Column 2 shows similar results when JHU data are used instead of CTP data. Although median errors are comparable with those listed in Column 1, an increase in mean AE is observed. Since the hyperparameters were selected using CTP data, this discrepancy reinforces the concept that the same data sources should be used to train and run any data-driven model. Column 3 shows the performance of EpiCovDA when weekly incidence forecasts (created using CTP data) are added (``aligned'') to the JHU truth and scored against JHU data. In this case, the performance is comparable to that of Column 1, both for the MAE and MedAE. The last two columns summarize the performance of the COVIDhub ensemble model when scored against JHU data (Column 4) or against the CTP data after alignment to this data source (Column 5). Column 3 is akin to the EpiCovDA forecasts that are actually submitted to the COVID-19 Forecast Hub. Comparing Columns 4 and 5 to Column 1 shows that the COVIDhub ensemble has better overall performance, clearly outperforming EpiCovDA on the 3- and 4-week ahead forecasts, has comparable performance to EpiCovDA on the 2-week ahead forecasts, but under-performs EpiCovDA for the 1-week ahead forecasts. In all cases, mean and median absolute errors are less than 3 deaths per 100,000 people. When aggregated nationally, the mean number of deaths over the 1-week ahead forecasting period was about 1.79 (median 1.71) per 100,000; over the 4-week ahead period the mean was 7.12 (median 7.06) per 100,000.

\begin{table}
\caption{\label{tab:point-score-comp} Comparison of death point forecasts generated with different data sources. Absolute errors in deaths are calculated per 100,000 population. The mean absolute error (MAE) and median absolute error (MedAE) are calculated over all 53 locations and forecast dates. (CTP) and (JHU) refer to the data sources used for forecasting and scoring, either The COVID Tracking Project \cite{CVDT} or Johns Hopkins CSSE \cite{JHUCD}, respectively. ``For alignment'' indicates that, after generation, the forecasts were aligned to the cumulative value from and scored by the indicated data source.}
\centering
\begin{tabu} to \linewidth {
			X[3,l]
		    	X[1,c]
		    	X[1,c]
		    	X[1,c]
			X[1,c]
			X[1,c]}
	\toprule
	\multicolumn{1}{c}{ } & \multicolumn{5}{c}{Model with Data Source} \\
	\cmidrule(l{3pt}r{3pt}){2-6}
Statistic & \rotatebox{90}{\shortstack{EpiCovDA (CTP)}} &    \rotatebox{90}{\shortstack{EpiCovDA (JHU)} } &  \rotatebox{90}{\shortstack{EpiCovDA (JHU  \\ for alignment only)}} &  \rotatebox{90}{\shortstack{COVIDhub Ensemble \\(as published, JHU)}} &  \rotatebox{90}{\shortstack{COVIDhub Ensemble \\ (CTP for alignment)}}\\
	\midrule
MAE, overall & 1.38 & 1.58 & 1.46 & 1.07 & 1.12\\
MedAE, overall & 0.62 & 0.67 & 0.64 & 0.52 & 0.51\\
\addlinespace
MAE, 1 wk & 0.42 & 0.50 & 0.45 & 0.46 & 0.59\\
MedAE, 1 wk & 0.23 & 0.24 & 0.24 & 0.24 & 0.25\\
\addlinespace
MAE, 2 wk & 0.86 & 1.02 & 0.92 & 0.82 & 0.90\\
MedAE, 2 wk & 0.52 & 0.52 & 0.52 & 0.46 & 0.45\\
\addlinespace
MAE, 3 wk & 1.54 & 1.78 & 1.65 & 1.25 & 1.28\\
MedAE, 3 wk & 0.90 & 0.96 & 0.94 & 0.69 & 0.68\\
\addlinespace
MAE, 4 wk & 2.70 & 3.01 & 2.82 & 1.74 & 1.73\\
MedAE, 4 wk & 1.55 & 1.63 & 1.62 & 0.92 & 0.91\\
	\bottomrule
\end{tabu}
\end{table}

A recent article by Cramer {\sl et al.} \cite{Cramer21} provides a comparative analysis of models submitted to the COVID-19 Forecast Hub for the COVIDhub ensemble, including EpiCovDA. As an ensemble model, the COVIDhub is expected to have (and has) more consistently accurate performance compared to individual model forecasts \cite{Solazzo13,Kioutsioukis14,Ray18,McGowan19,Reich19,Ray20,Cramer21}. The scores presented in \cite{Cramer21} apply to slightly different versions of EpiCovDA than discussed here (see the Supplementary Materials for how these versions evolved) and the JHU data are used as truth. Moreover, the results of \cite{Cramer21} apply to forecasts created in real-time with data that had not been backfilled, whereas the results presented in this article describe forecasts made with data that became available in November. Nevertheless, as of 02/05/2021, the evaluation of \cite{Cramer21} places EpiCovDA in the middle of the 27 evaluated models, and its  performance appears to be similar to that of MOBS-GLEAM\_COVID \cite{Chinazzi20} and the IHME-SEIR models \cite{IHME20}, which are more complex in nature and use a broader range of input data \cite{Cramer21}. 

\section{Discussion}
\label{Sec:Discussion}

EpiCovDA is a minimalist mechanistic epidemiological model that provides short-term forecasts of case counts or deaths. Although it builds on the work of \cite{Lega20}, it is the combination of ICC curves with variational data assimilation that makes this approach unique. The model consists of a core forecaster for case counts (primary output), supplemented by a linear regression module with delay that estimates deaths (secondary output). It is a local model which, in the case of COVID-19, works well at the state level. With large numbers of county-level cases, we also expect EpiCovDA to provide valuable forecasts at that smaller level of granularity. We use the word {\it minimalist} here to emphasize that estimating one future incidence trajectory of the disease involves evaluating a small number of parameters (4), and that the model input data are of the same nature as its output; in particular, case counts are predicted solely from case counts. Such structural simplicity is an advantage when faced with an emerging disease. 

EpiCovDA has four core parameters and fewer than 20 hyperparameters (reviewed in Appendix \ref{sec:Hyper}).
If necessary, decisions to change hyperparameter selections from their default values may be guided by direct comparison between forecasts and observed data. The use of \textit{variational} data assimilation increases the computational efficiency of the approach by replacing typical MCMC sampling with a sequence of searches in a 4-dimensional parameter space.
Moreover, combining data assimilation with a simple method for identifying priors directly from existing case reports (as described in Section \ref{sec:Priors}) imply that independent knowledge of epidemiological parameters is not required. Even if sufficient data are not available at first, rough estimates of the contact rate of the disease and of its basic reproduction number are sufficient to create an initial set of priors, which can then be refined as more epidemiological reports are published. Similarly, priors may be later revised to account for the presence of more transmissible variants. Importantly, fitting epidemiological data to an ICC curve alleviates parameter identifiability issues that often affect the performance of mechanistic models. 

By construction, the model produces forecasts that are consistent, both in magnitude and trends with its input data. The use of short-term (3 and 5 days) and longer-term (14 days) data assimilation windows allows EpiCovDA to react to mitigation efforts, as long as their effect is reflected in epidemiological reports. It is however implicitly assumed that current trends will continue for the duration of the forecasting period and, as a consequence, forecasts are run weekly, so that they can evolve with, and adapt to, changes in the dynamics of the disease. Nevertheless, because of the simplicity of the approach, which amounts to estimating parameter values by fitting ICC curves to the data and noisy versions thereof, forecasts are not computationally onerous. For instance, predictions for the US and 52 ``states'' run in about 5 minutes  on a MacBook Pro (2.3 GHz i5 processor, 16 GB RAM). This includes generating all of the probabilistic forecasts (i.e. repeating the VDA procedure 50 times per ``state'' and finding the optimal linear regression and delay for deaths forecasts) at all locations. Similarly, analyzing one month of data to estimate the priors takes a few minutes and only needs to be done once.

The results presented in this article show that EpiCovDA produces good forecasts for up to 4 weeks into the future. We present results on cumulative counts, which are obtained by adding incidence predictions to the cumulative estimate for the day each forecast is made. When reported in this fashion, weekly or multi-week incidence predictions benefit from ``averaging'' daily fluctuations over the corresponding periods of time. Nevertheless, when our 4-week ahead projection captures the truth, it means that the model correctly predicted the general trend a month out, even if week-to-week incidence estimates might not have been as accurate.

The goal of EpiCovDA is not to estimate the actual number of people infected, but to provide a probabilistic forecast of future counts, given recent incidence reports. As a consequence, the model cannot be used to assess the future prevalence of a disease unless essentially all existing cases are being reported. In addition, EpiCovDA provides short-term predictions, as opposed to long-term scenarios. The former may be used to guide public health decisions such as ordering personal protective equipment, staffing hospitals and clinics, deciding where to run vaccine trials, or whether curfews or strong control measures should be put in place to prevent forecasted surges. The latter often provide a rationale for longer-term policy decisions, such as shutting down businesses and schools for long periods of time, in order to ``flatten the curve.''

Because of its simplicity and minimal data requirements, EpiCovDA may easily be adapted to forecast the unfolding of other outbreaks, and be transported to other locations. The model is most useful early, when little information is known about an emerging disease; once sufficient data are available, including which mitigation efforts are being put in place and vaccine efficacy, more complex models are likely to outperform the present approach. However, when there is uncertainty or little to no information about an outbreak (as was the case for COVID-19 from May to October 2020), EpiCovDA can be put into use to provide relatively good forecasts that can guide the initial public health response. Additionally, once the priors and hyperparameters have been chosen, the model does not require significant post-forecast human adjustments and can therefore be run on a large scale with limited personnel resources. 

Finally, EpiCovDA's layered structure lends itself to the inclusion of other modules (e.g. for hospitalizations), and to the coupling of single forecasting units into a global network, for instance to revise local forecasts on the basis of global mobility or policy data. This may be accomplished by appropriately training a graph neural network and such work is currently in progress by our team. Of course, since an initial set of case count reports is needed to produce forecasts, the model is not designed to predict \textit{where} or \textit{when} the next disease might emerge. 

\appendix
\section{Model hyperparameters}
\label{sec:Hyper}
EpiCovDA has a small number of hyperparameters, whose values play an important role in the performance of the model. These parameters were selected according to the following guiding principles, and are listed below. First, simplicity: the best choice is often the most natural one; second, computational effectiveness: samples that are too large increase computational time without significant improvement in accuracy; third, performance: when the previous two criteria did not obviously lead to specific parameter values, the latter were chosen as to improve the overall accuracy of the forecast.

\medskip
\noindent List of hyperparameters.
\begin{itemize}
\item The values used to initialize the parameter search. Currently, $(\beta_0,\gamma_0) = (\langle \beta_o\rangle,\langle \gamma_o\rangle)$,
$N_0$ is 1/3 of the state population, and $\kappa_0 = 1 + 100/N_0$.
\item The range (3, 5, or 14 days) of ${\mathcal K} = [k_m, k_M] \cap \mathbb{N}$ and the number $n_i$ of different intervals $\mathcal K$ used to build the ensemble forecast. Currently $n_i = 3$.
\item The region $\mathbf{a} \times \mathbf{b}$ that defines admissible values of $N$ and $\kappa$. As mentioned above, the only current restriction is that $N \ge C_{k_M}$.
\item The number $n_o$ of pseudo-observations used to make a forecast. Currently $n_o = 50$ for each interval $\mathcal K$.
\item The parameters used in the smoothing procedure, currently a 7-day moving window applied twice, used to estimate $S_k$.
\item The value of $\textbf{C}_{k_{m-1}}$, currently set at $\mathbf{C}_{k_m-1} = \sum_{j = 1}^{k_m -1} S_j$.
\item The variance $\sigma_k^2 $ of the noise added to $S_k$ to generate pseudo-observations; currently $\sigma_k^2 = S_k$.
\item The initial condition $C_{k_M} = \sum_{j = 1}^{k_M} S_j$ used to make the forecasts.
\item The parameters used in the augmentation procedure: the values $\mut$ for the mean and $\sigmat = \zeta \cdot \max\{\mut,v_t\}$ for the variance, the value of $\zeta$, and the number of forecasts $n_f$ added to the ensemble in this augmentation step. Currently $\mut = \frac{1}{N_f} \sum_{i=1}^{N_f} I_i(t)$, $v_t = \frac{1}{N_f-1}\sum_{i=1}^{N_f} (I_i(t) - \mut)^2$, $\zeta$ is as defined by Equations \eqref{eq:zeta_1} and \eqref{eq:zeta_2}, and $n_f = N_f = n_i \cdot n_o$.
\item The number $N_c$ of data points used in the linear regression between case counts and deaths. Currently, the default is $N_c = 10$.
\item The bounds on the delay $\tau$ between case counts and deaths, currently set at 0 and 21 days.
\end{itemize}
\medskip
\noindent Parameters whose selection was guided by forecast accuracy are likely to depend on the quality of the input data stream. For instance, re-running case count and death forecasts using the JHU data as input and truth leads to a drop in performance (compare Columns 1 and 2 of Table 1), likely due to differences in which weekend data are reported by JHU in comparison to the COVID Tracking Project. As previously mentioned, we initially used the COVID Tracking Project data because it provided early case counts for all states when we started working on this project. 

Because different public dashboards use different data sources, are updated at different times, and potentially handle backfill in different ways, it is important to (i) identify hyperparameter values that lead to optimal performance once sufficient data are available, and (ii) indicate which data stream is considered as the ``truth'' for a particular instance of the model. Tuning these hyperparameters can be computationally expensive. For example, increasing the range of $\mathcal K$ will typically increase the computation time for each forecast. Additionally, in order to identify optimal hyperparameter combinations, weekly forecasts should be created and scored for each set. Once hyperparameters are chosen however, they can remain fixed for all future forecasts, unless a drop in performance is noticed.

\section*{Acknowledgments}
We thank Matt Biggerstaff, Michael Johansson, Nick Reich, as well as the members of and contributors to the COVID-19 Forecast Hub, for fostering a stimulating open-science community centered on COVID-19 forecasting in the US, from which we have greatly benefited. We acknowledge, and are grateful for, partial financial support from NIH grant GM084905 (HRB) and NSF grant DMS-RAPID-2028401 (HRB \& JL). 
Finally, we thank Matti Morzfeld for his comments on the first draft of this manuscript.

\clearpage

\end{document}


\maketitle
This document contains the following information.
\begin{itemize}
    \item Development and evolution of EpiCovDA
    \item Smoothing procedure
    \item State-level cumulative cases forecasts
    \item State-level cumulative deaths forecasts
    \item Selected incident case and death forecasts
    \item Interval score results
\end{itemize}
\section{Model development and evolution}
\label{sec:EpiCovDA_evol}
We began developing EpiCovDA as COVID-19 started to spread worldwide, and the model has been continuously evolving since then. For each version, hyperparameters were tuned first to achieve minimization of the overall MAE calculated from available data, followed by improvement on calibration. Three main versions were considered, which mostly differed in the definition of the prior for the parameters. Version 1 used $\kappa = 1$ and a Gaussian prior on $\beta$, $\gamma$, and $N$. Version 2 used $R_0$ and $\beta N$ as parameters. This choice was motivated by the form of Equation (2.1) of the main text, in which ${\mathcal I}$ can be written as a function of $C/N$ with parameters $\beta N$ and $R_0$. It did not lead to any improvement, probably due to the uncertainty on $N$. Version 3 is the current version of the model, with a Gaussian prior on $\beta$ and $\gamma$, and a uniform prior on $N$ and $\kappa$. We also discovered by trial and error that using all of the data available, including irregular weekend reports, led to better predictions. Finally, we found that initializing the parameter search at $N_0$ equal to 1/3 of the state population and $\kappa_0 = 1 + 100/N_0$, allowed the optimizer to explore a wide range of values and led to realistic optimal parameter choices, with values away from the selected initial conditions.

\section{Data Smoothing} \label{app:smoothing}
This section details the smoothing introduced in the \textit{Pseudo-Observations} section. This smoothing procedure was previously described in \cite{Biegel20}. Suppose that incidence data are available from day $1$ through day $M$, and let $G_k$ be the true reported incidence on day $k$. Assuming $M \ge 12$,  the smoothed data on day $k$, $S_k$, is calculated as follows.

\noindent For $k = 1,2,3:$ 
$$S_k = \frac{1}{3+k} \sum_{j = 1}^{k+3} \left(\frac{1}{3+j}\sum_{i = 1}^{j+3} G_i\right), $$
and for $k = 4,5,6:$
$$  S_k =  \frac{1}{7}\left(\sum_{j = k-3}^{3}\left( \frac{1}{3+j} \sum_{i=1}^{j+3} G_i \right)\right) +   \frac{1}{7}\left(\sum_{j = 4}^{k+3}\left( \frac{1}{7} \sum_{i = j-3}^{j+3} G_i \right)\right).$$
For $k = 7,8,\hdots, M-5$:
$$S_k = \frac{1}{7} \sum_{j = k-3}^{k+3} \left(\frac{1}{7} \sum_{i = j-3}^{j+3} G_i \right)=\frac{1}{49} \sum_{j = k-3}^{k+3} (7+j - k) G_j.$$
For $k = M-5,M-4,M-3$:
$$S_k = \left( \frac 1 7 \sum_{j = k-3}^{M-3} \left( \frac 1 7 \sum_{i = j - 3}^{j + 3} G_i\right) \right) + \left( \frac 1 7 \sum_{j = M-2}^{k+3} \left( \frac{1}{M-j + 4} \sum_{i = j -3}^M G_i\right)\right). $$
For $k = M-2,M-1,M$:
$$S_k = \frac{1}{36} \sum_{j = M-5}^M \left(\sum_{i = M-5}^M G_i\right).$$

\section{State-level Forecasts}
\label{fig:EpiCovDA_case_append}
This section provides plots of cumulative case and deaths forecasts at the state level and illustrates how model performance varies with case counts and location.

\subsection{Cumulative case forecasts}
We show below the cumulative case forecasts for each of the 50 states, D.C., and Puerto Rico. The black curves indicate the true values as reported by the COVID Tracking Project \cite{CVDT}. The widest shaded regions correspond to the central 95\% prediction intervals. The smaller shaded regions correspond to the central 50\% prediction intervals. The darker colored curves in shaded regions are the median forecasts. COVID-19 data was provided by The COVID Tracking Project at {\sl The Atlantic} under a CC BY 4.0 license.

\bigskip

\begin{center}
 \includegraphics[width=0.32\linewidth]{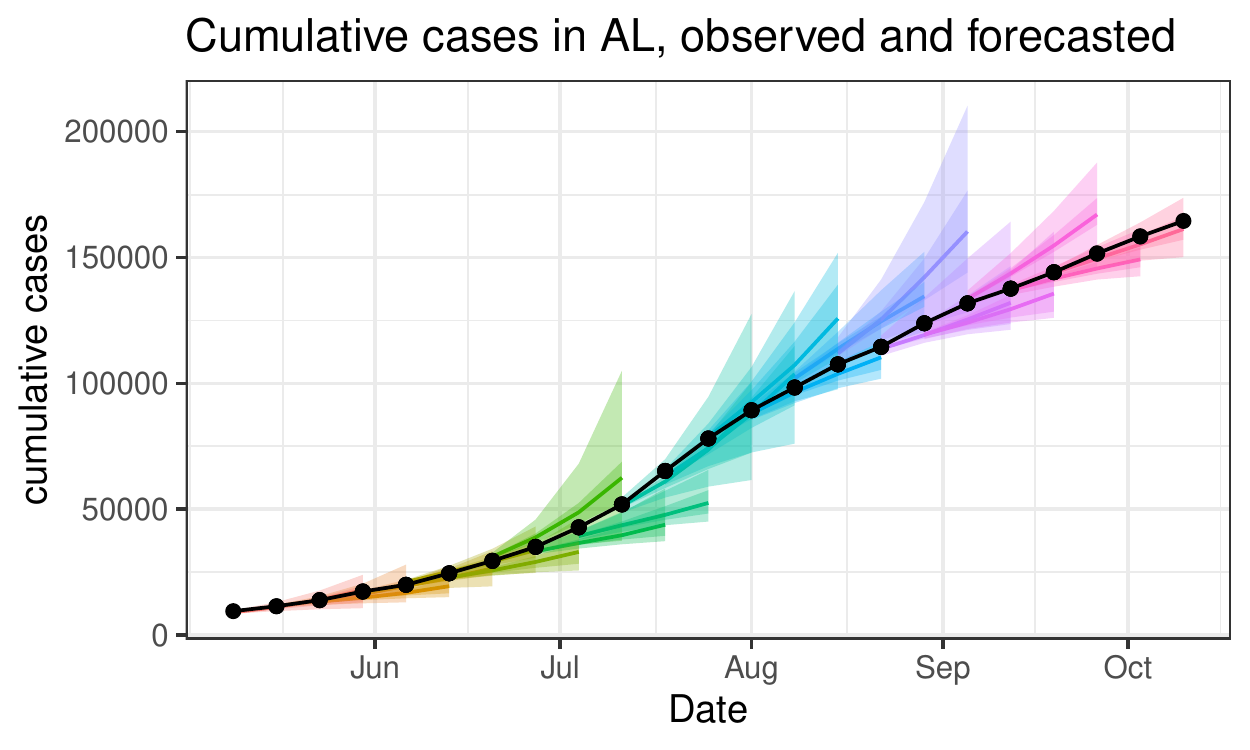}
 \includegraphics[width=0.32\linewidth]{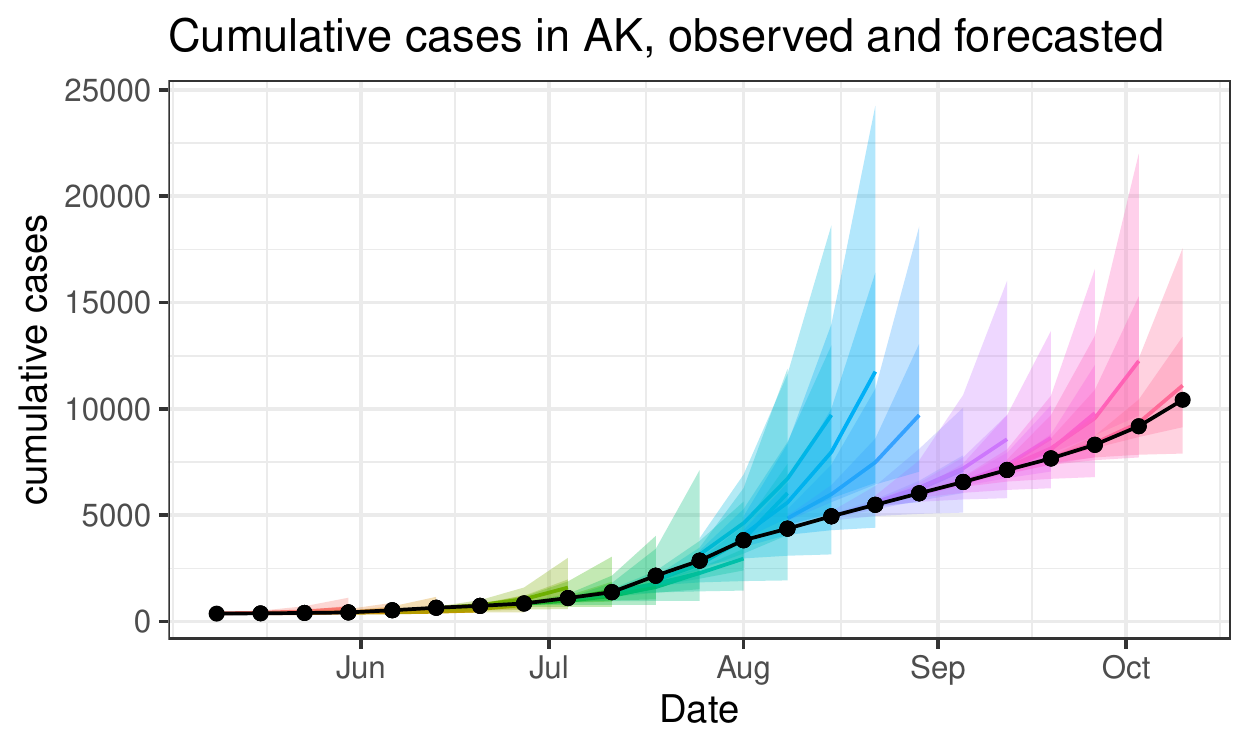}
 \includegraphics[width=0.32\linewidth]{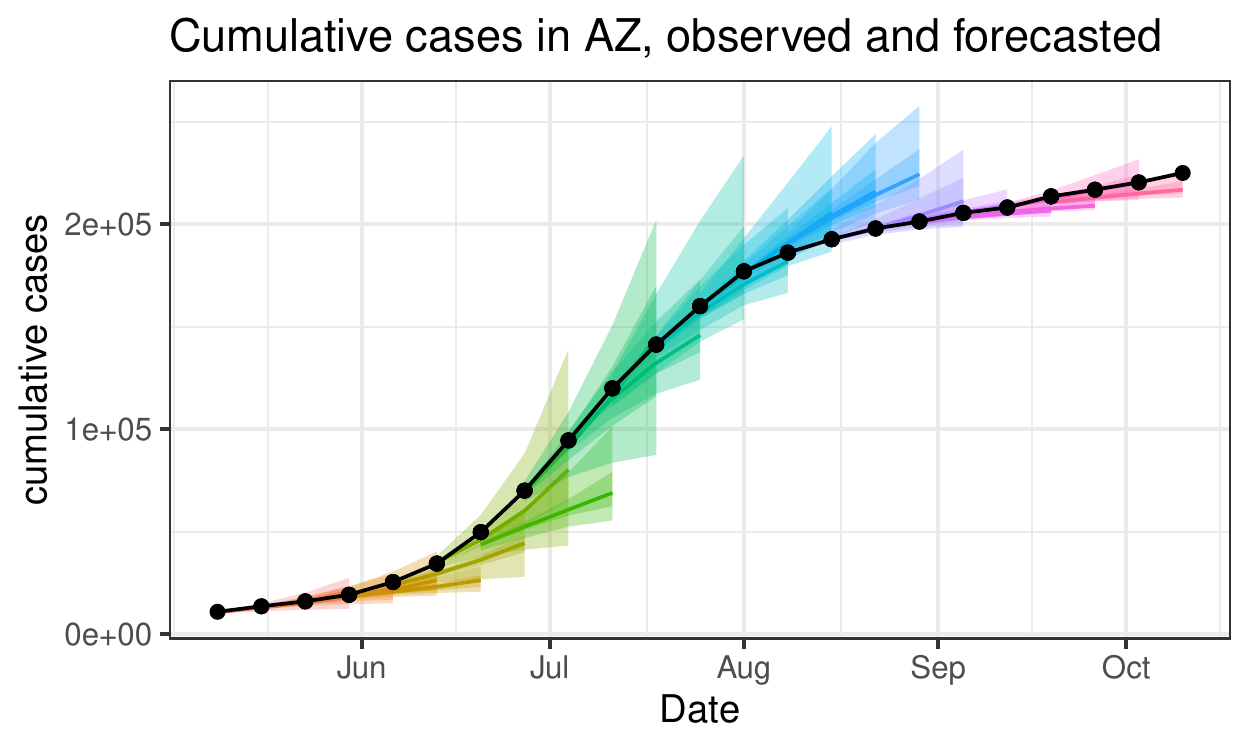}

 \bigskip

 \includegraphics[width=0.32\linewidth]{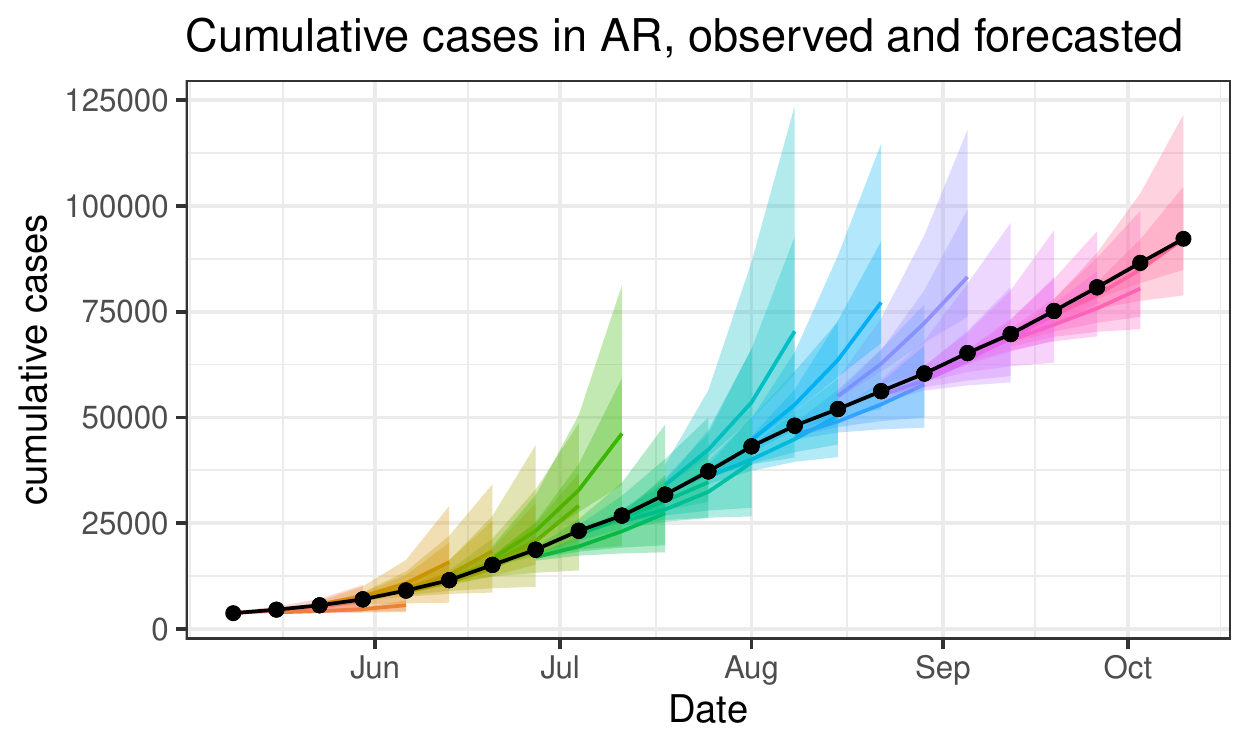}
 \includegraphics[width=0.32\linewidth]{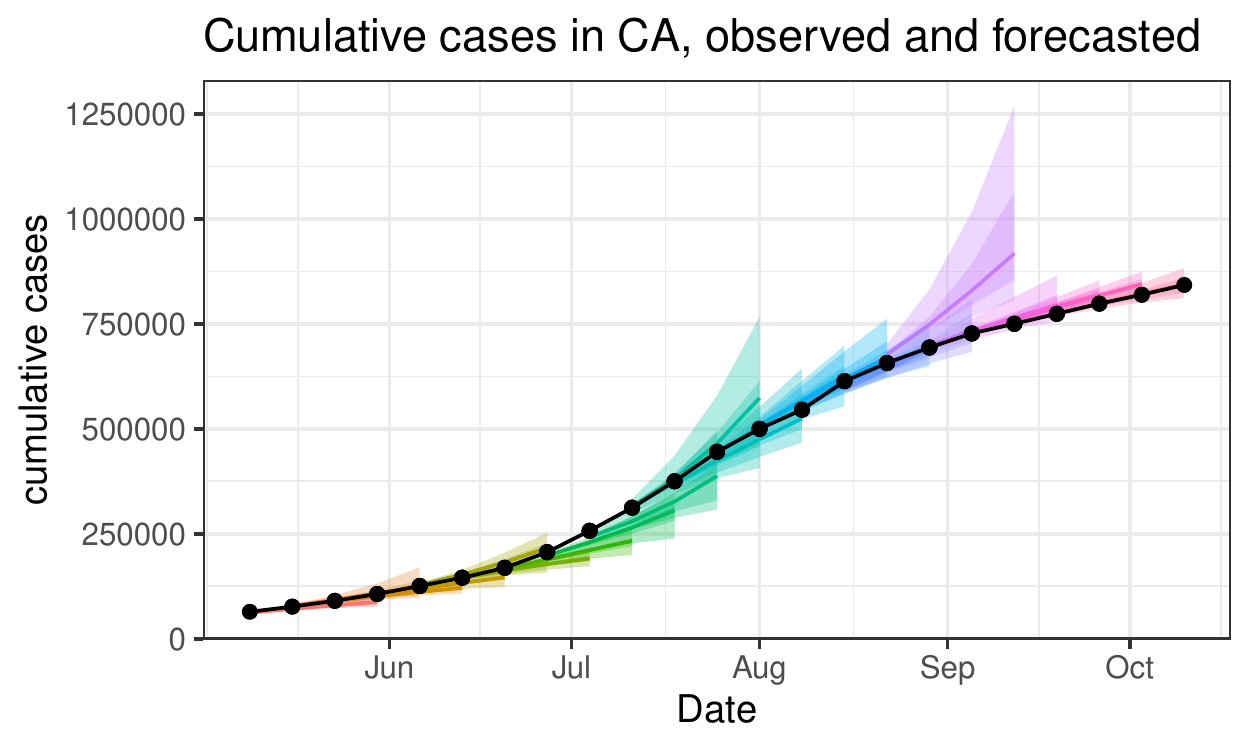}
 \includegraphics[width=0.32\linewidth]{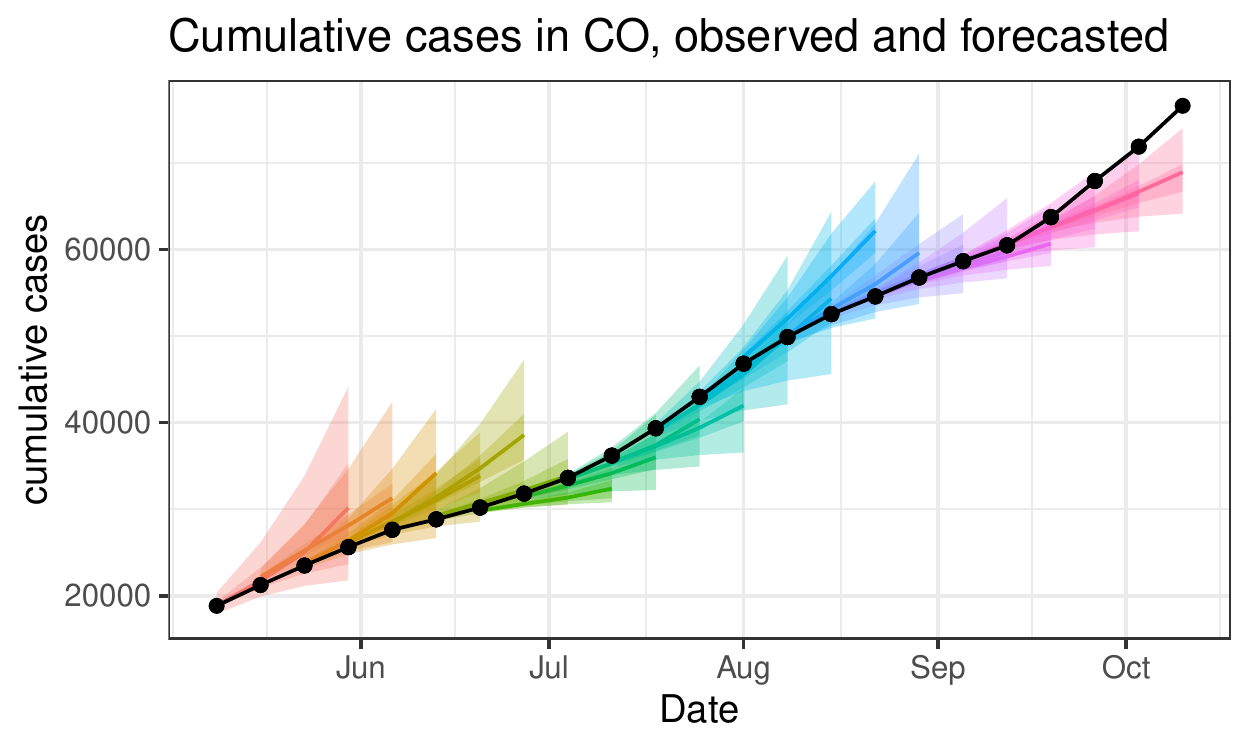}

 \bigskip

 \includegraphics[width=0.32\linewidth]{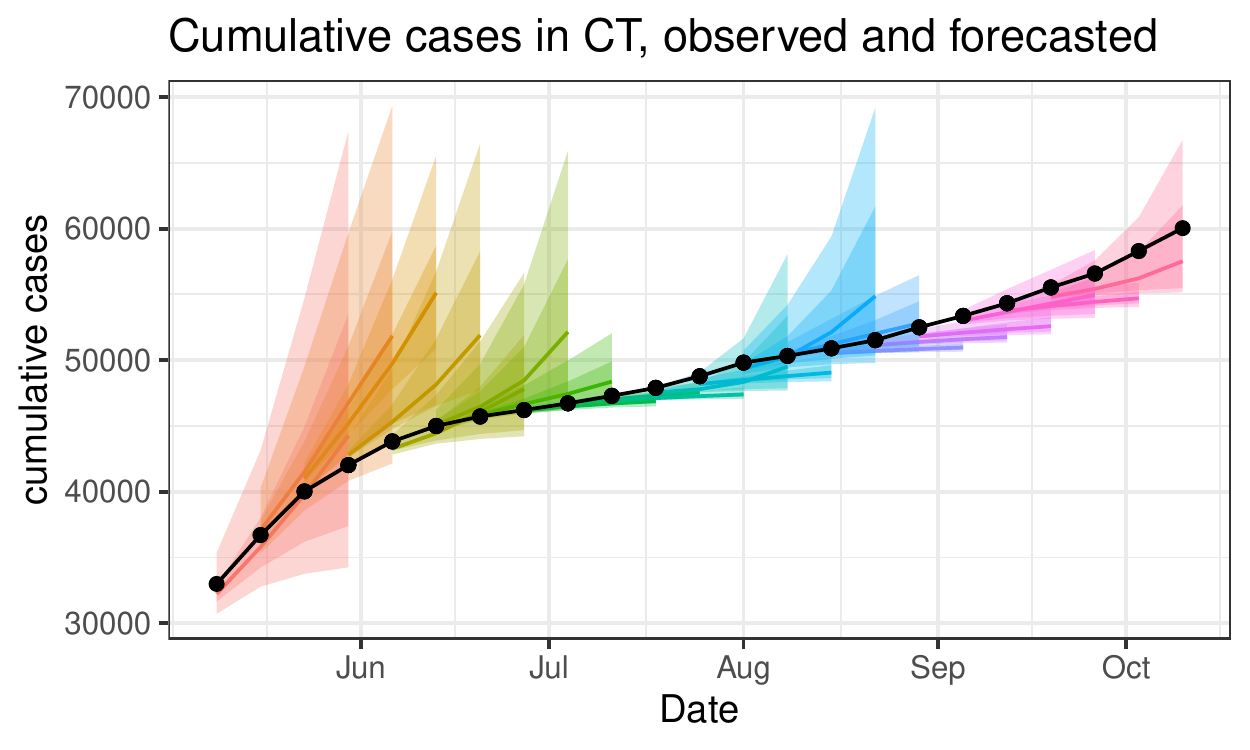}
 \includegraphics[width=0.32\linewidth]{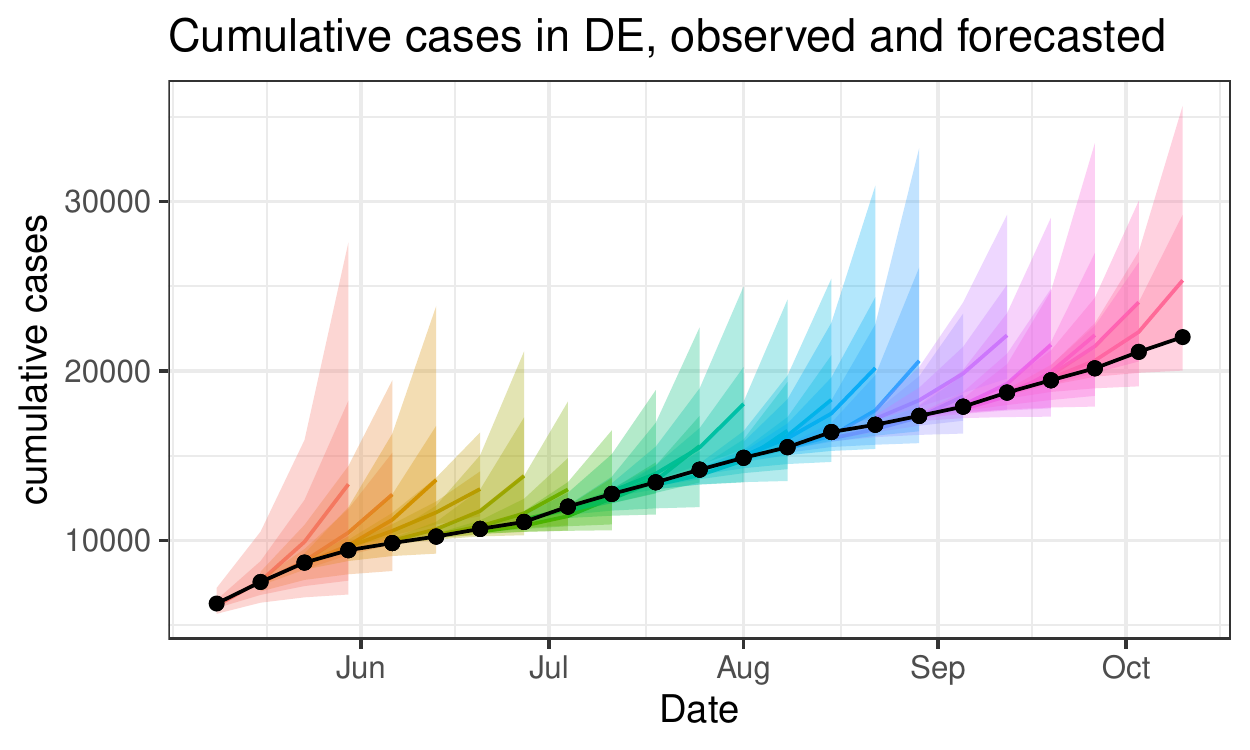}
 \includegraphics[width=0.32\linewidth]{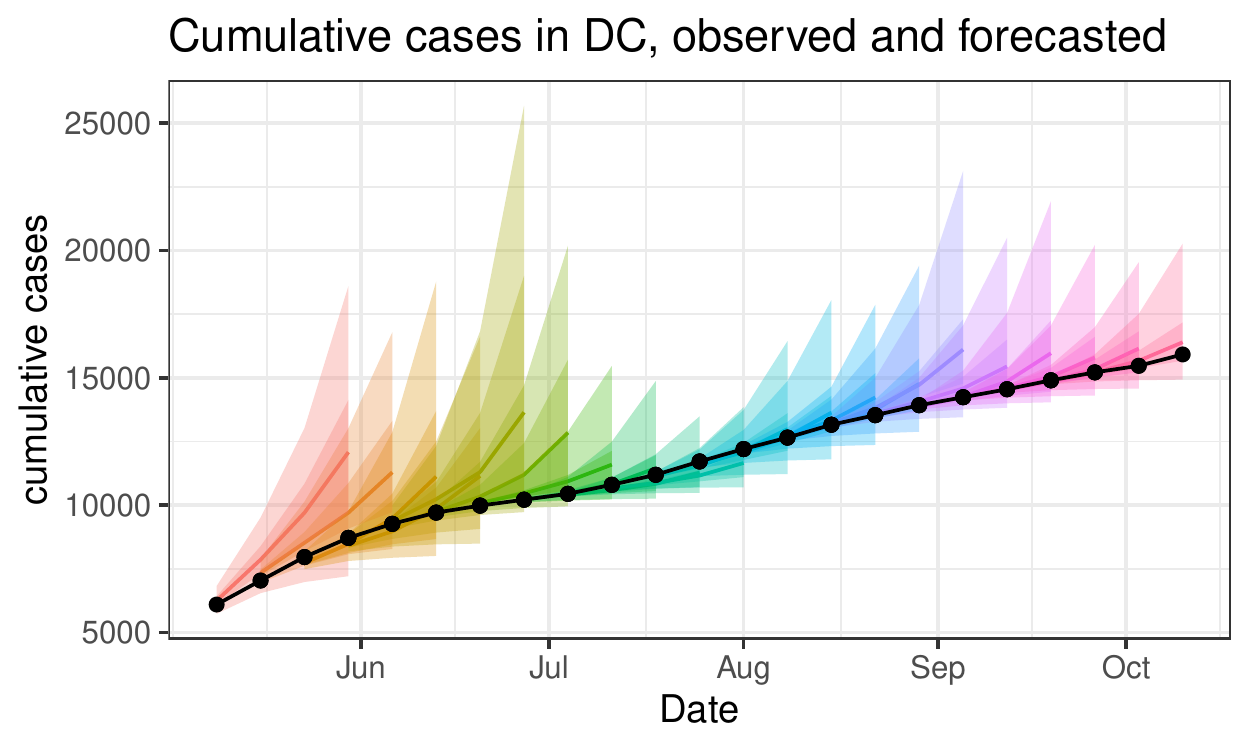}

 \bigskip

 \includegraphics[width=0.32\linewidth]{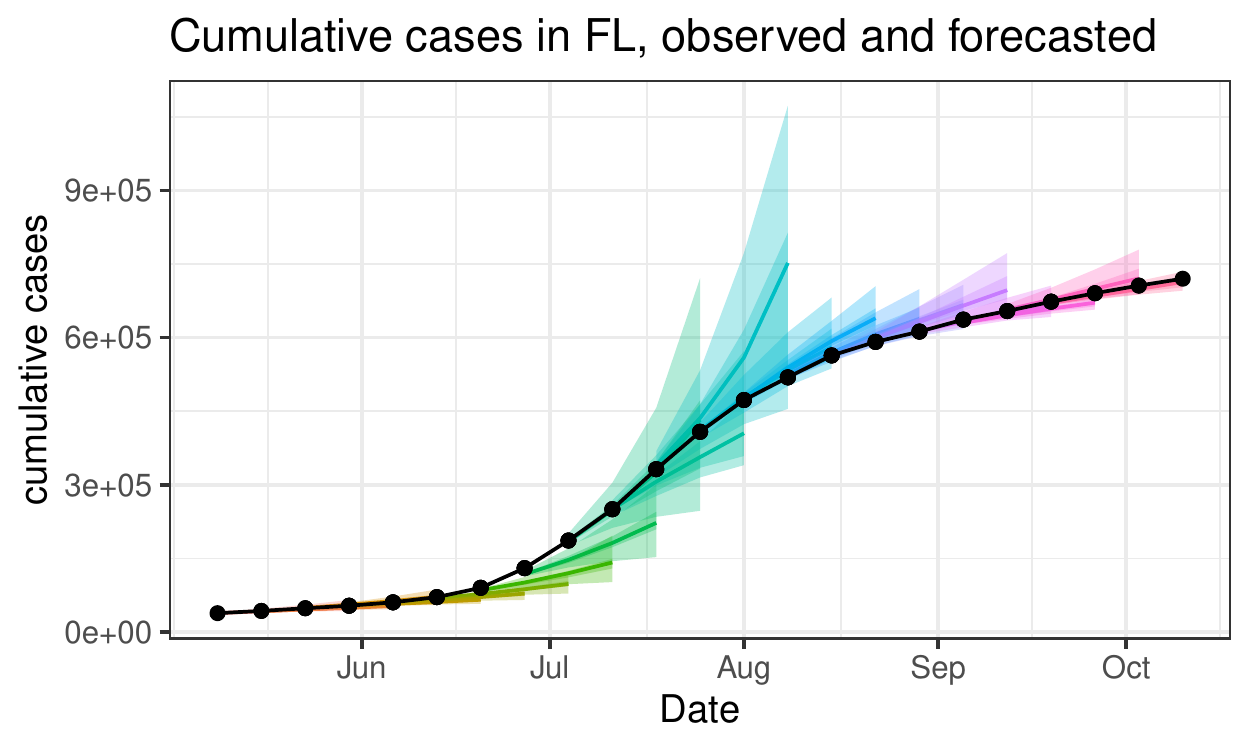}
 \includegraphics[width=0.32\linewidth]{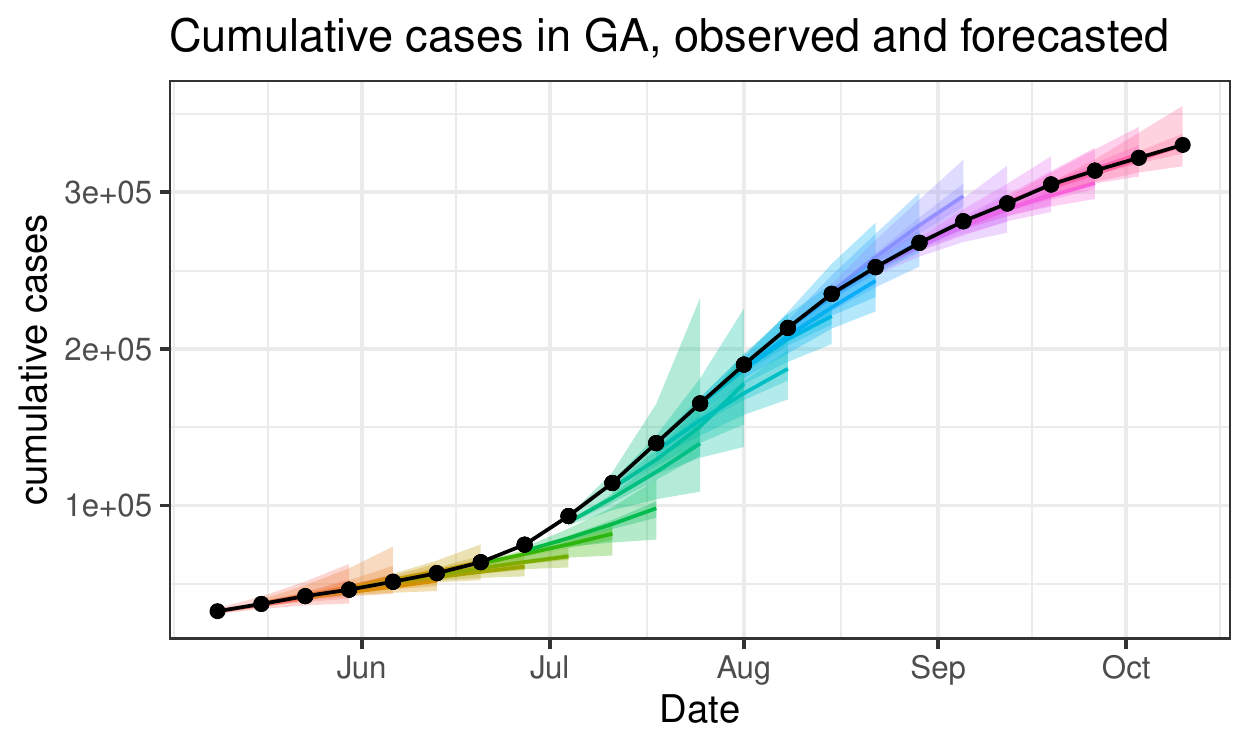}
 \includegraphics[width=0.32\linewidth]{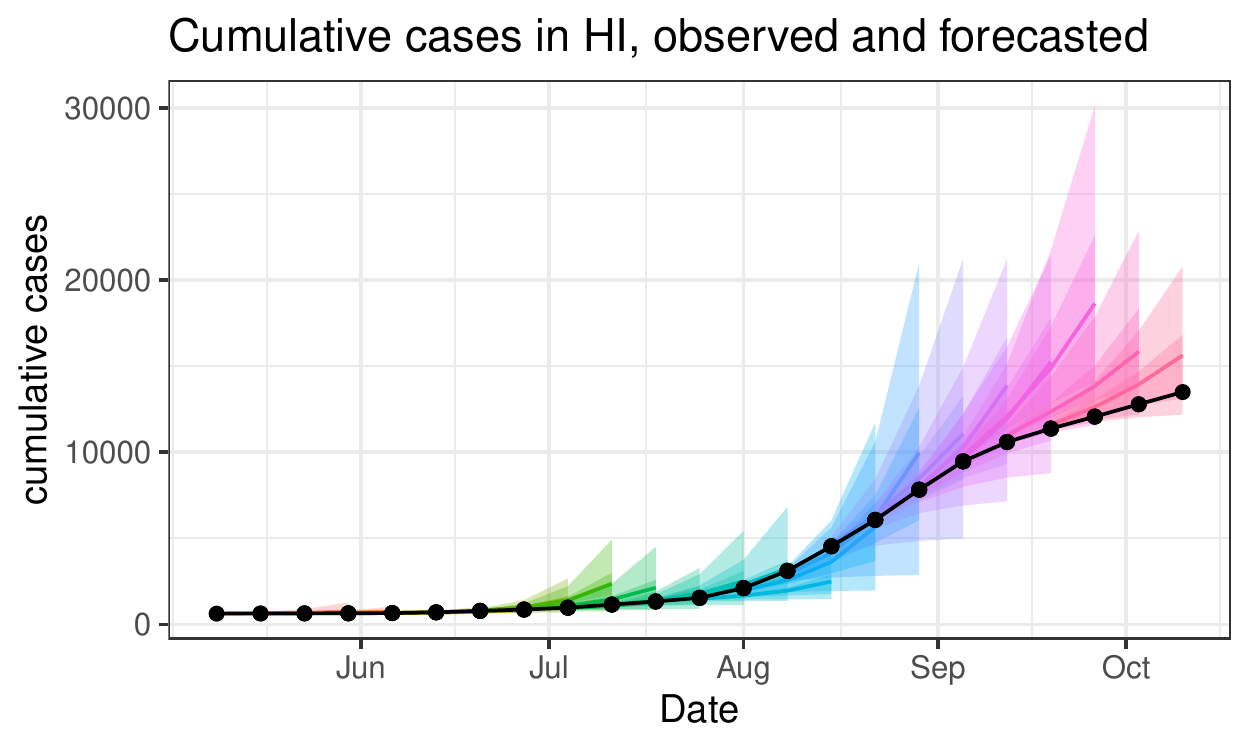}

 \bigskip

 \includegraphics[width=0.32\linewidth]{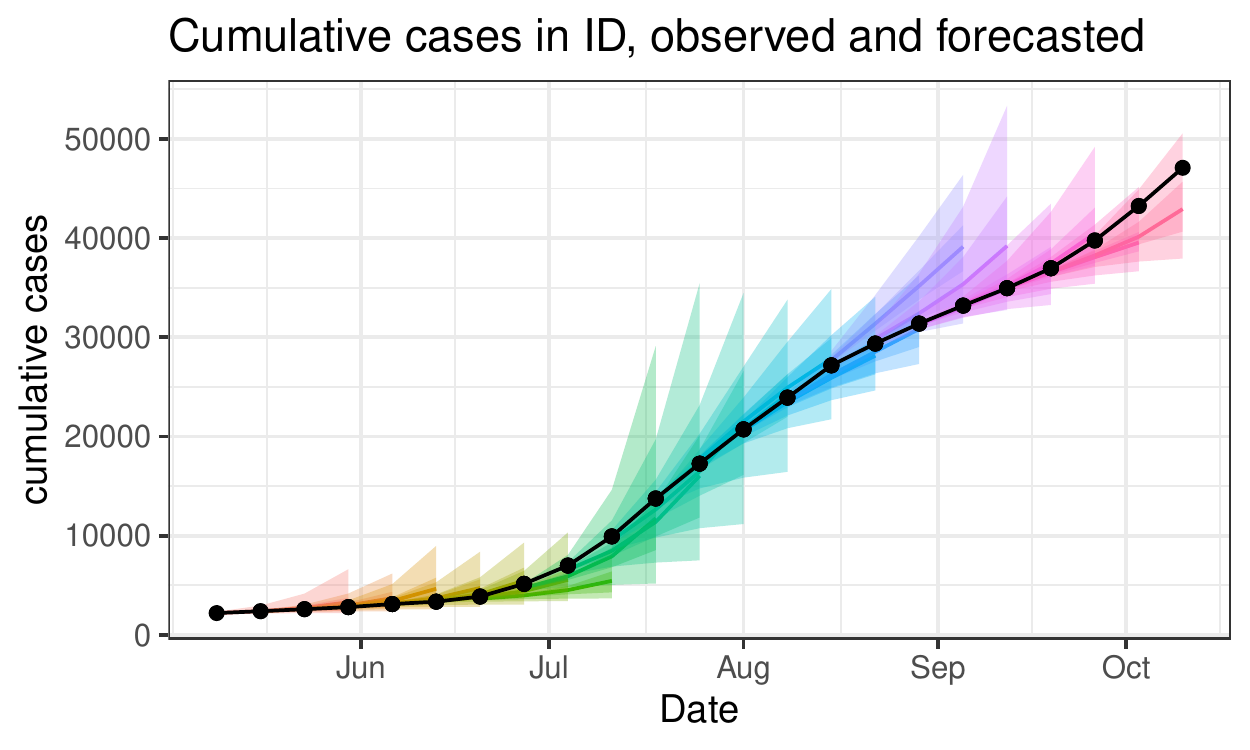}
 \includegraphics[width=0.32\linewidth]{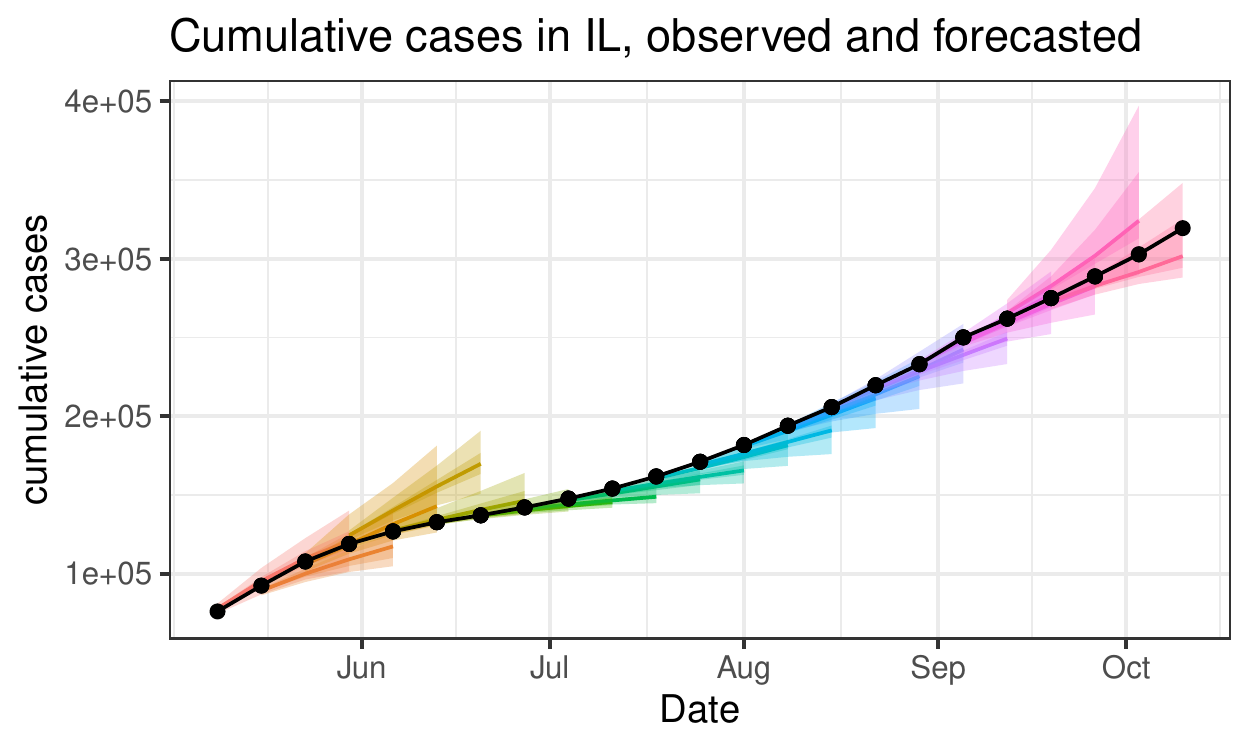}
 \includegraphics[width=0.32\linewidth]{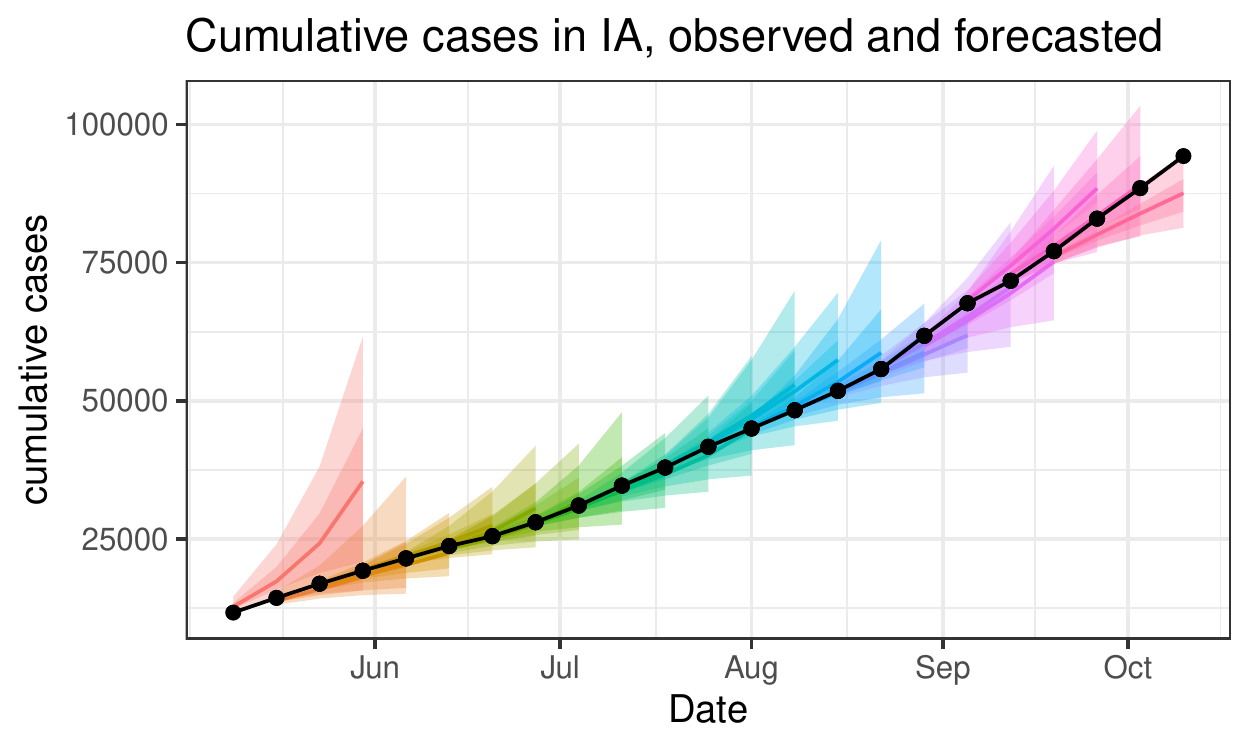}

 \bigskip

 \includegraphics[width=0.32\linewidth]{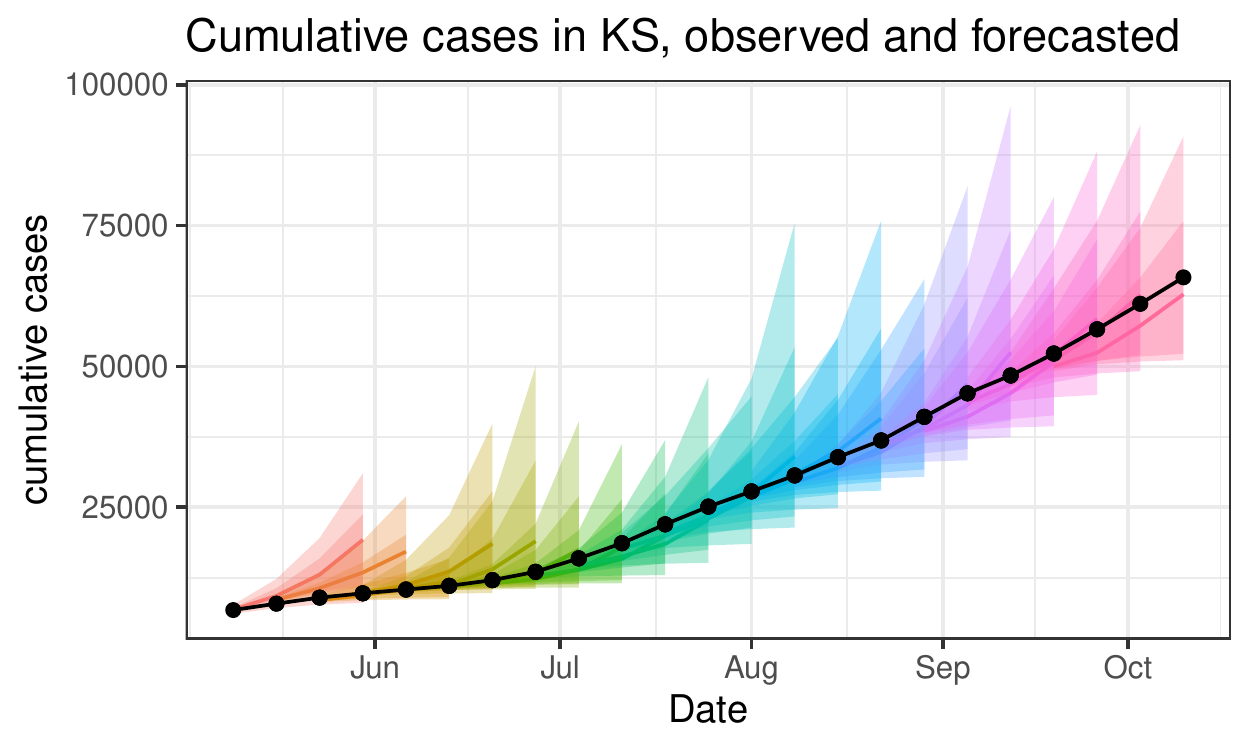}
 \includegraphics[width=0.32\linewidth]{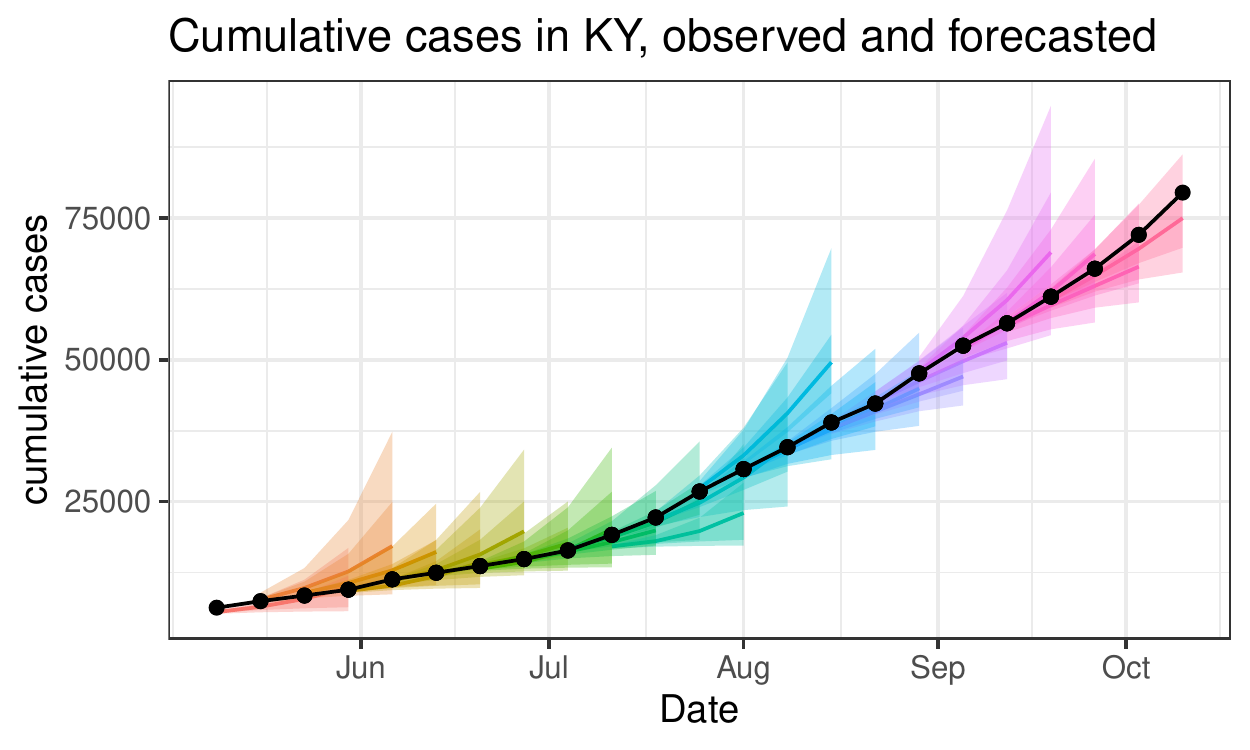}
 \includegraphics[width=0.32\linewidth]{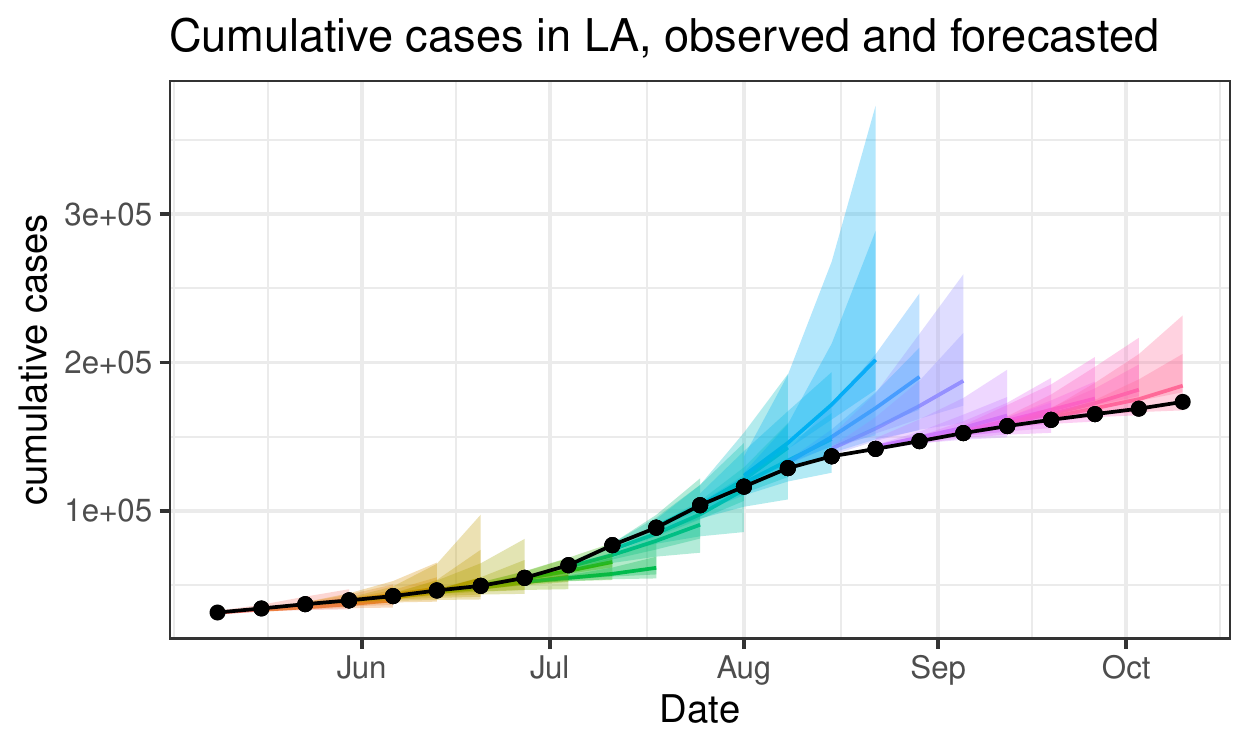}

\bigskip

 \includegraphics[width=0.32\linewidth]{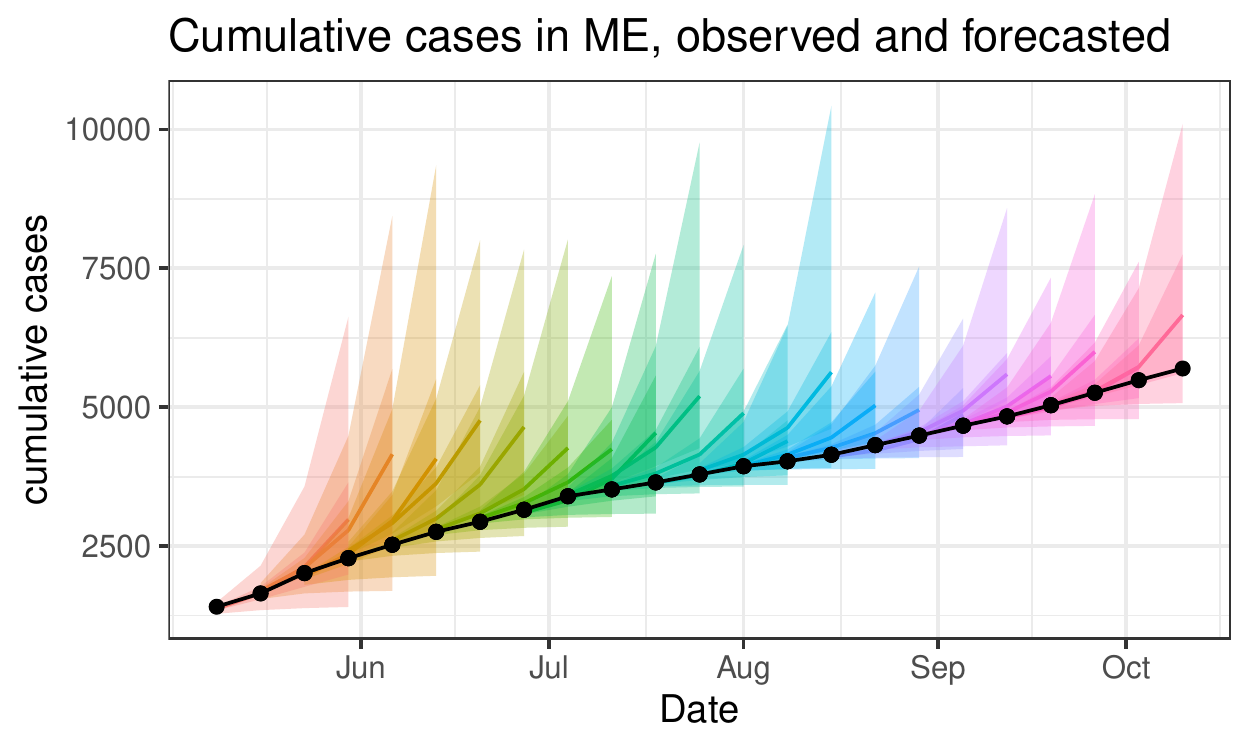}
 \includegraphics[width=0.32\linewidth]{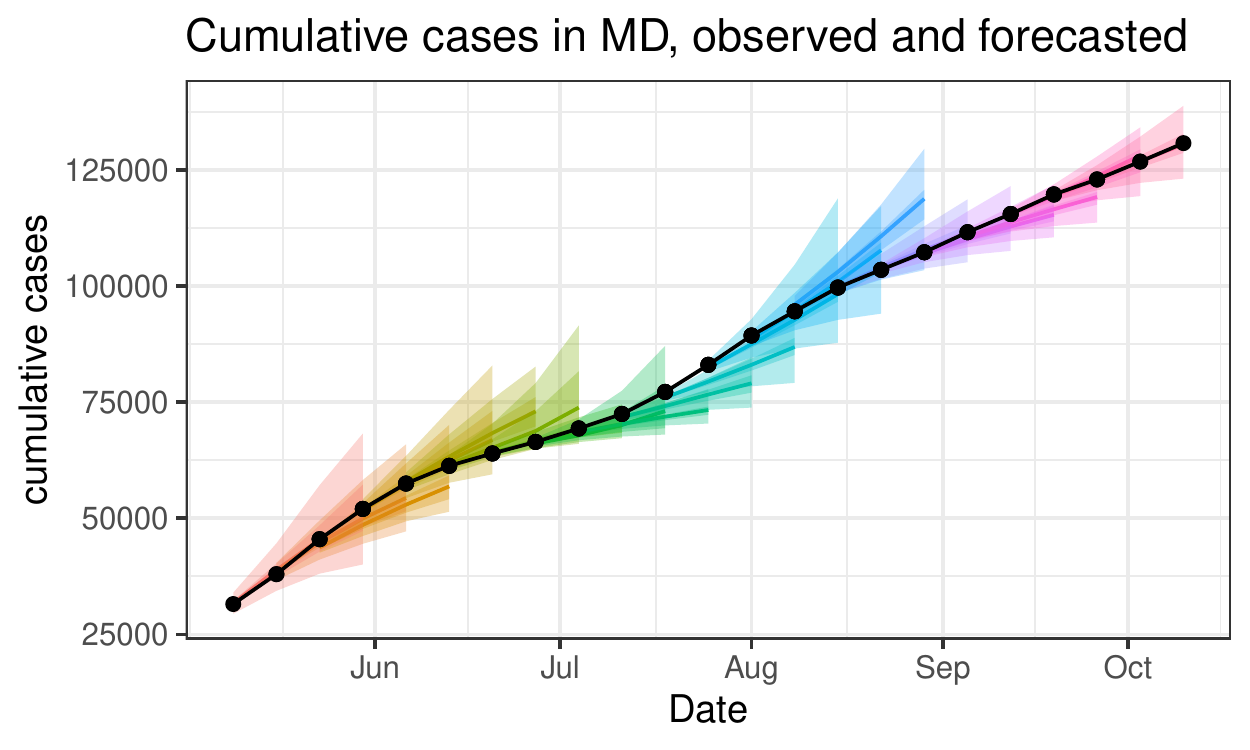}
 \includegraphics[width=0.32\linewidth]{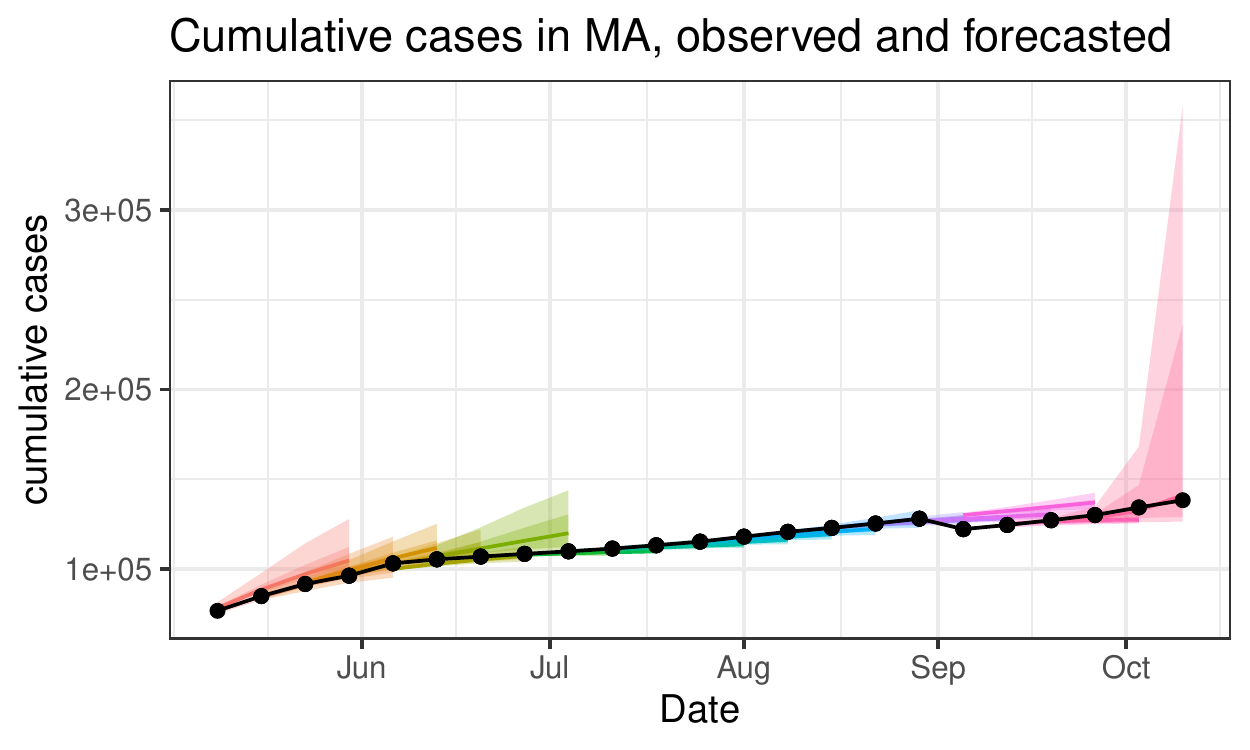}

 \bigskip

 \includegraphics[width=0.32\linewidth]{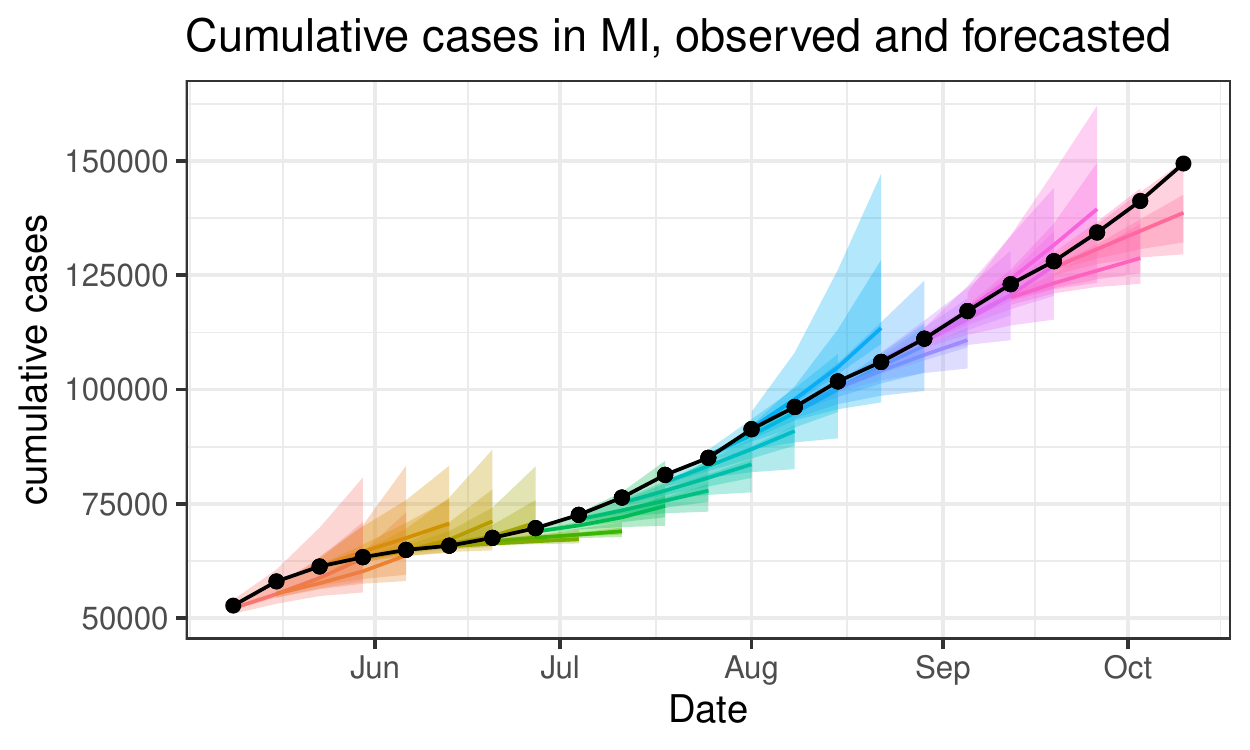}
 \includegraphics[width=0.32\linewidth]{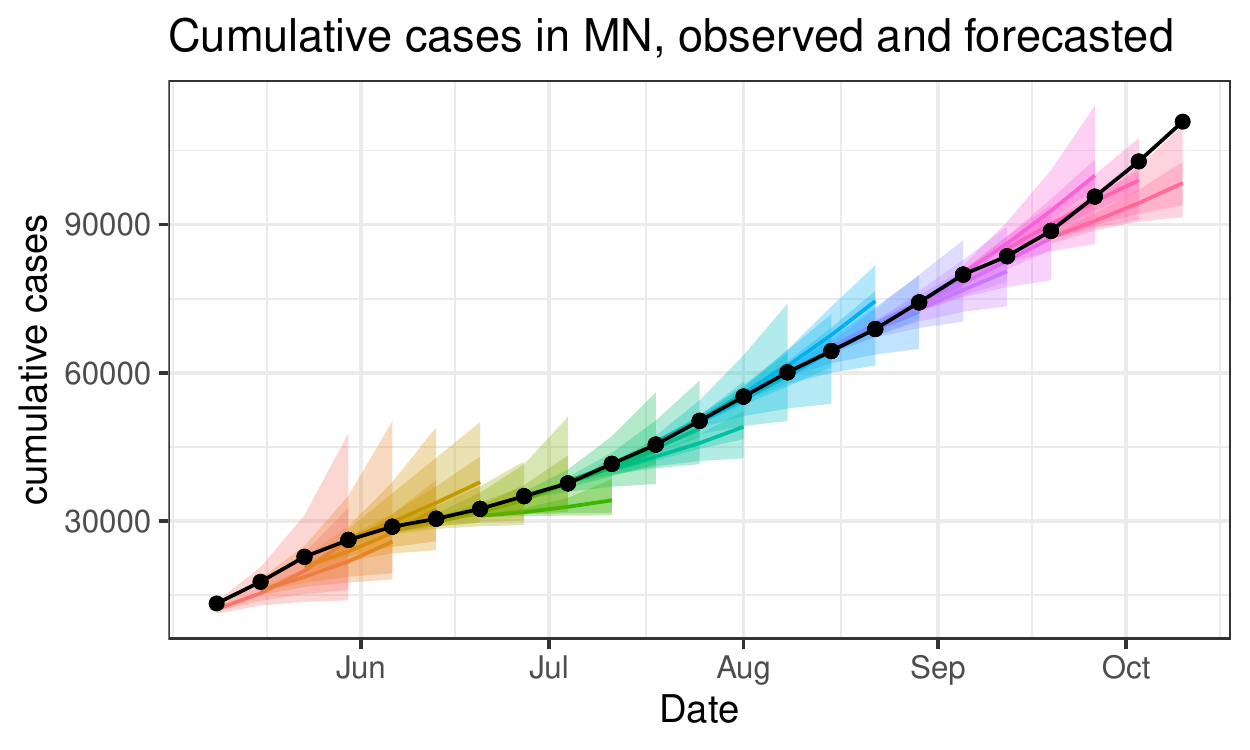}
 \includegraphics[width=0.32\linewidth]{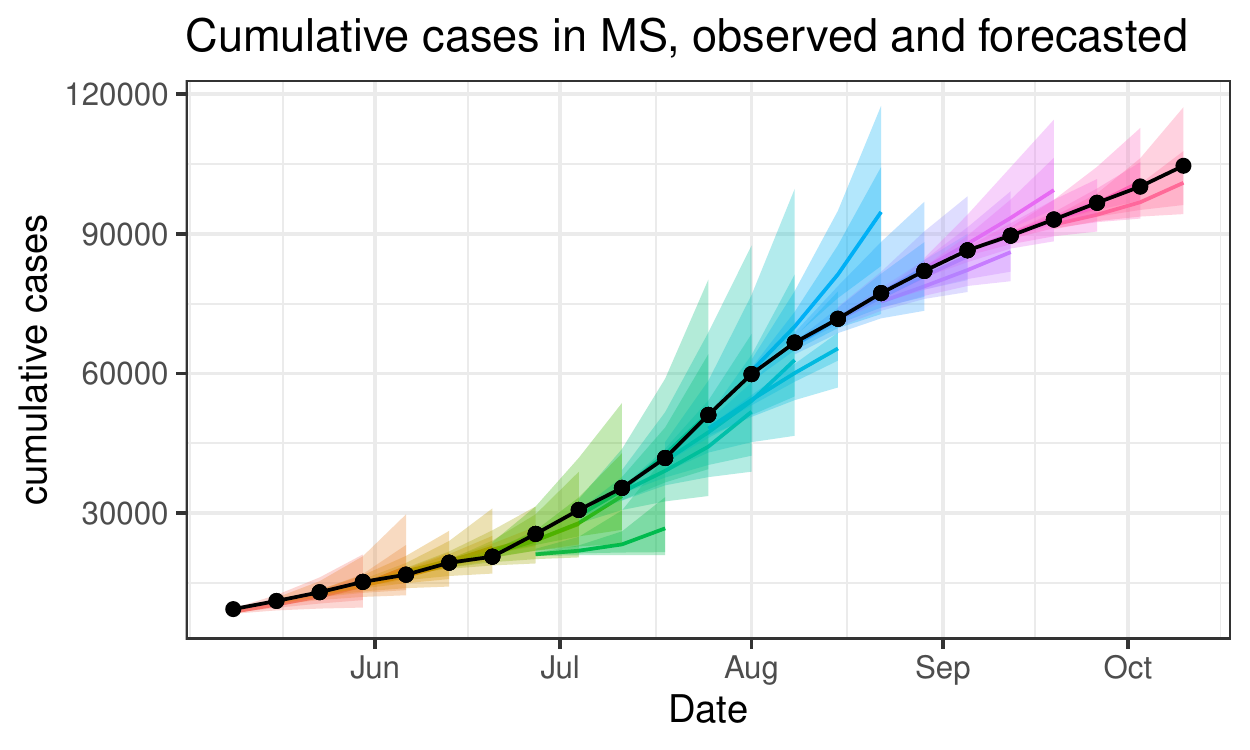}

 \bigskip

 \includegraphics[width=0.32\linewidth]{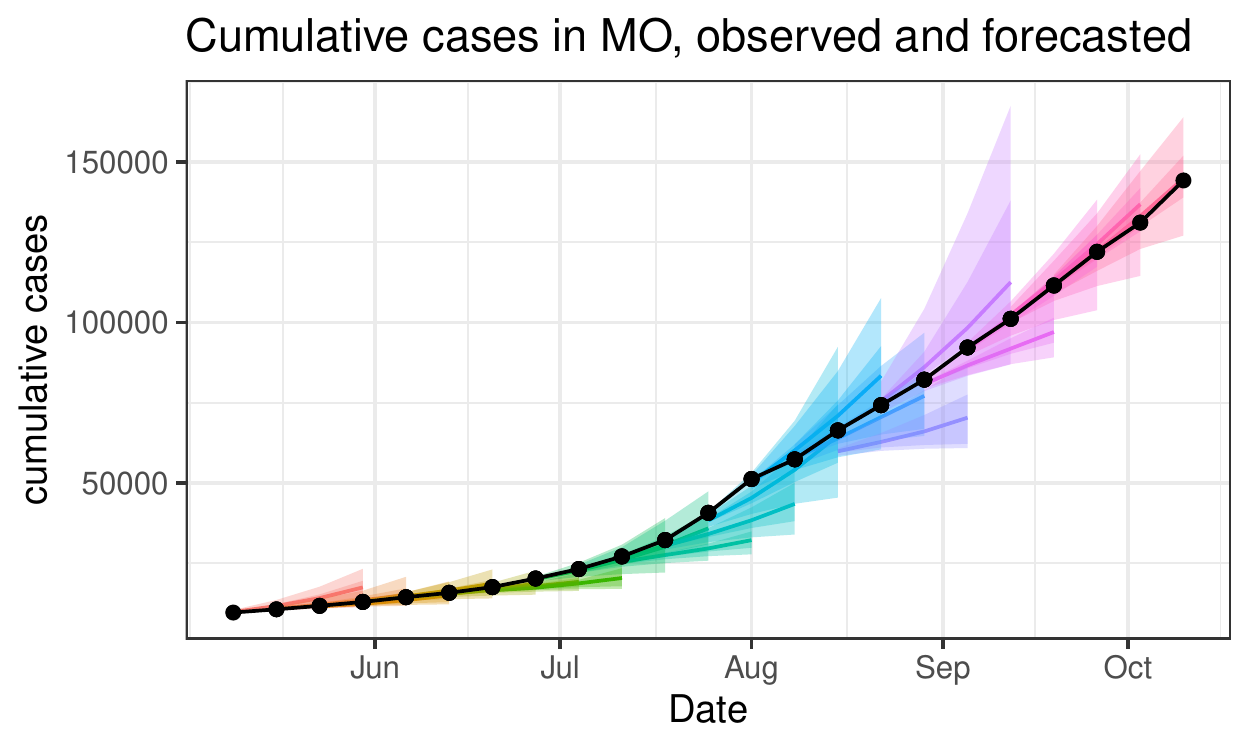}
 \includegraphics[width=0.32\linewidth]{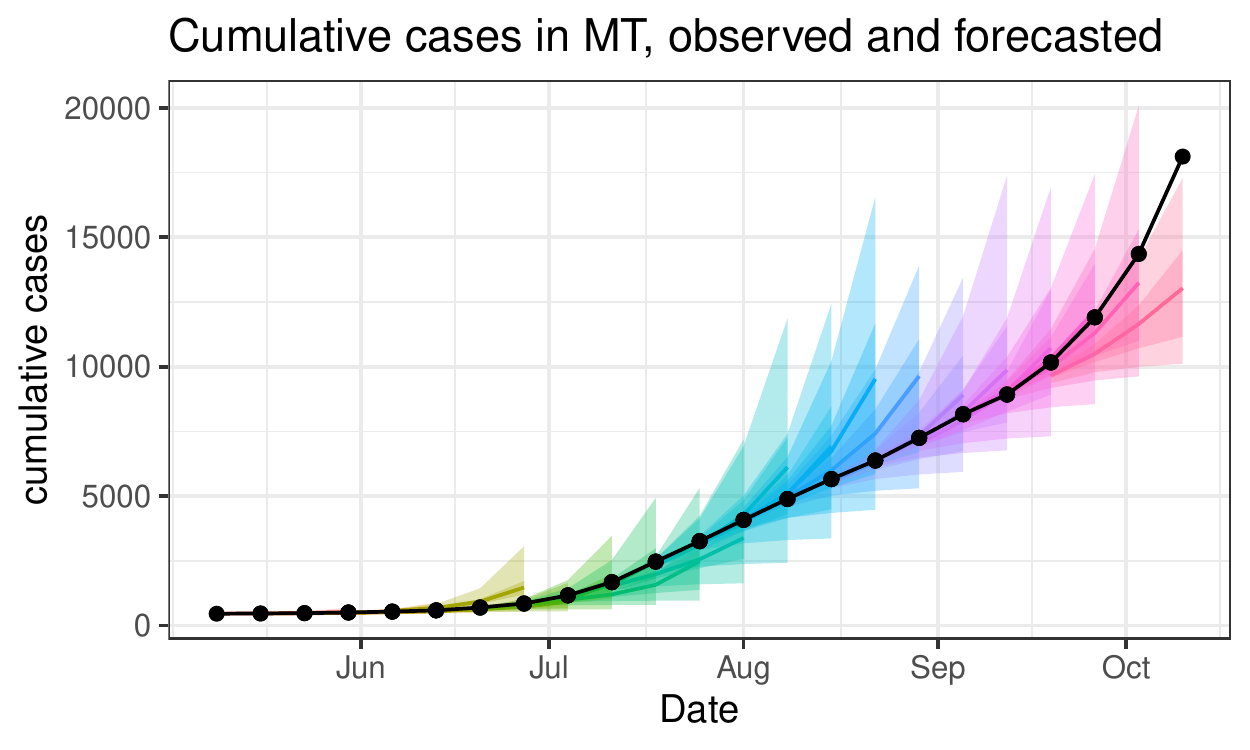}
 \includegraphics[width=0.32\linewidth]{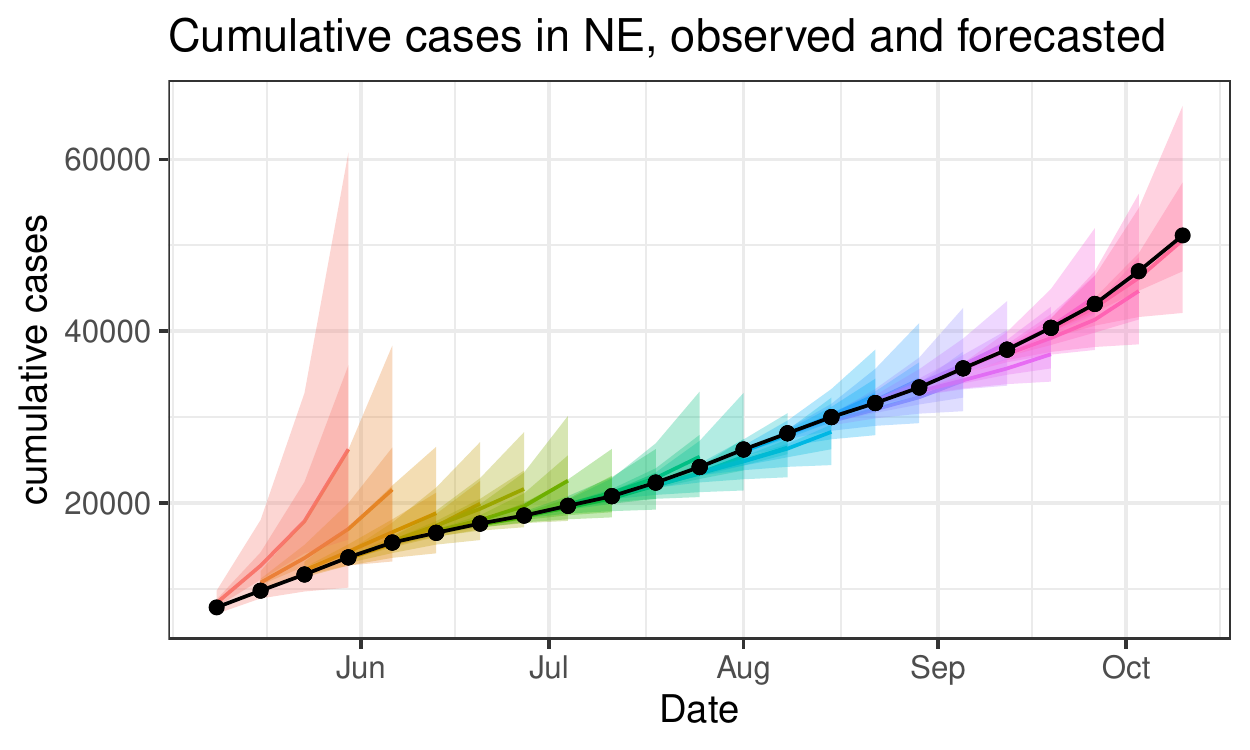}

\bigskip

 \includegraphics[width=0.32\linewidth]{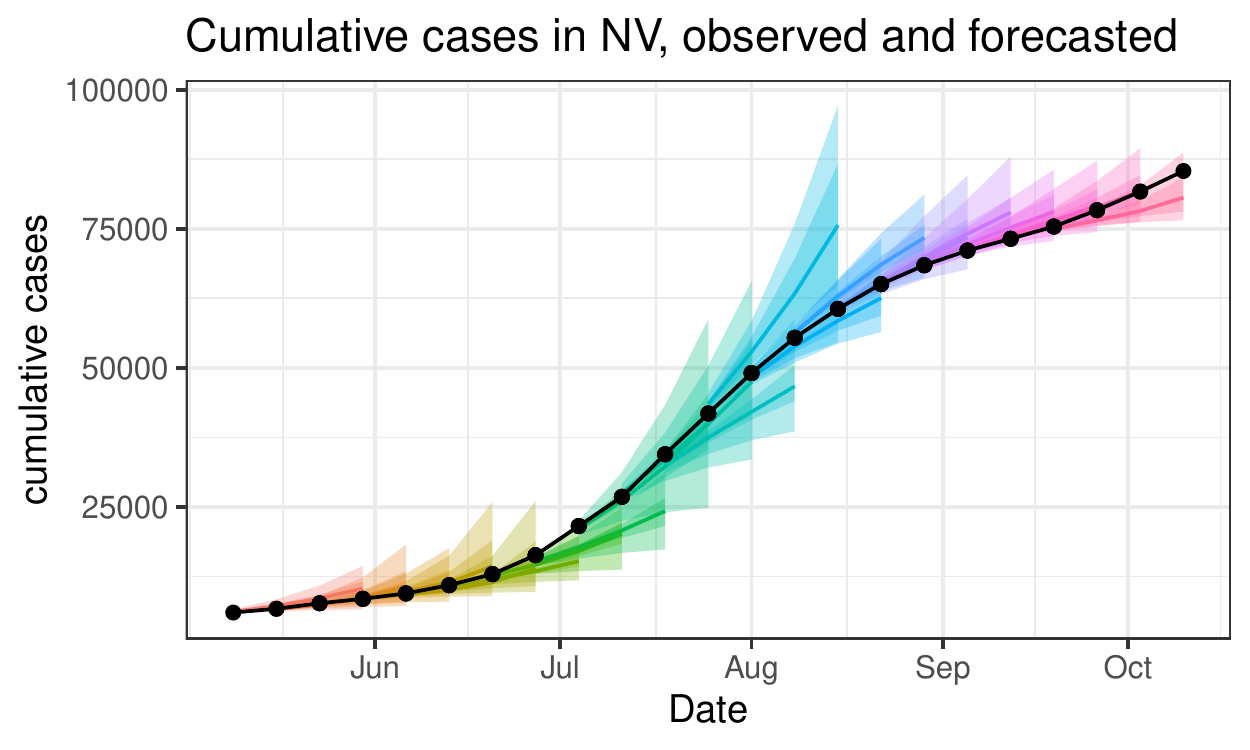}
 \includegraphics[width=0.32\linewidth]{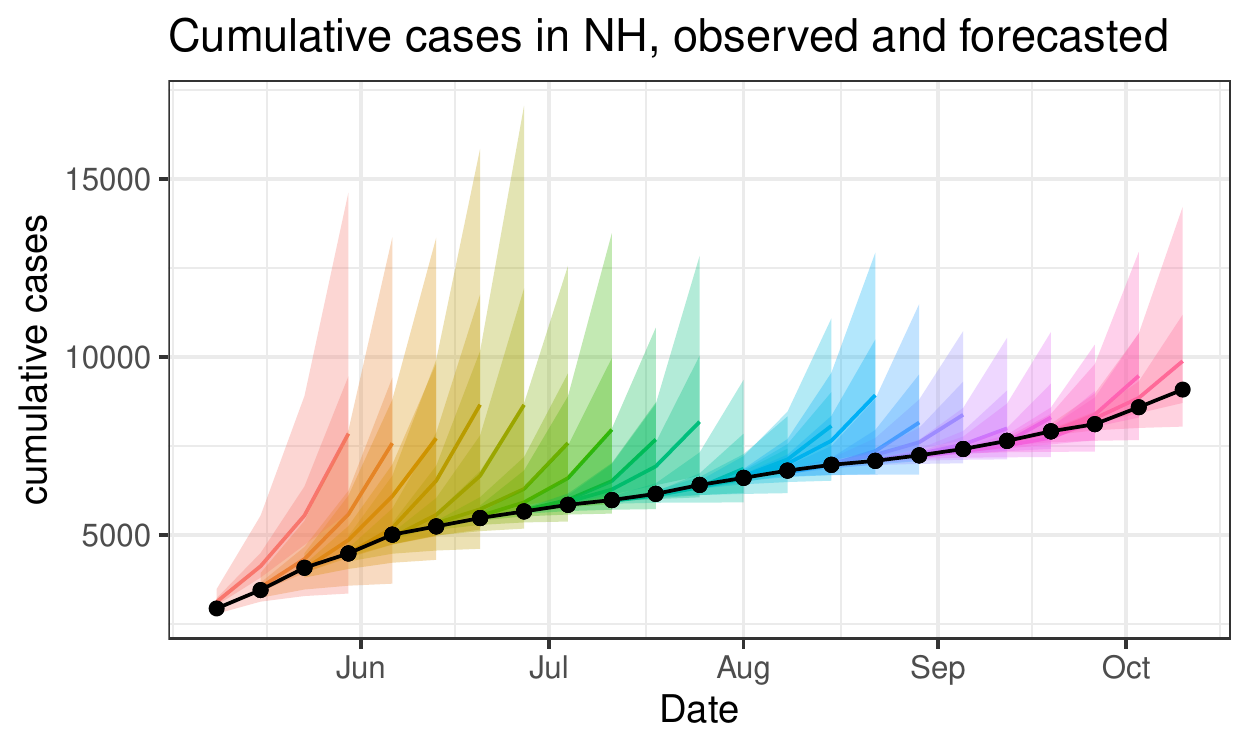}
 \includegraphics[width=0.32\linewidth]{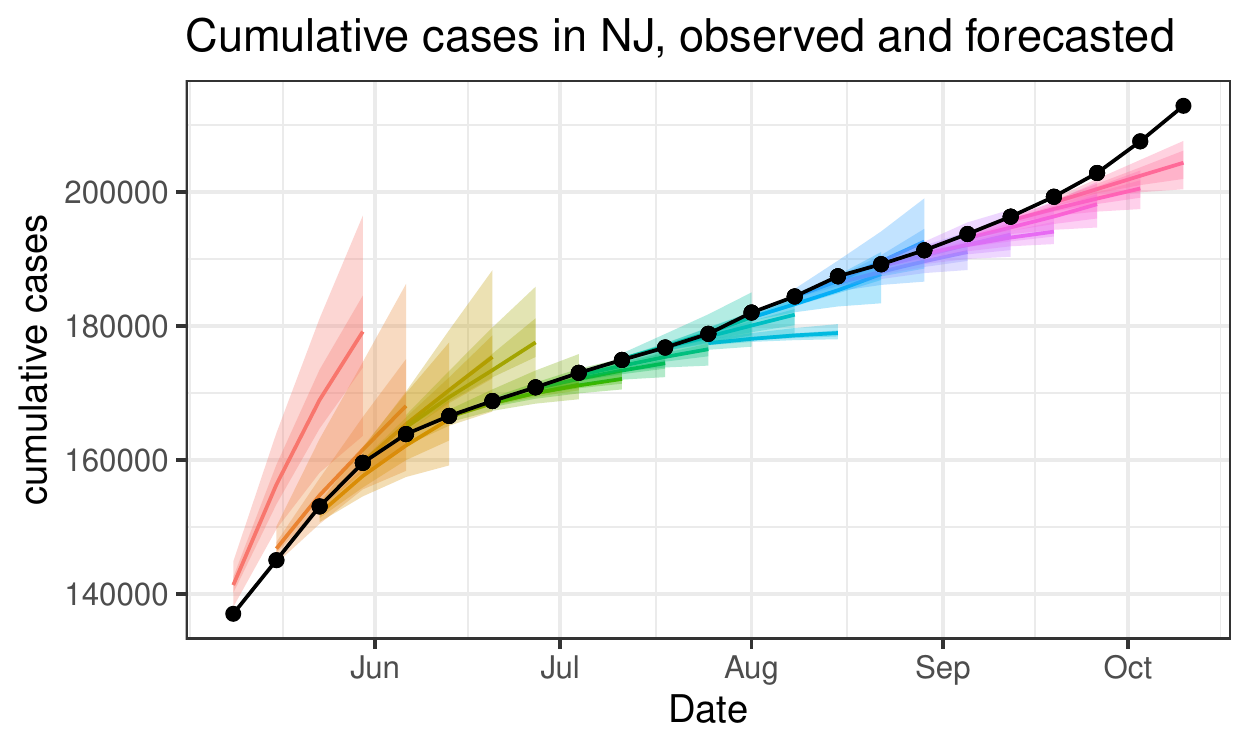}

 \bigskip

 \includegraphics[width=0.32\linewidth]{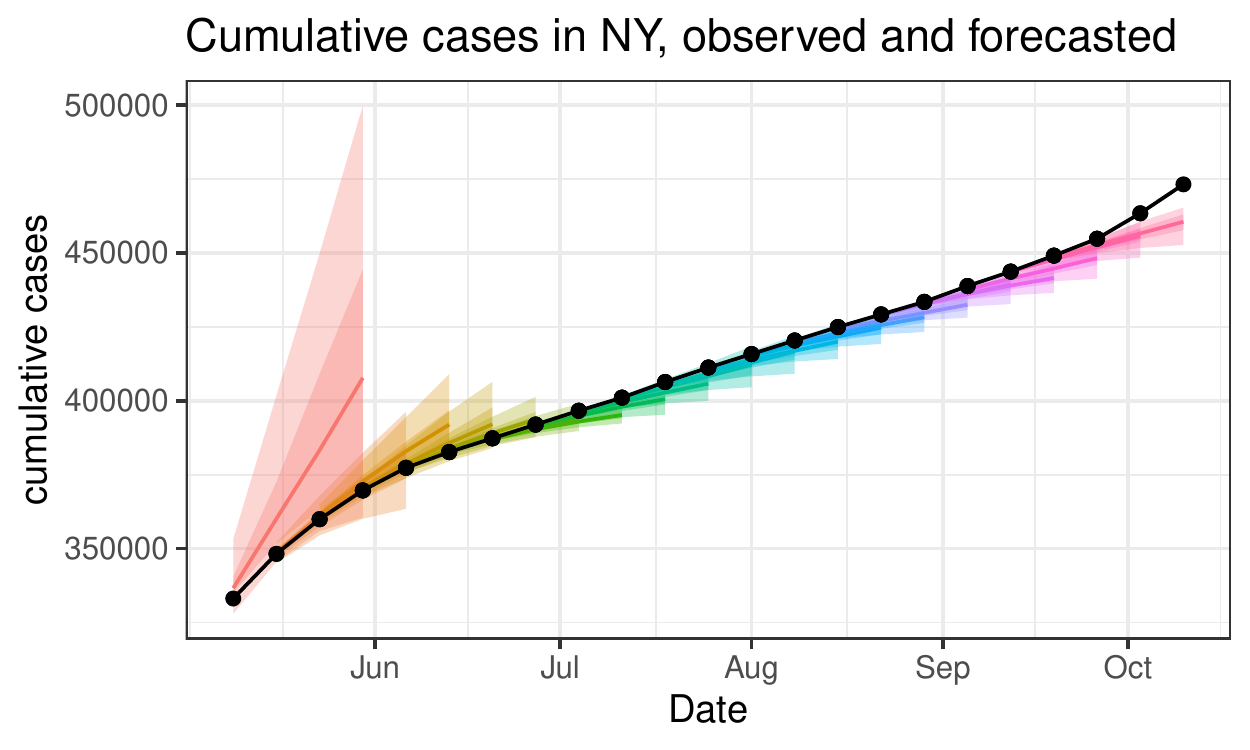}
 \includegraphics[width=0.32\linewidth]{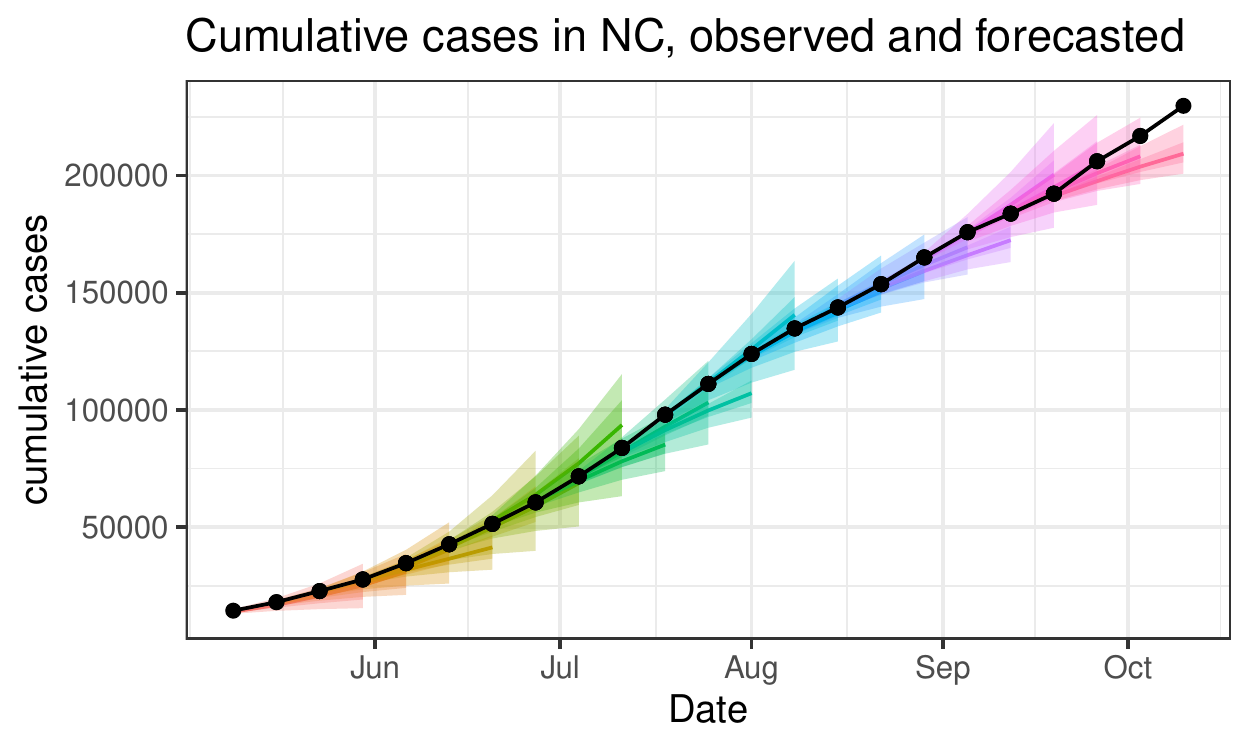}
 \includegraphics[width=0.32\linewidth]{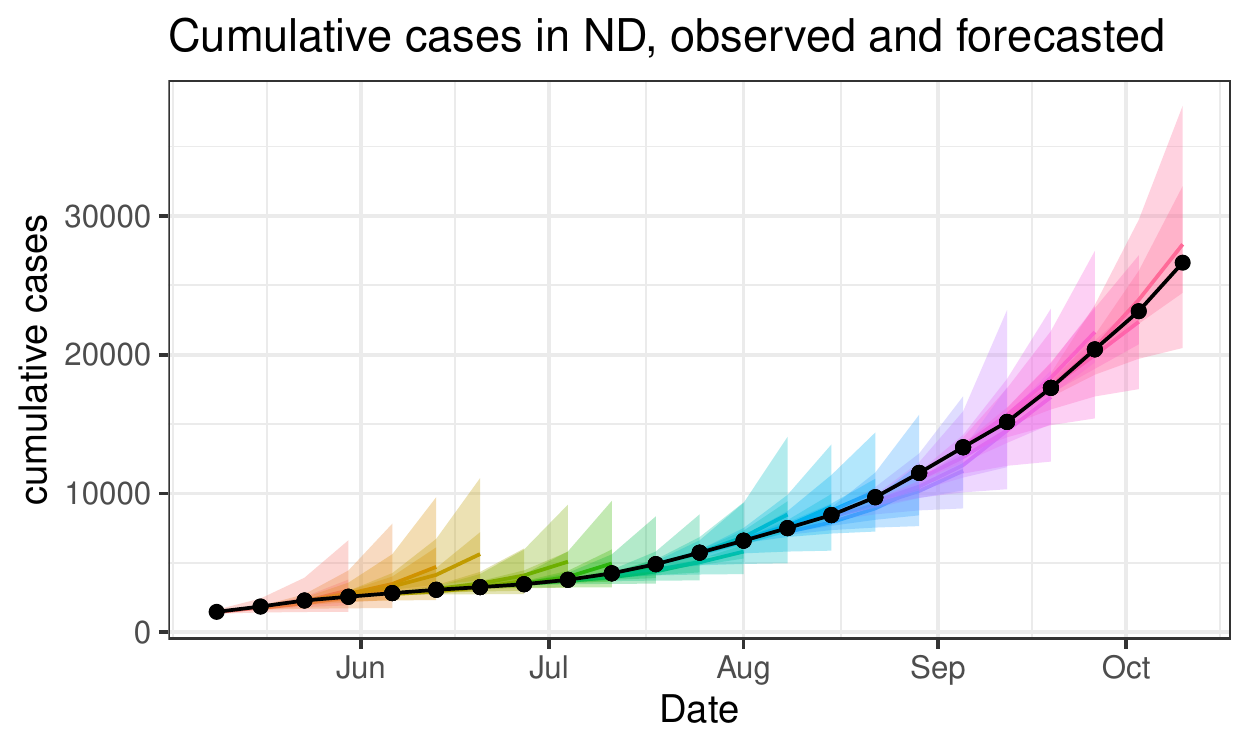}

 \bigskip

 \includegraphics[width=0.32\linewidth]{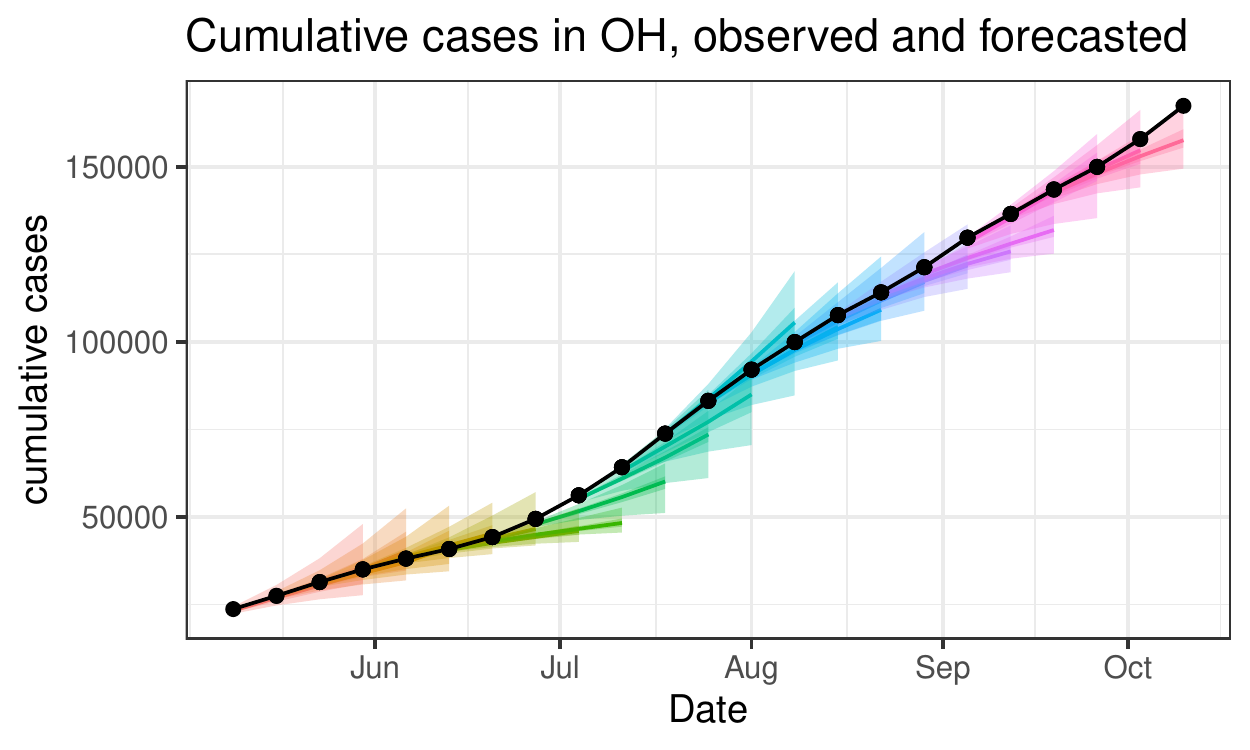}
 \includegraphics[width=0.32\linewidth]{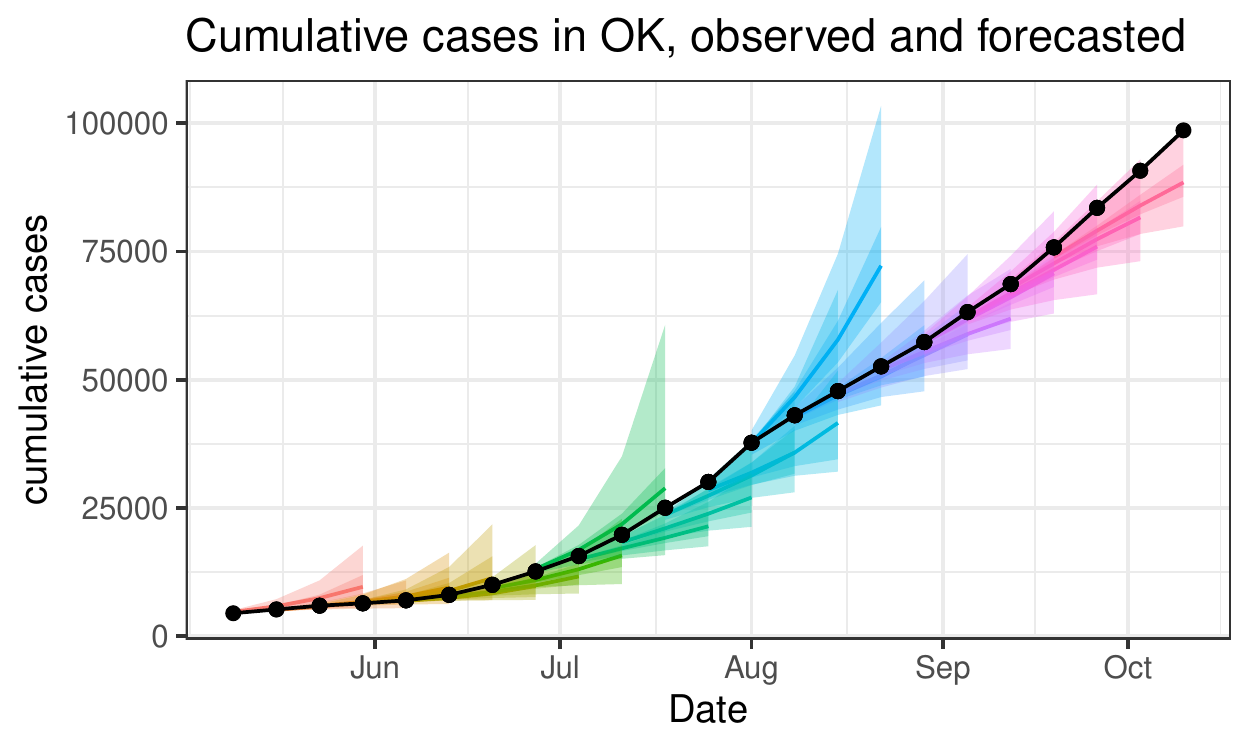}
 \includegraphics[width=0.32\linewidth]{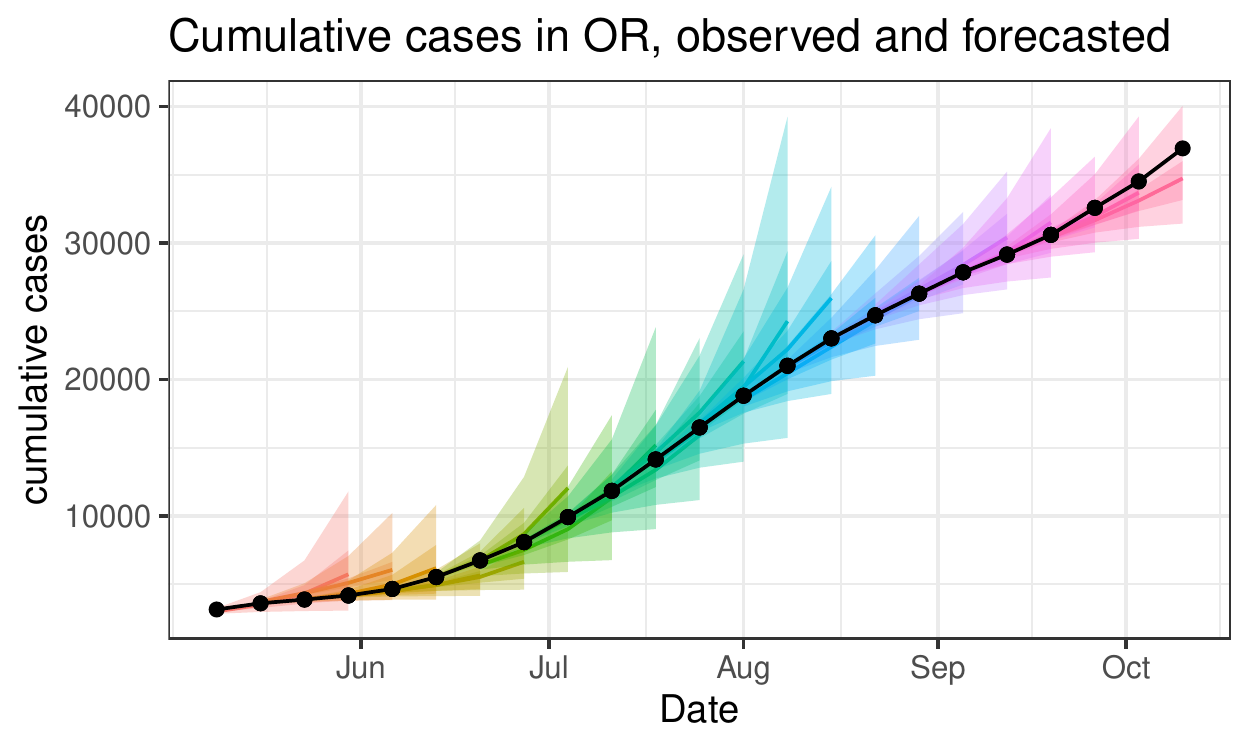}

\bigskip

 \includegraphics[width=0.32\linewidth]{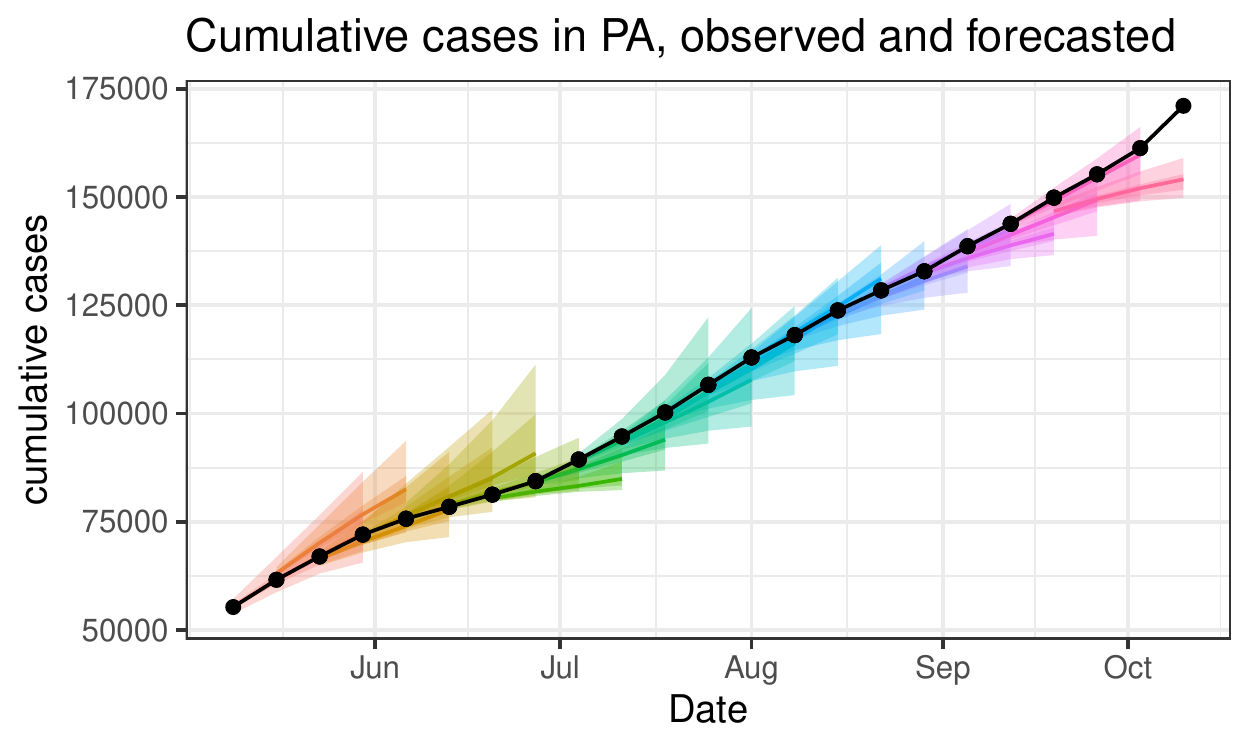}
 \includegraphics[width=0.32\linewidth]{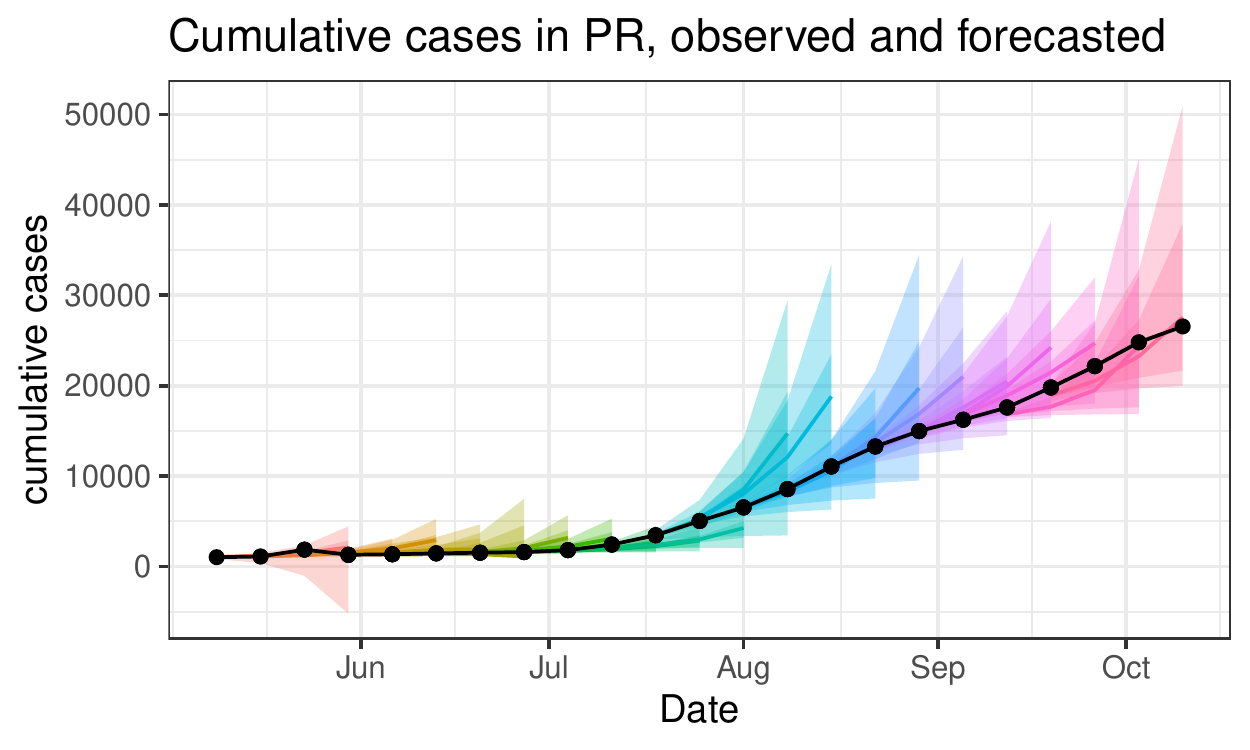}
 \includegraphics[width=0.32\linewidth]{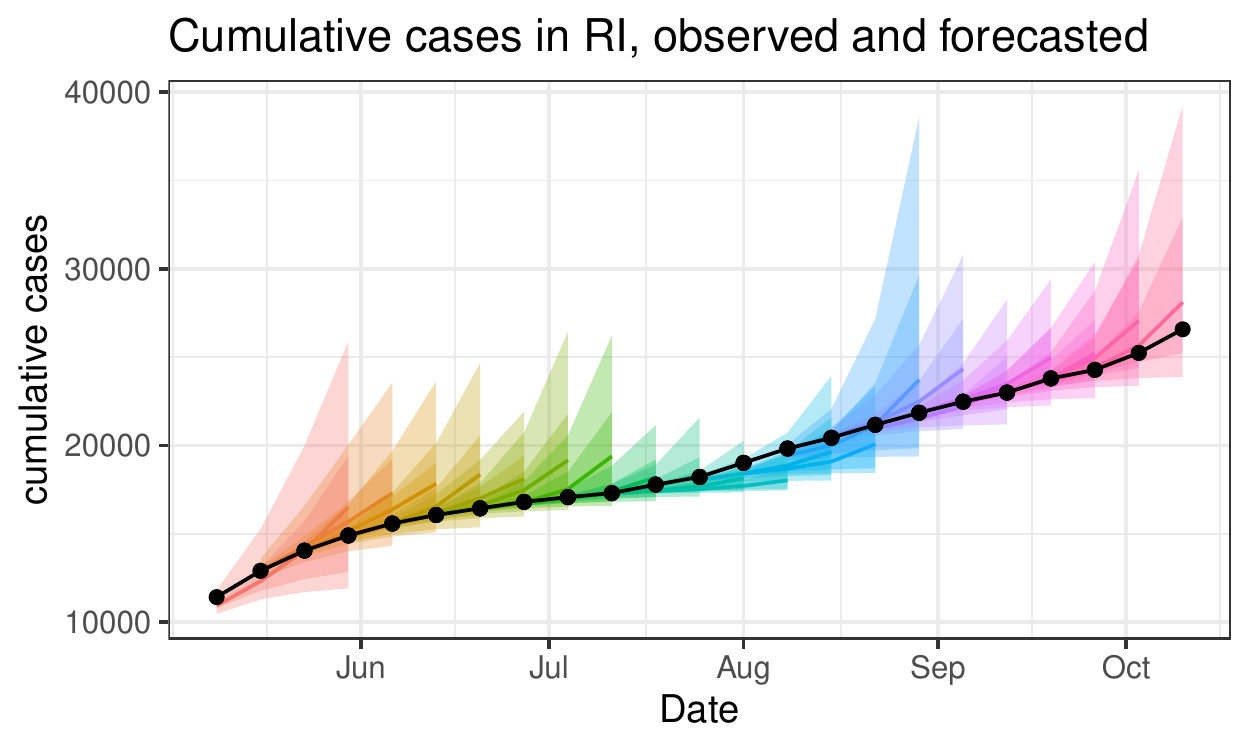}

 \bigskip

 \includegraphics[width=0.32\linewidth]{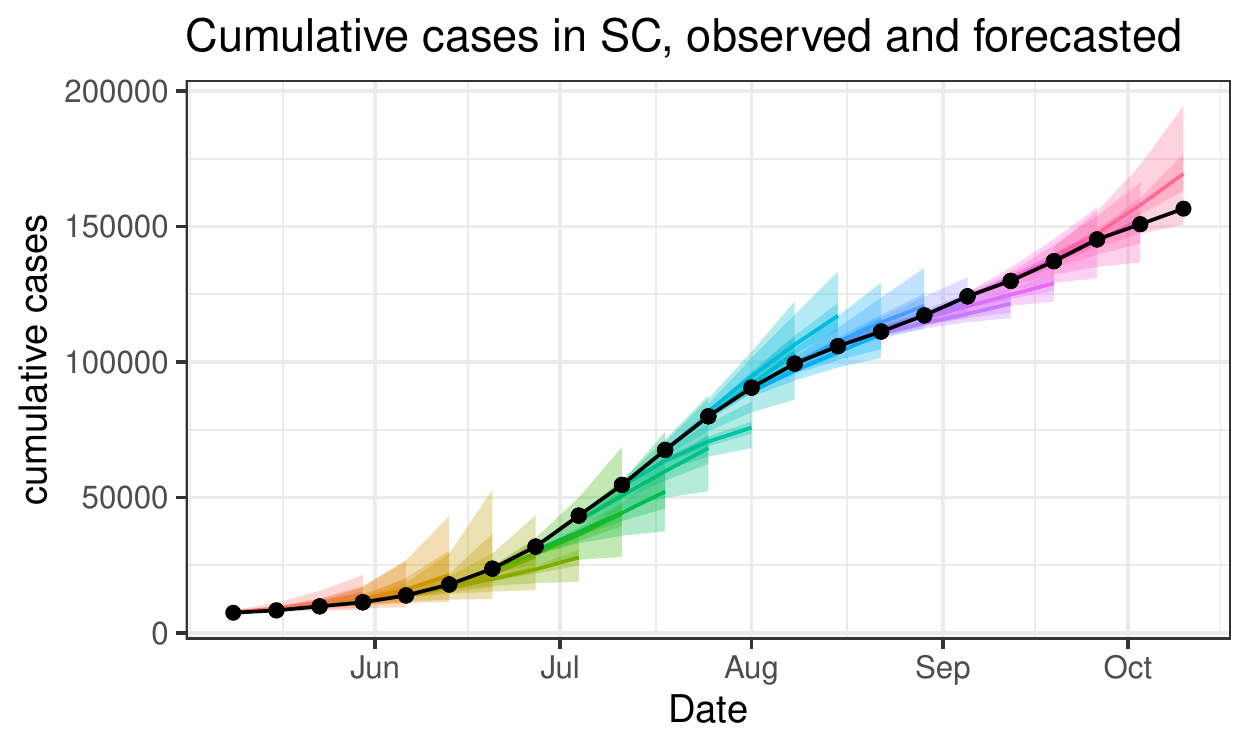}
 \includegraphics[width=0.32\linewidth]{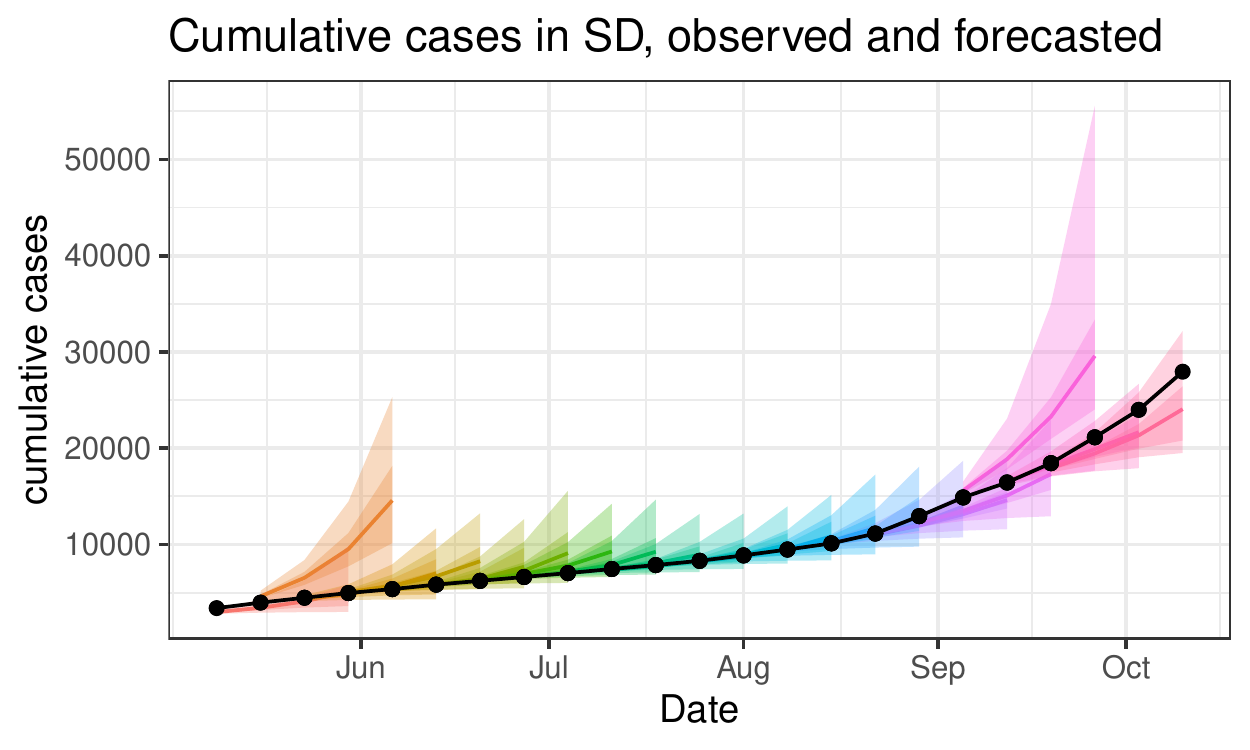}
 \includegraphics[width=0.32\linewidth]{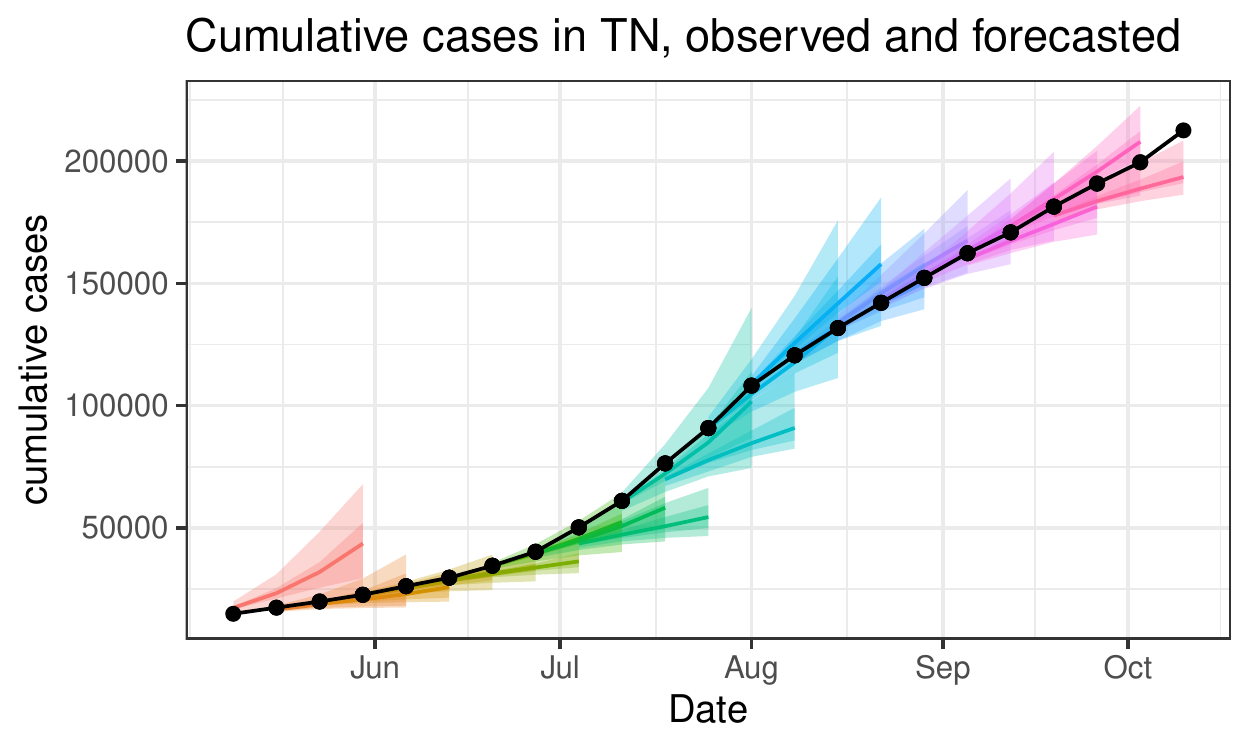}

 \bigskip

 \includegraphics[width=0.32\linewidth]{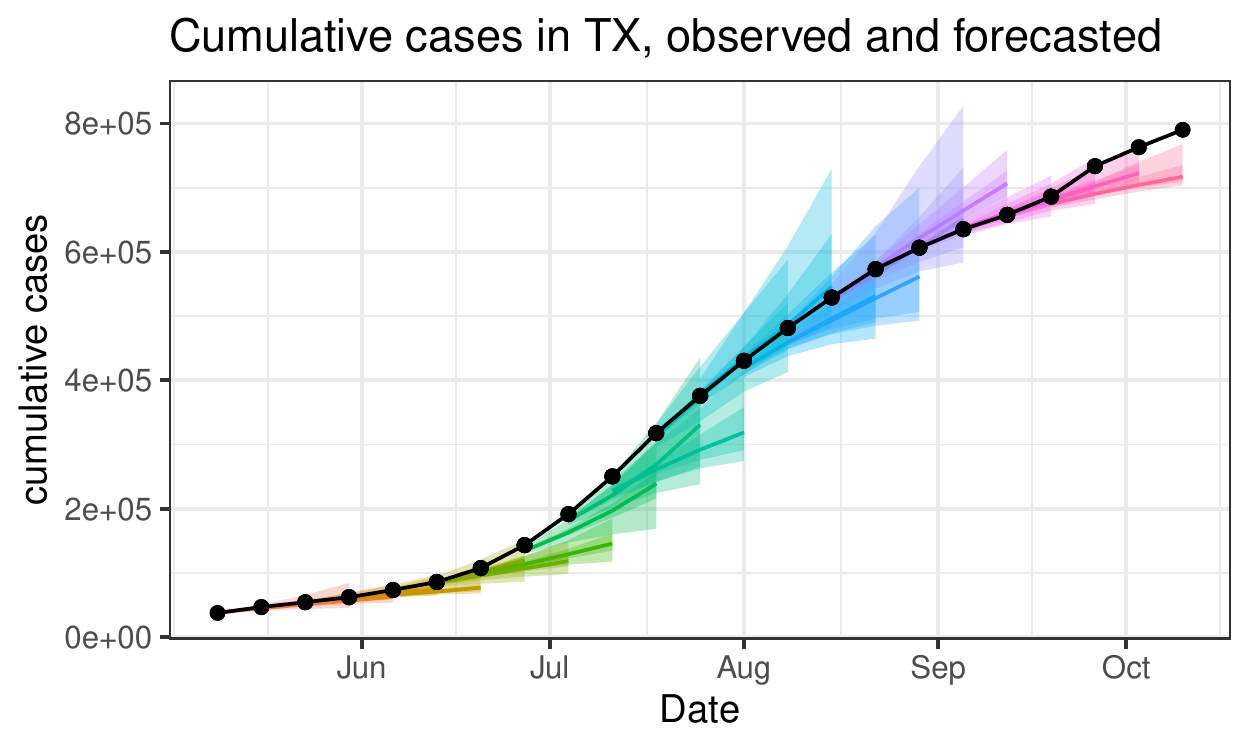}
 \includegraphics[width=0.32\linewidth]{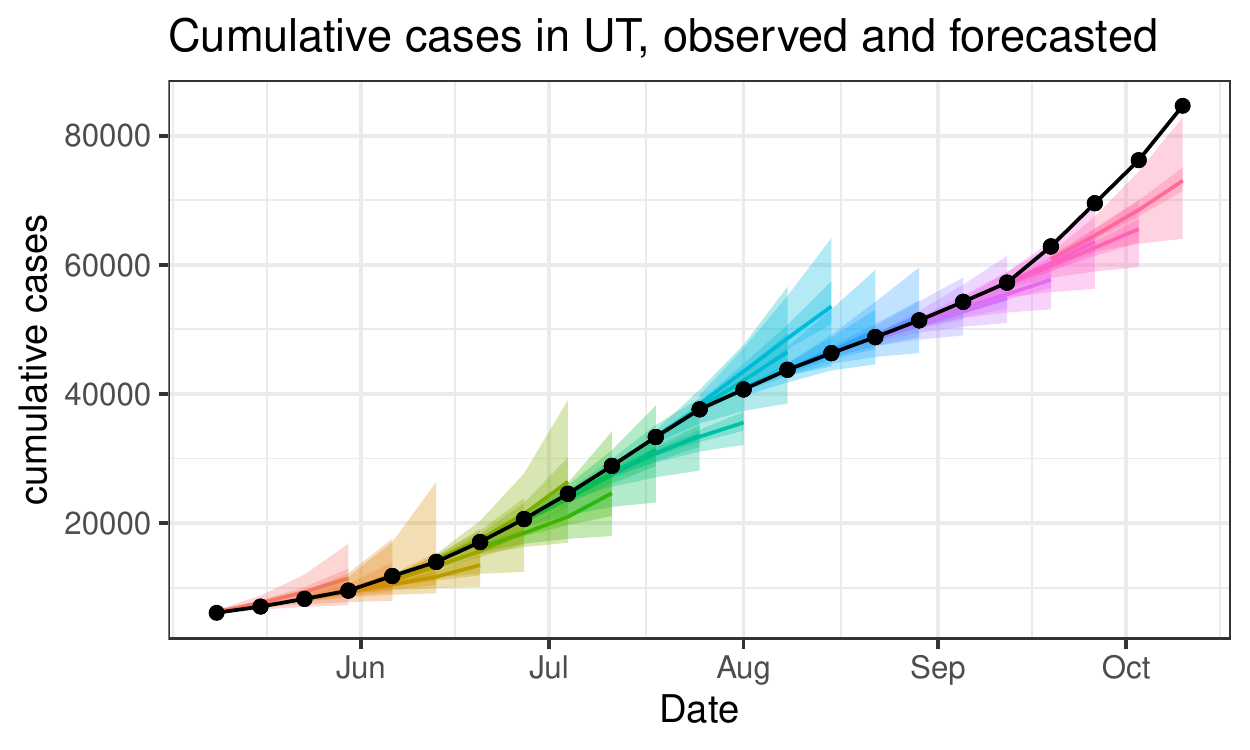}
 \includegraphics[width=0.32\linewidth]{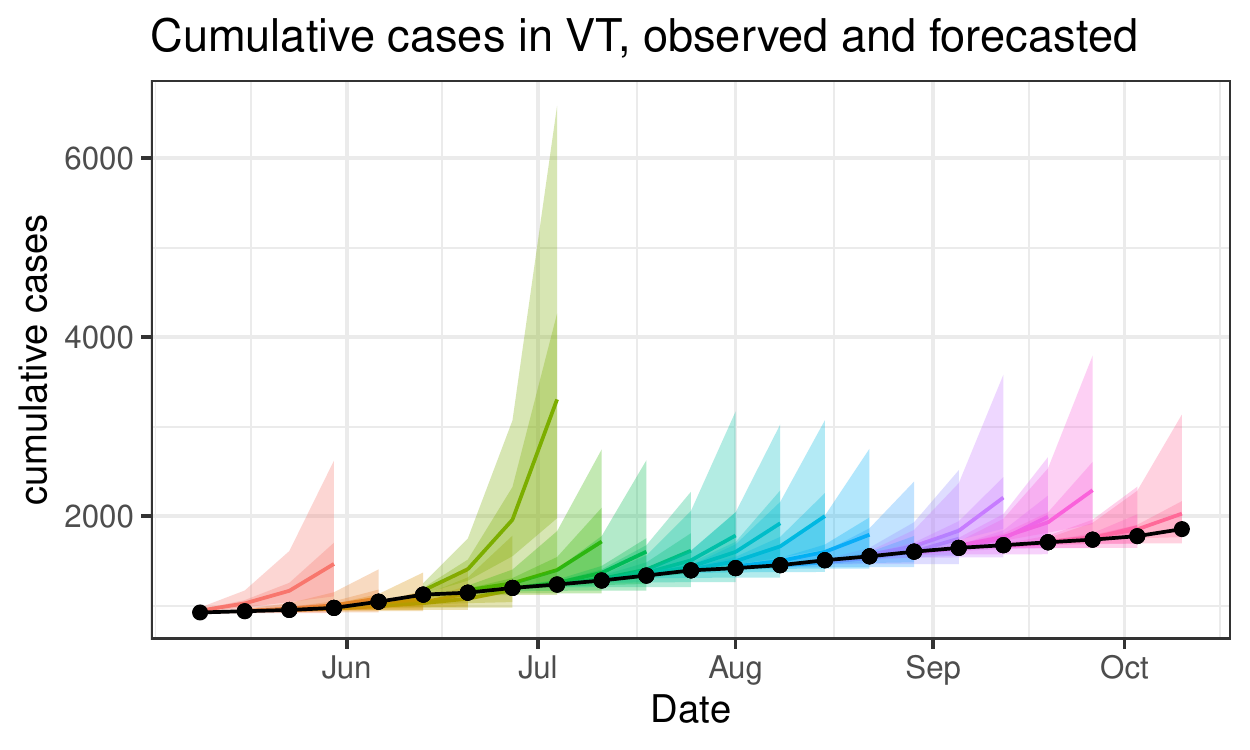}

 \bigskip

 \includegraphics[width=0.32\linewidth]{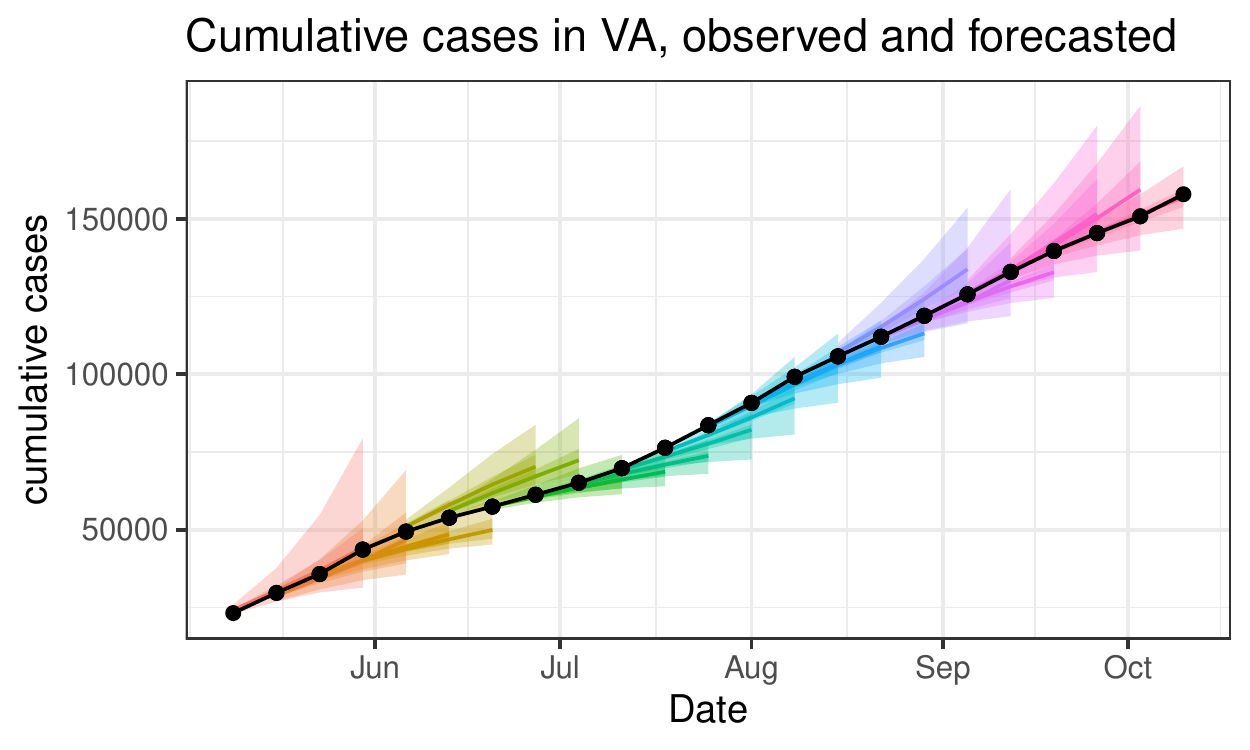}
 \includegraphics[width=0.32\linewidth]{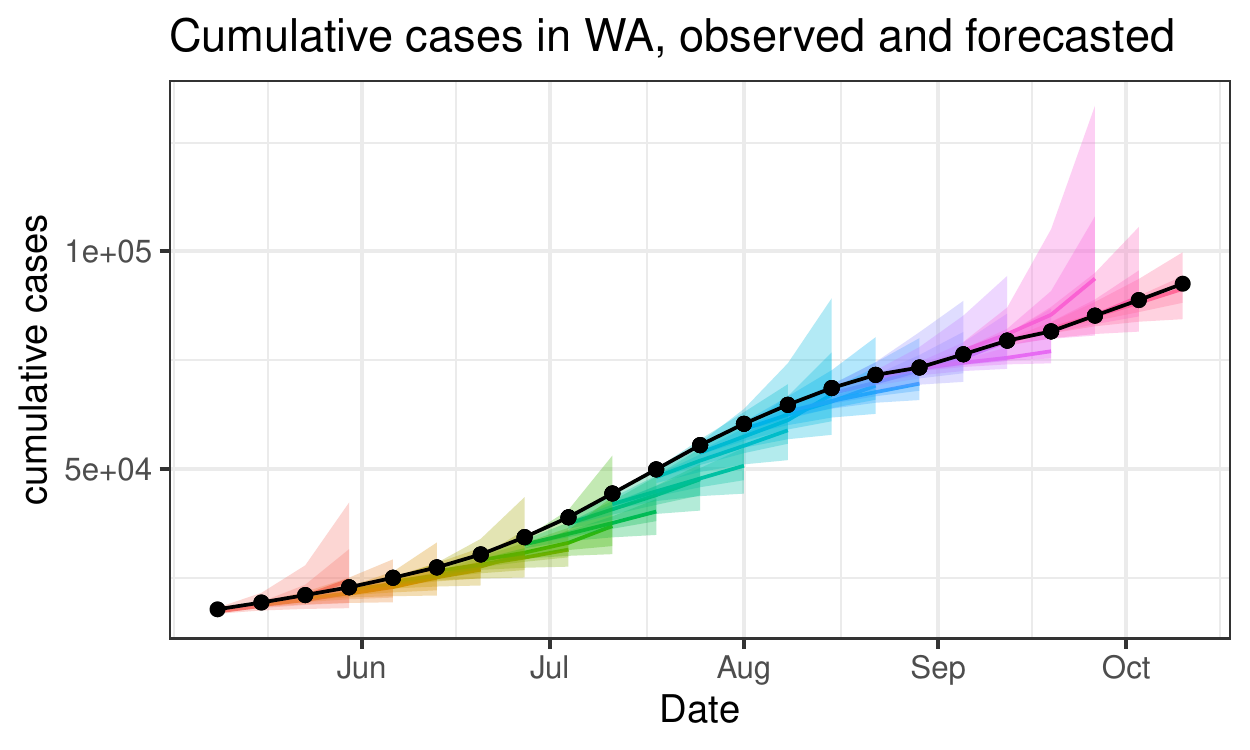}
 \includegraphics[width=0.32\linewidth]{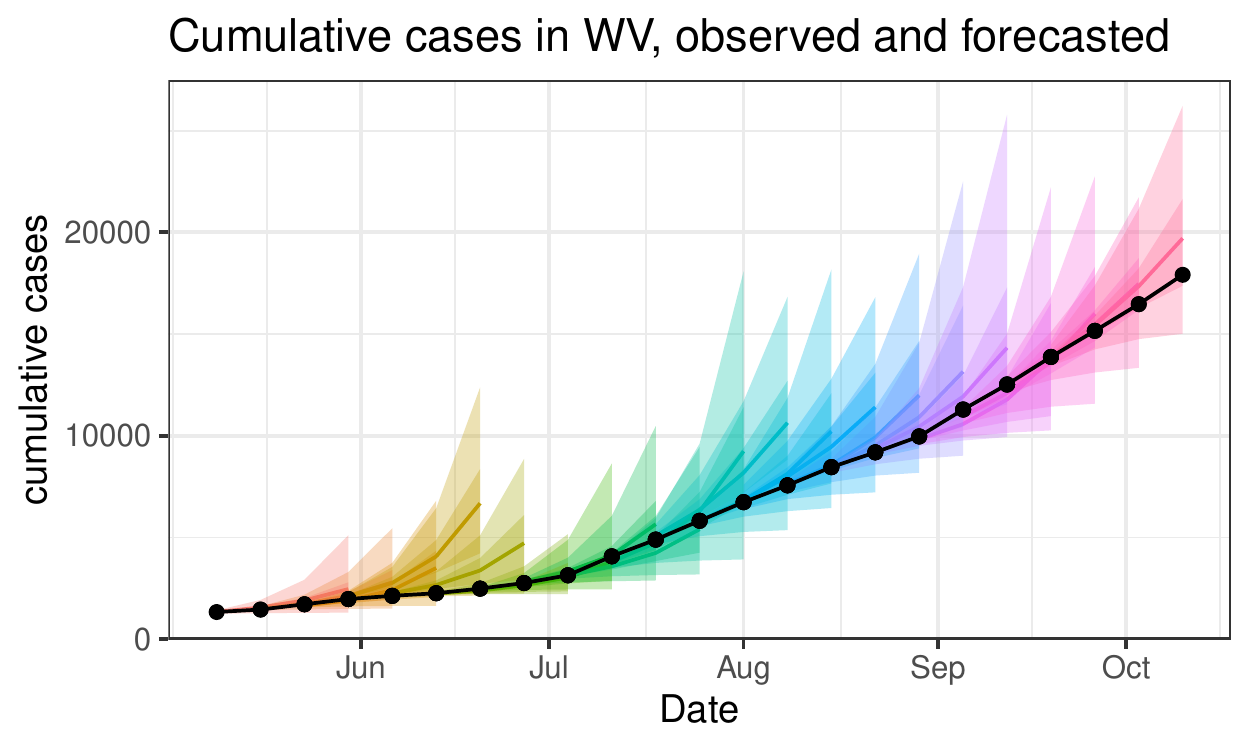}

 \bigskip

 \includegraphics[width=0.32\linewidth]{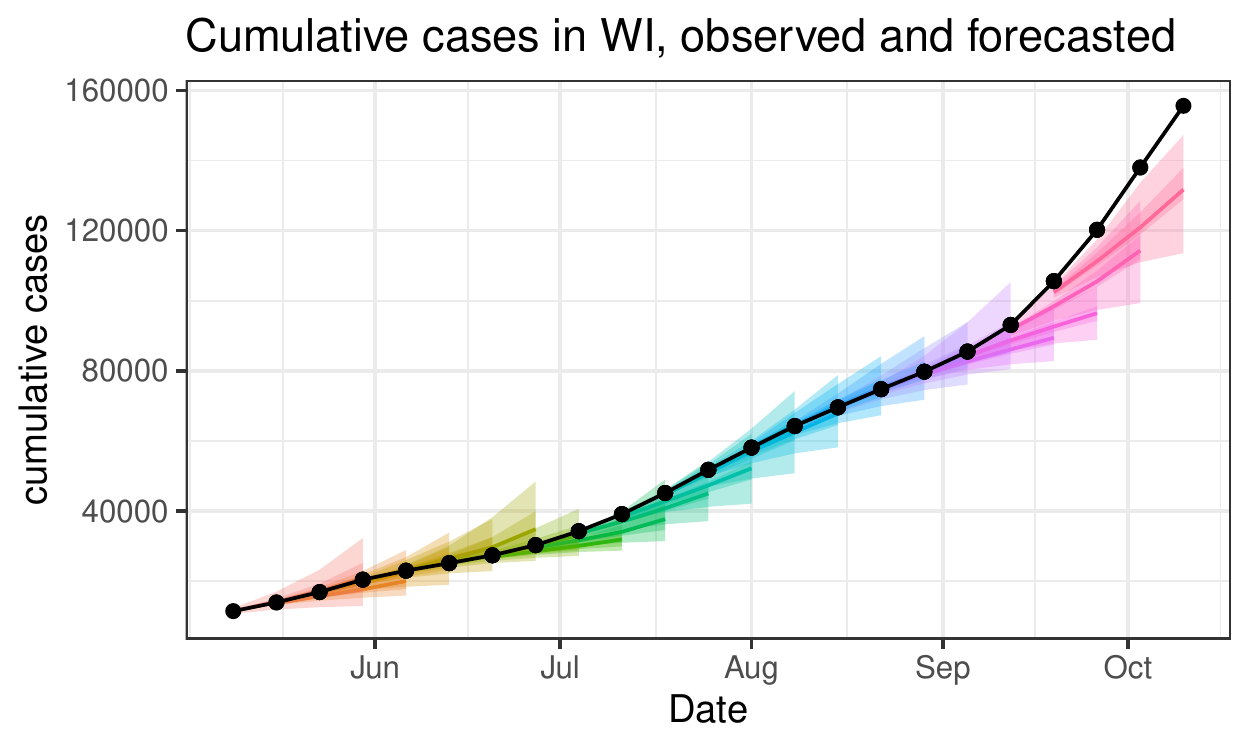}
 \includegraphics[width=0.32\linewidth]{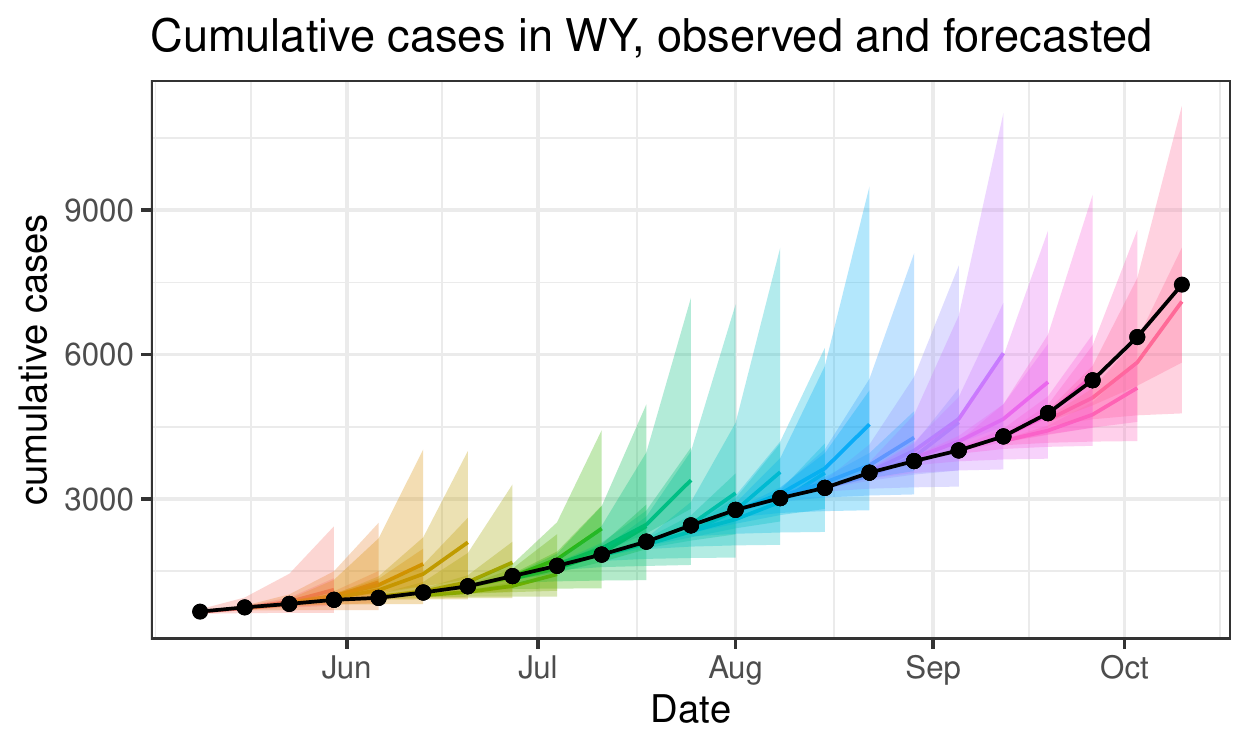}
 \includegraphics[width=0.32\linewidth]{COVIDfigs/US_case_quant_forecasts.pdf}
\end{center}

\subsection{Cumulative death forecasts}
\label{fig:EpiCovDA_death_append}
 We show below the cumulative death forecasts for each of the 50 states, D.C., and Puerto Rico. The black curves indicate the true values as reported by the COVID Tracking Project \cite{CVDT}. The widest shaded regions correspond to the central 95\% prediction intervals. The smaller shaded regions correspond to the central 50\% prediction intervals. The darker colored curves in shaded regions are the median forecasts. COVID-19 data was provided by The COVID Tracking Project at {\sl The Atlantic} under a CC BY 4.0 license.
 
 \bigskip
 
\begin{center}

\includegraphics[width=0.32\linewidth]{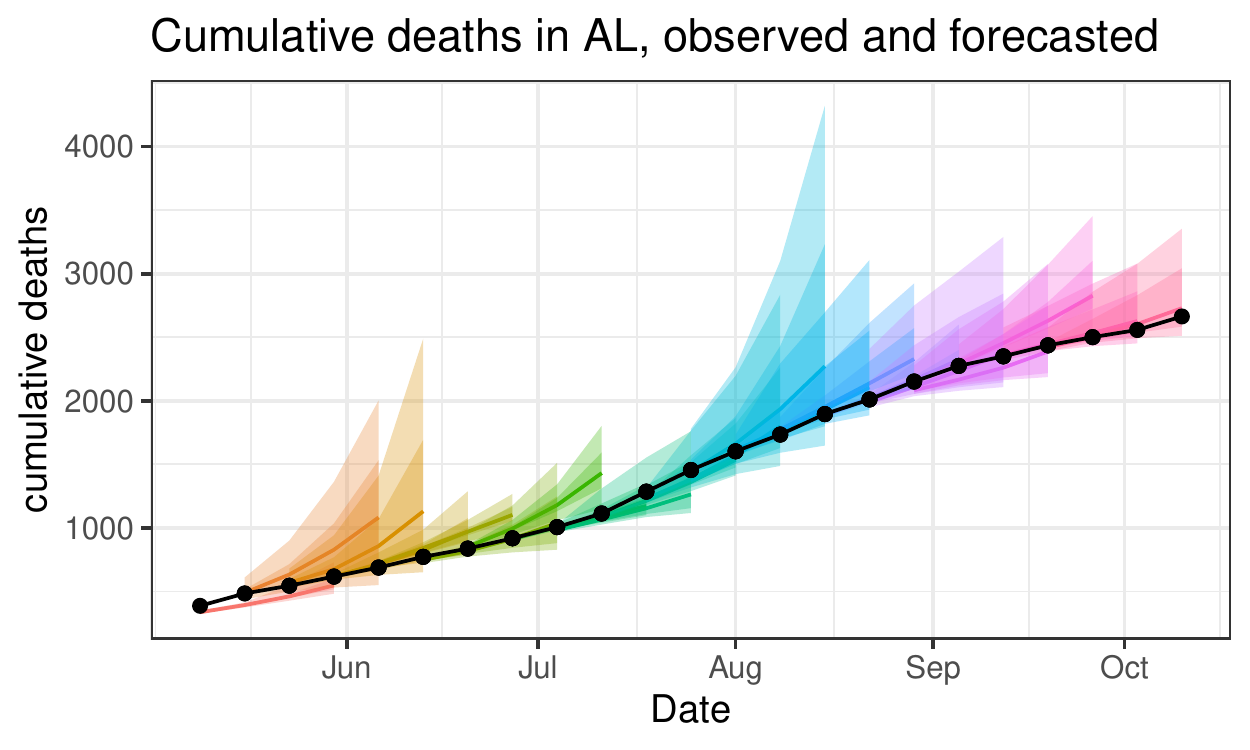}
 \includegraphics[width=0.32\linewidth]{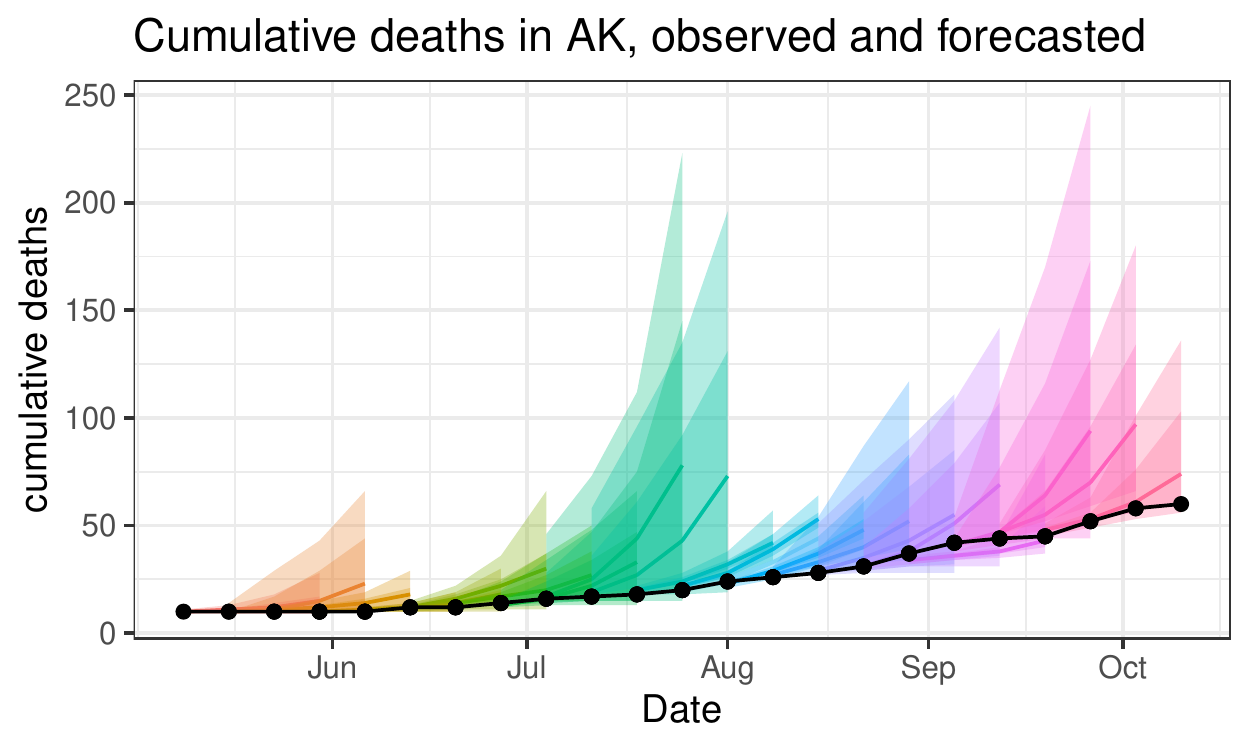}
 \includegraphics[width=0.32\linewidth]{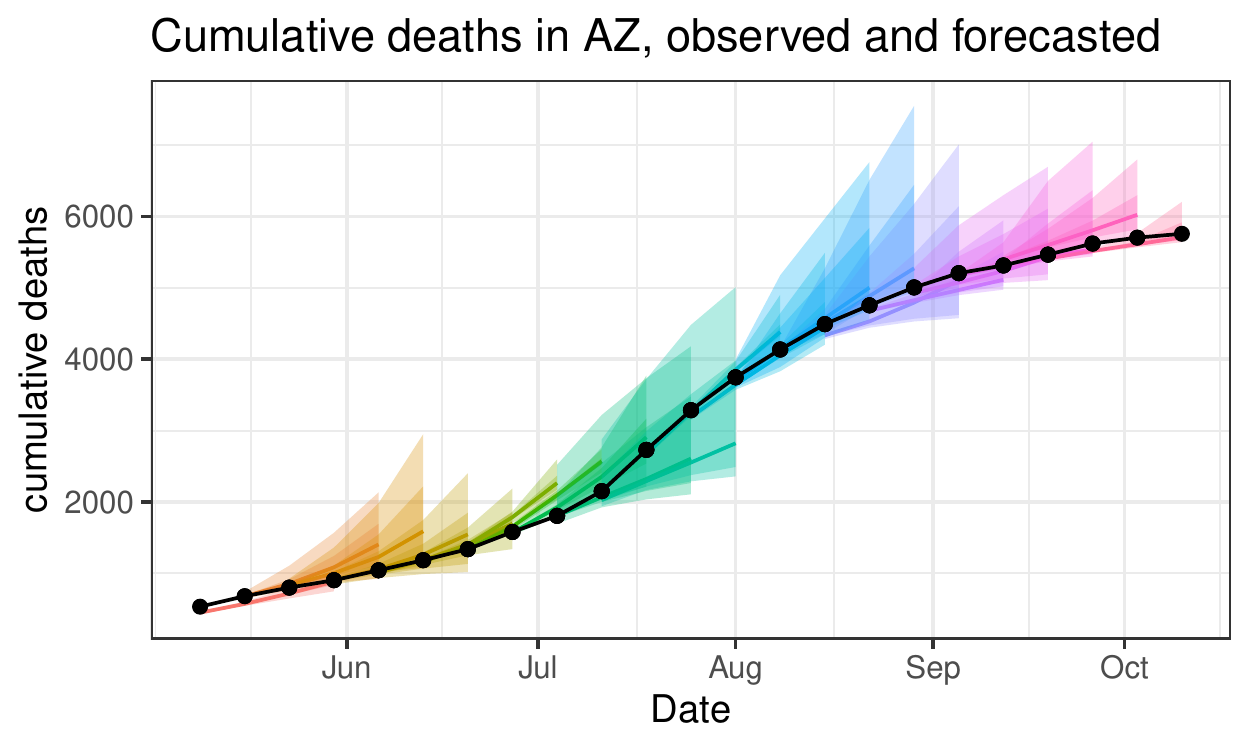}

 \bigskip

 \includegraphics[width=0.32\linewidth]{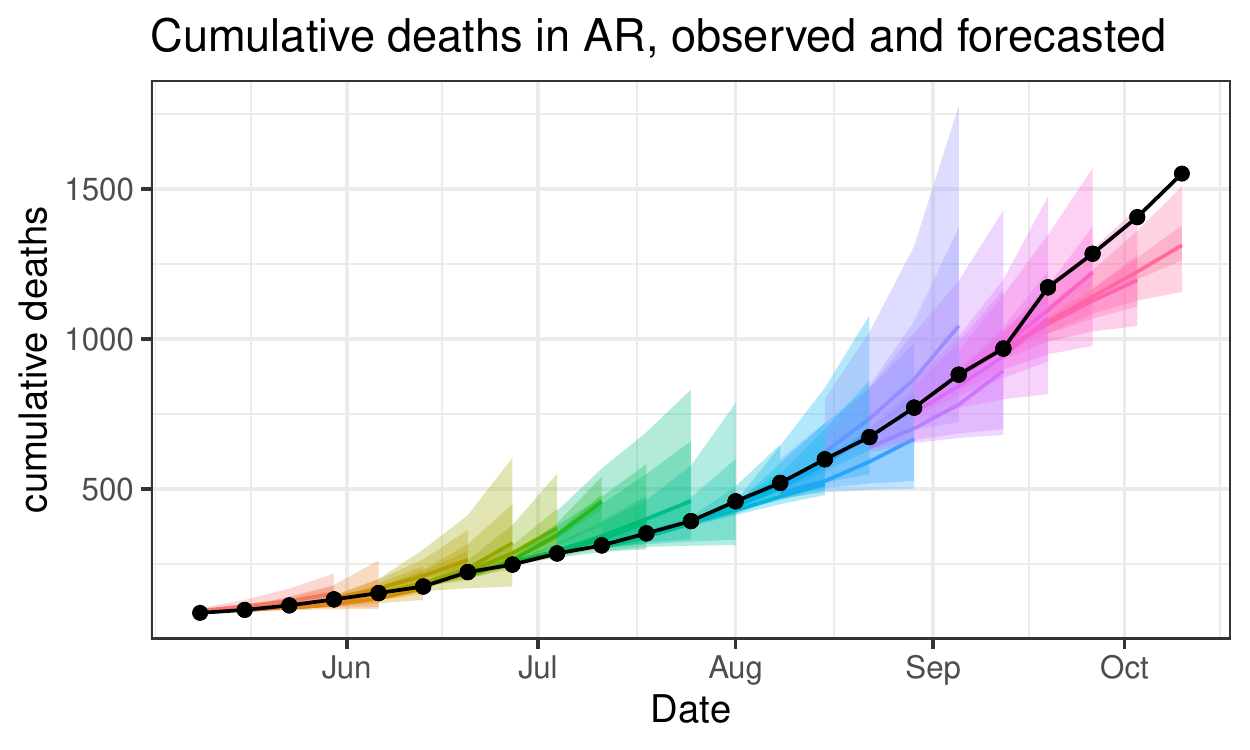}
 \includegraphics[width=0.32\linewidth]{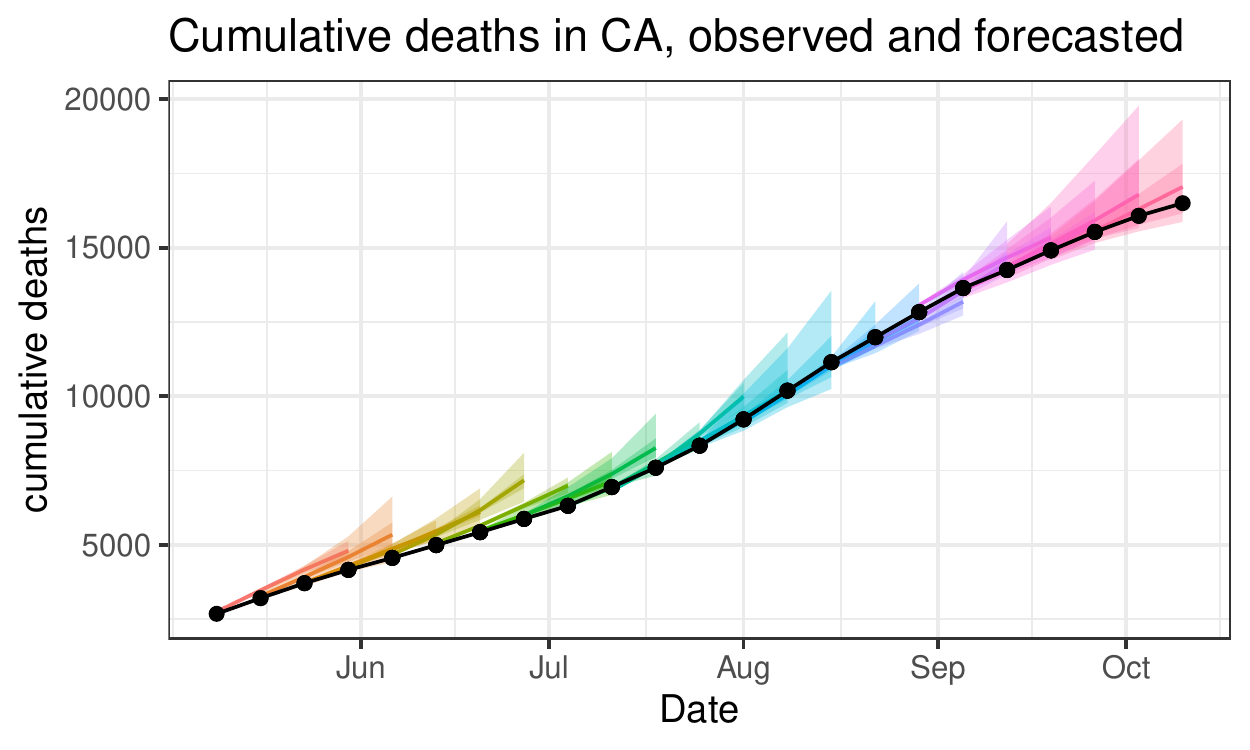}
 \includegraphics[width=0.32\linewidth]{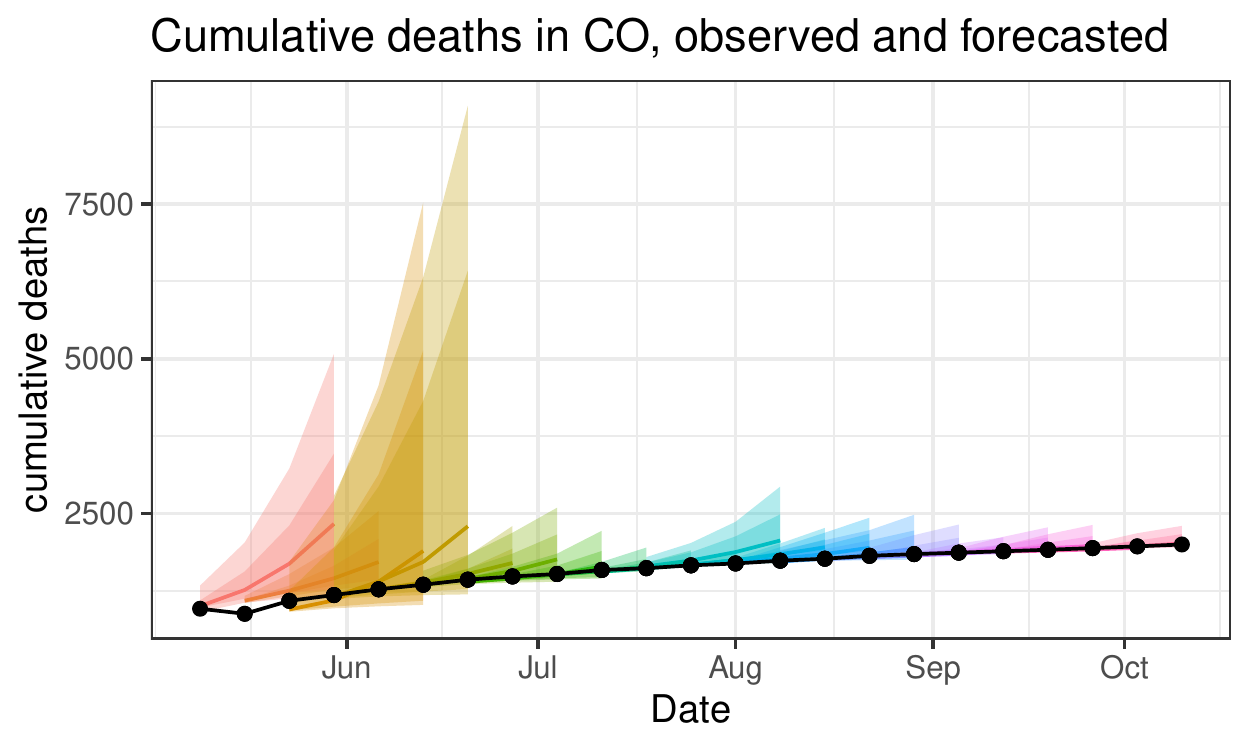}

 \bigskip

 \includegraphics[width=0.32\linewidth]{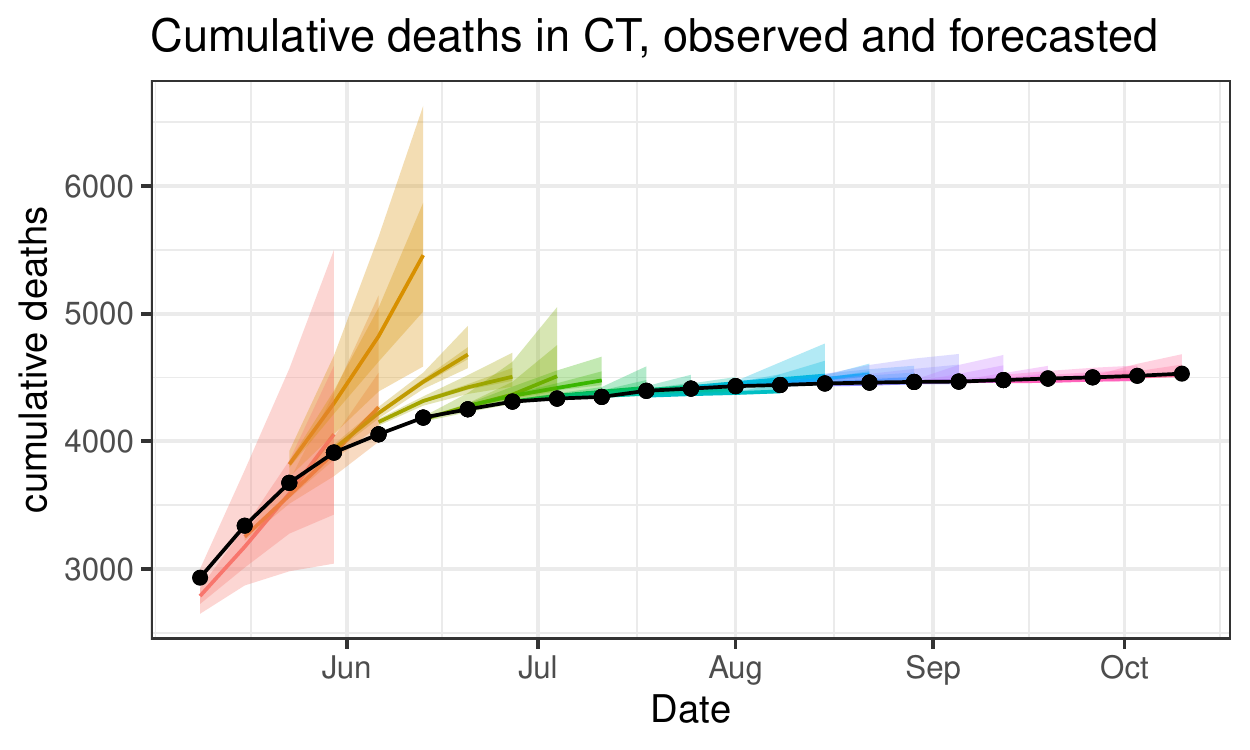}
 \includegraphics[width=0.32\linewidth]{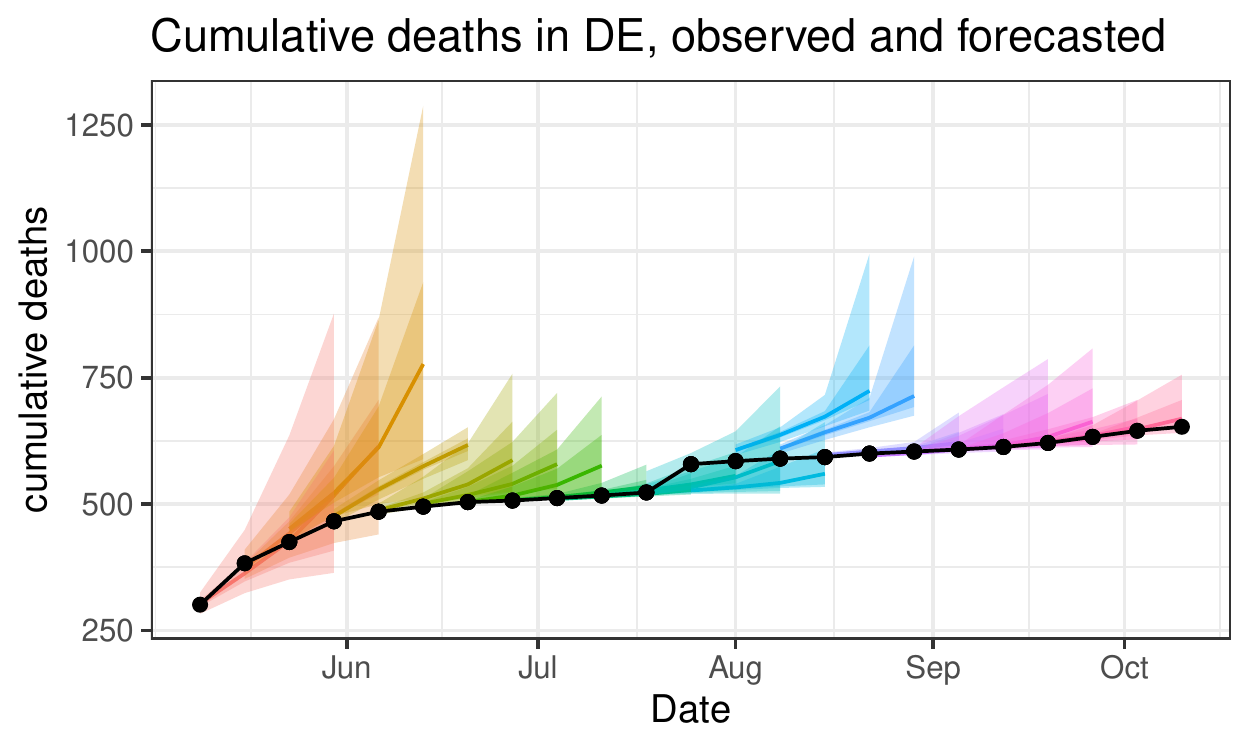}
 \includegraphics[width=0.32\linewidth]{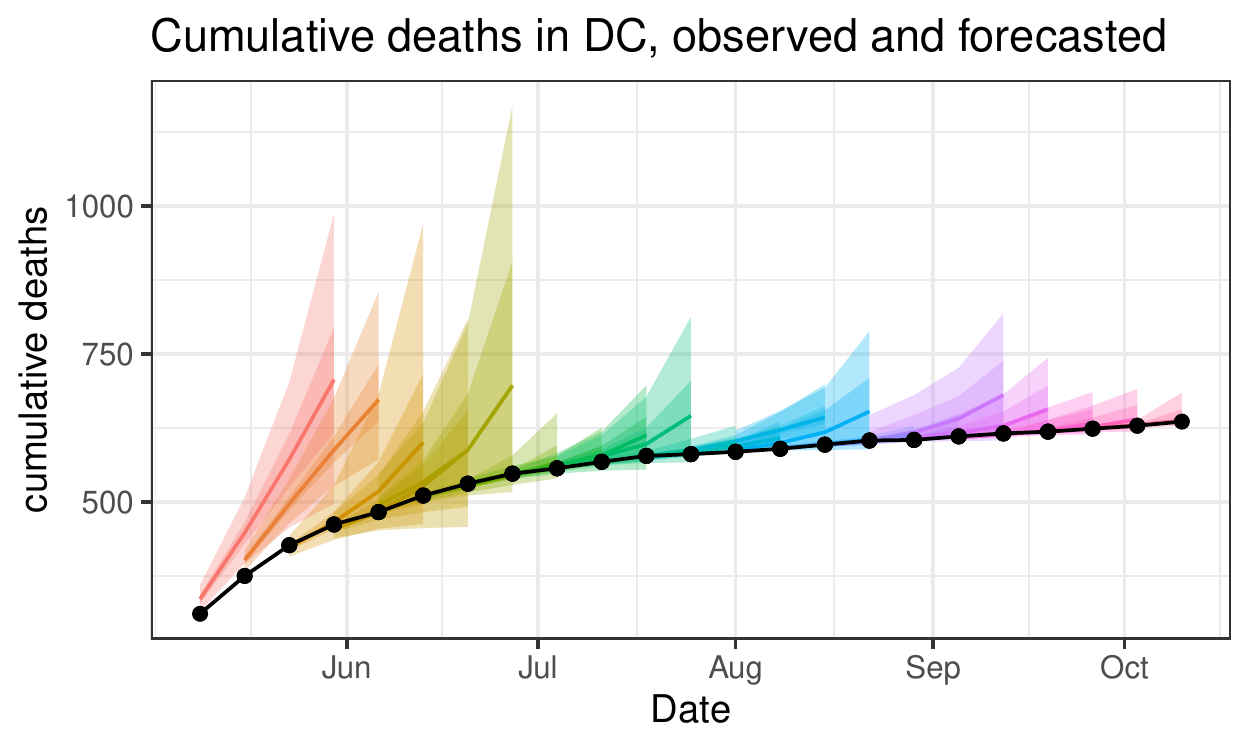}

 \bigskip

 \includegraphics[width=0.32\linewidth]{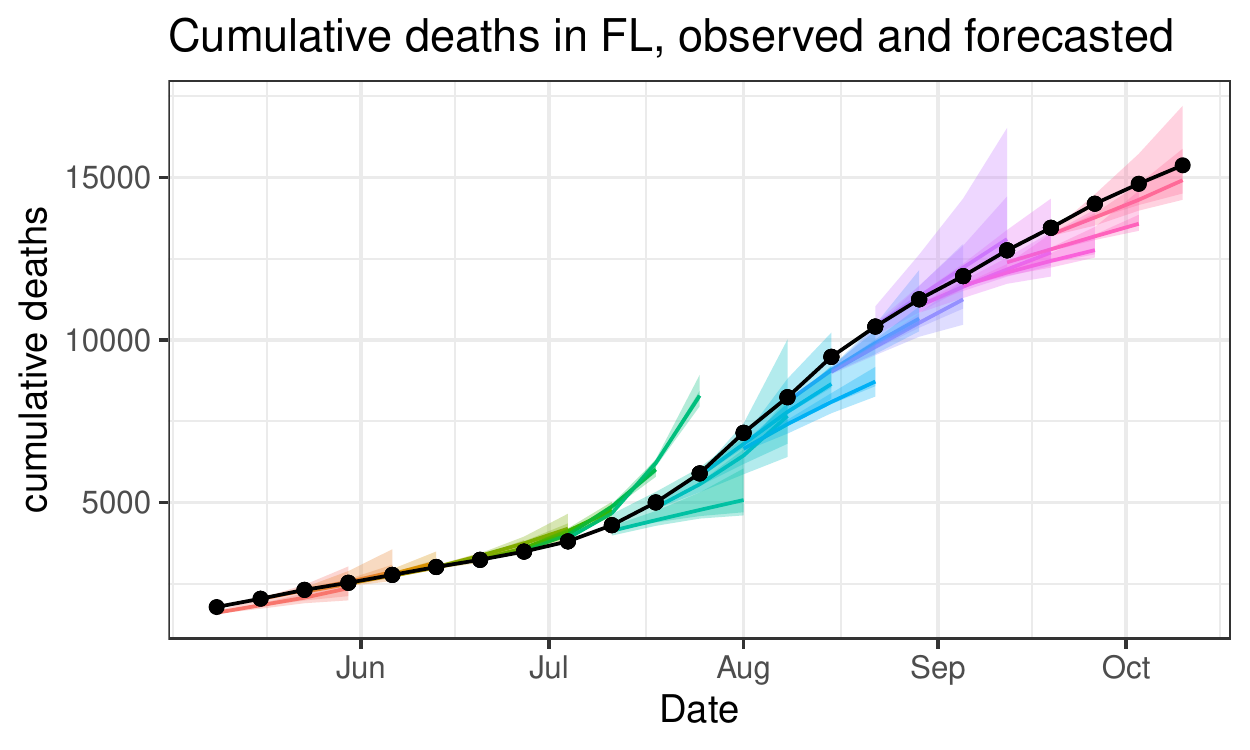}
 \includegraphics[width=0.32\linewidth]{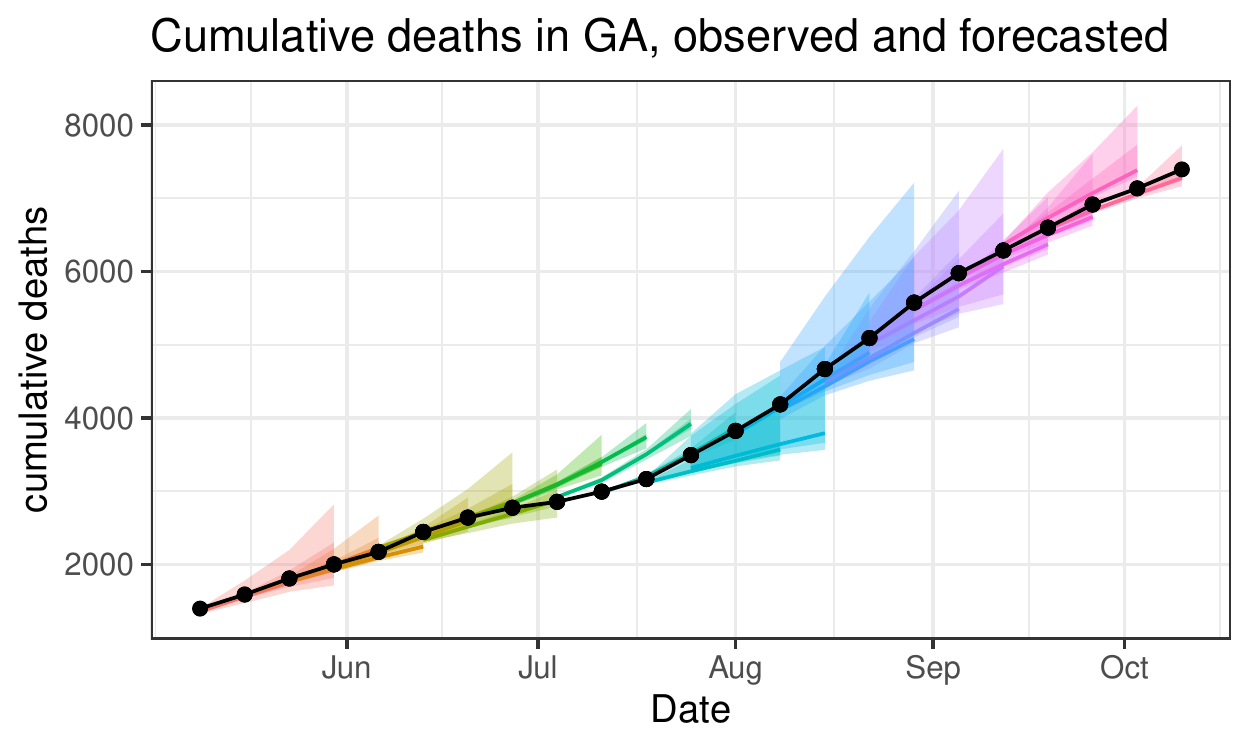}
 \includegraphics[width=0.32\linewidth]{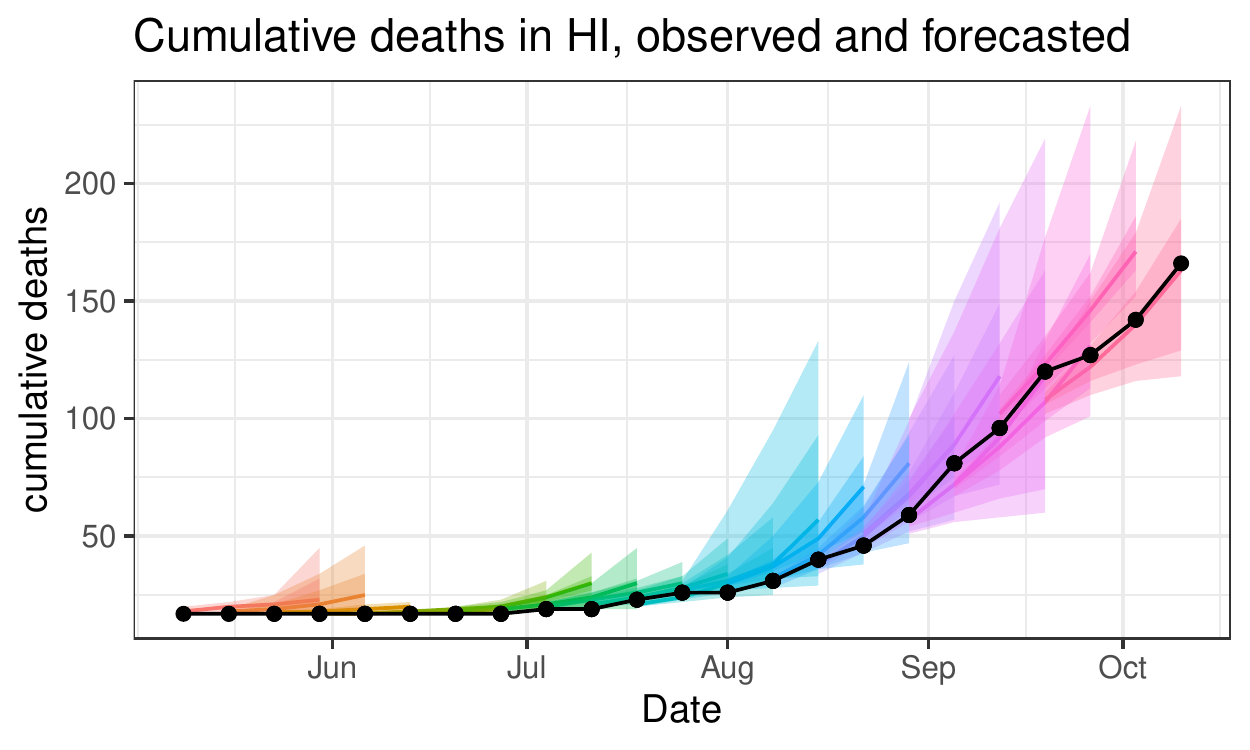}

 \bigskip

 \includegraphics[width=0.32\linewidth]{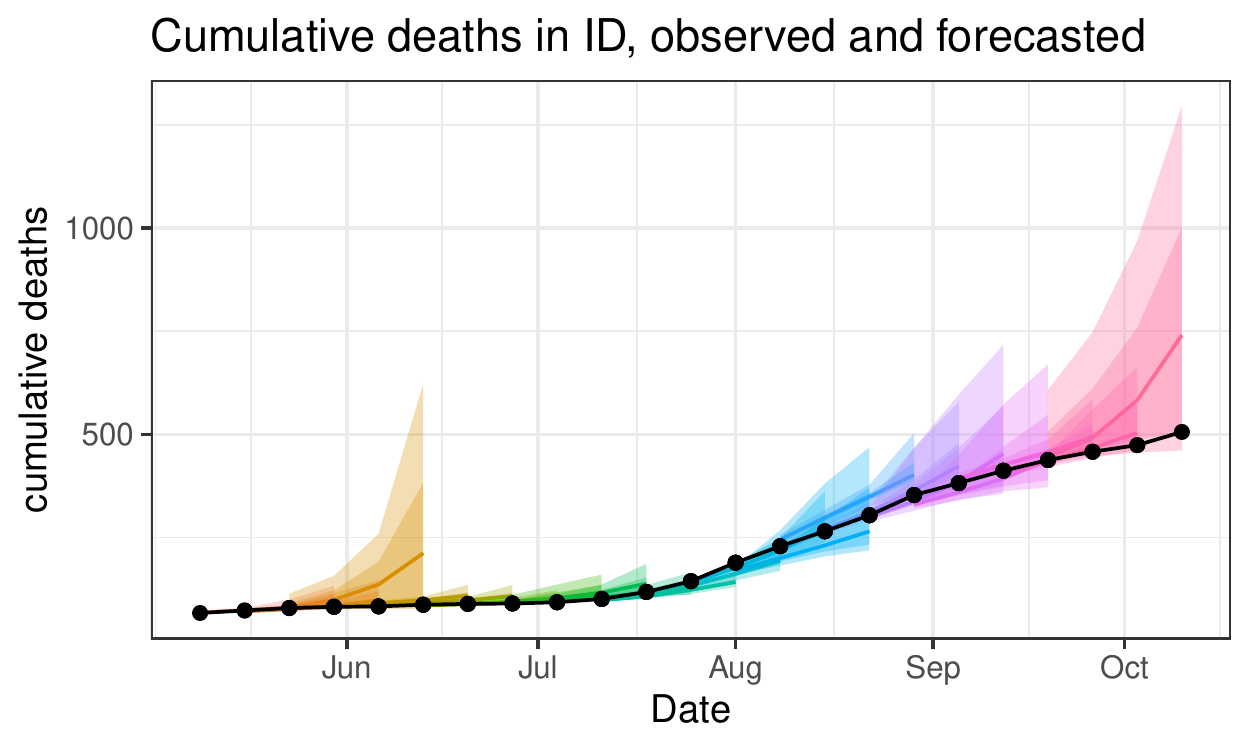}
 \includegraphics[width=0.32\linewidth]{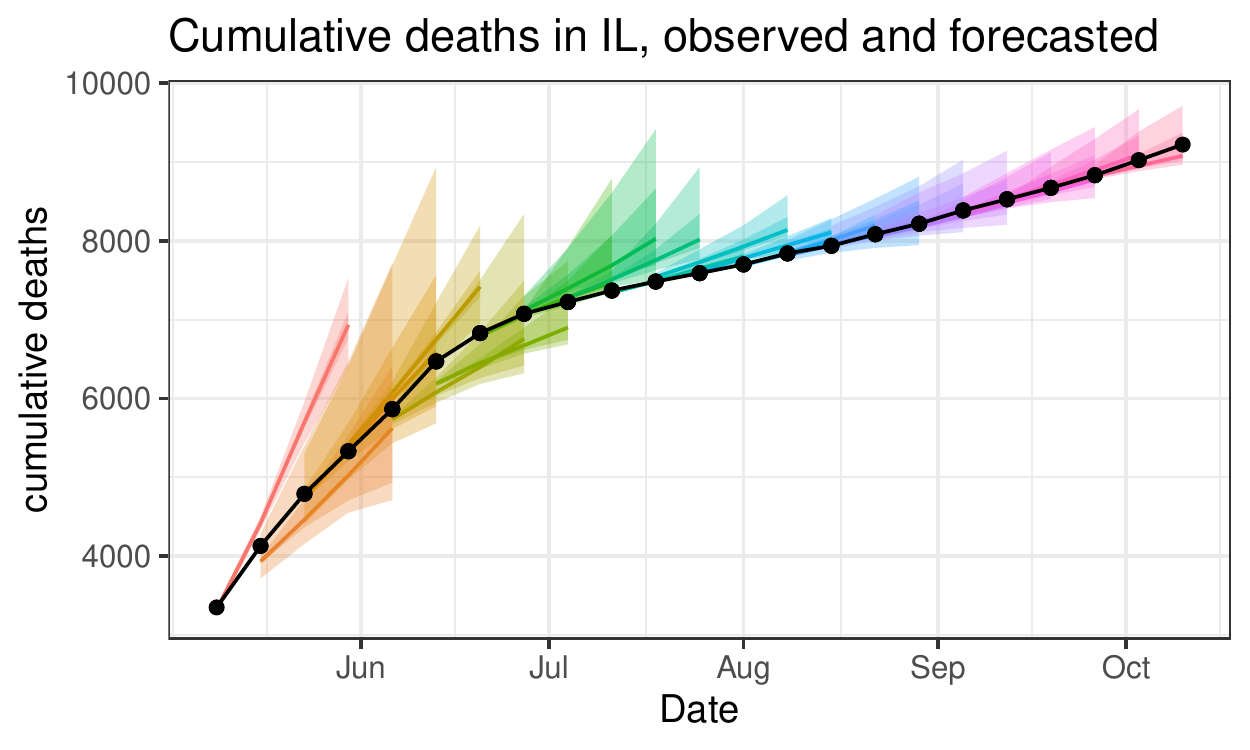}
 \includegraphics[width=0.32\linewidth]{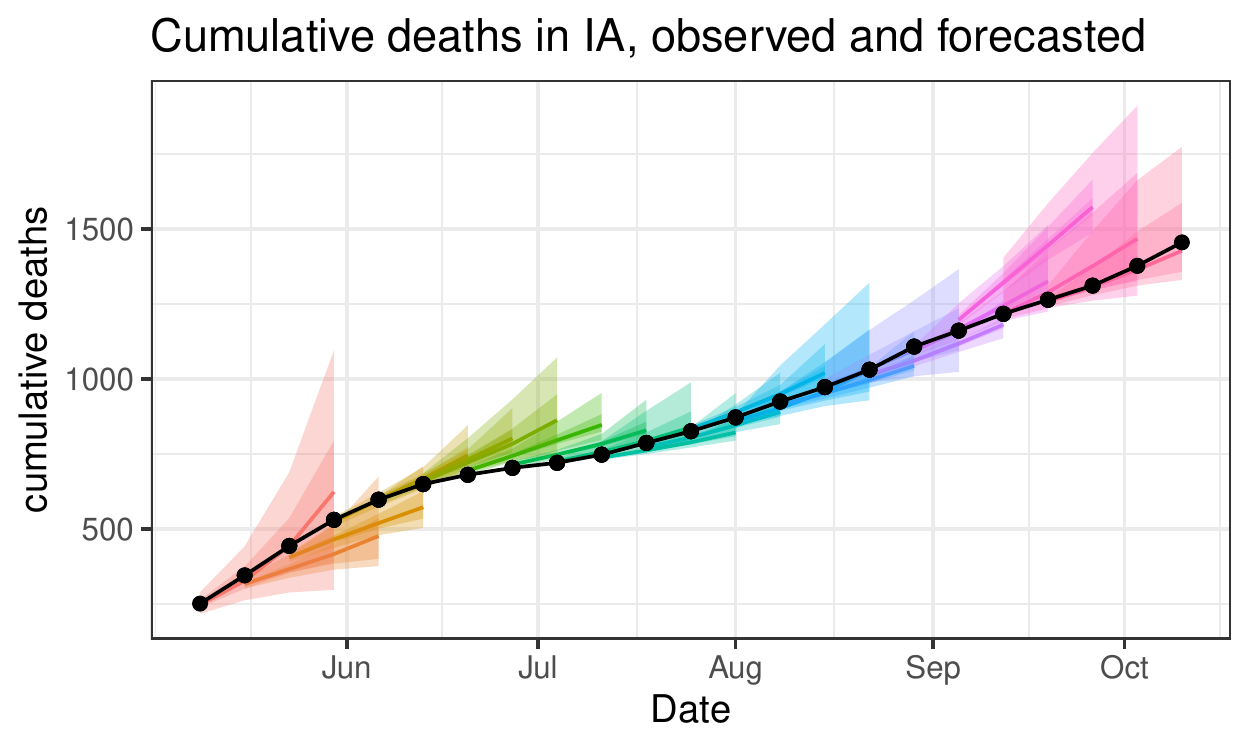}

 \bigskip

 \includegraphics[width=0.32\linewidth]{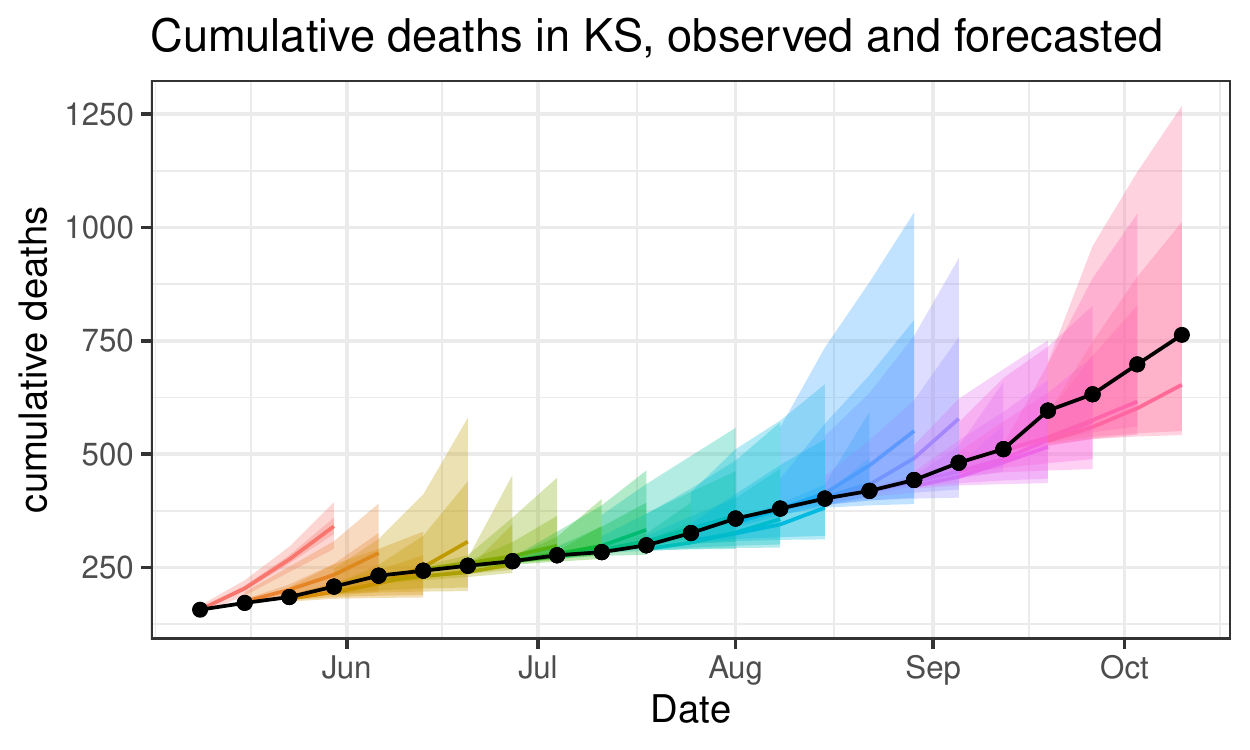}
 \includegraphics[width=0.32\linewidth]{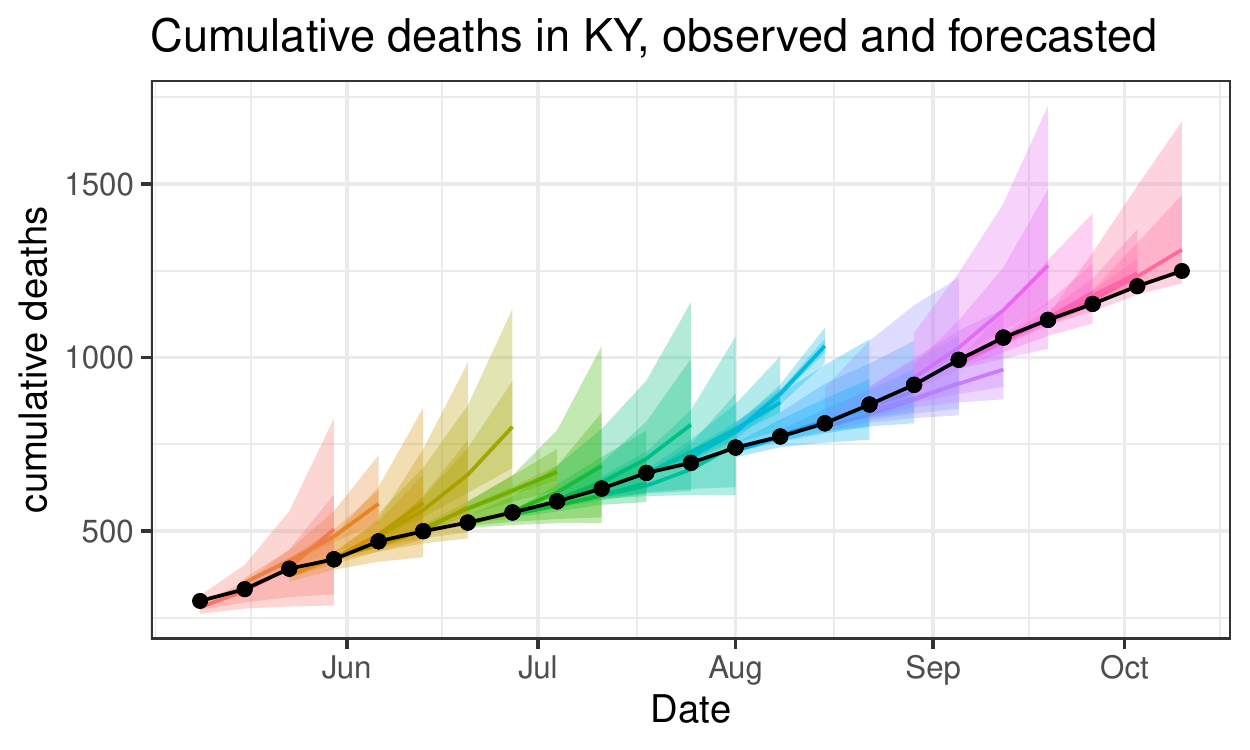}
 \includegraphics[width=0.32\linewidth]{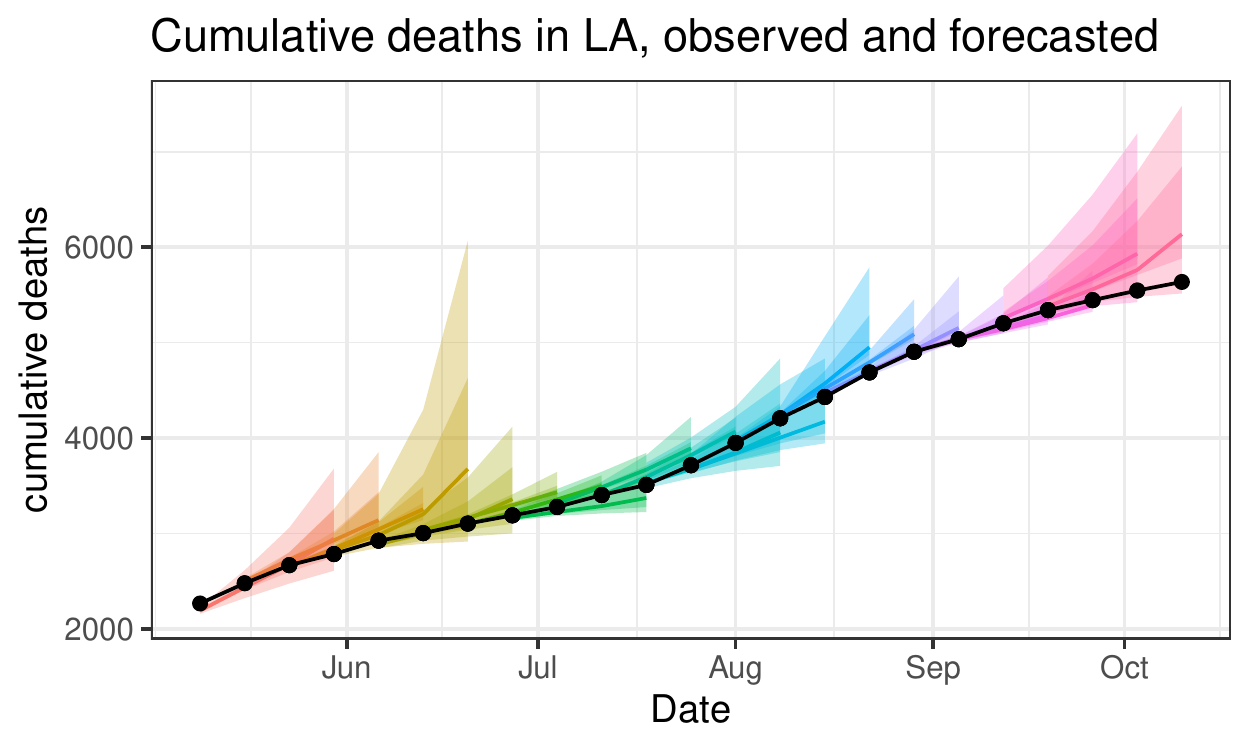}

\bigskip

 \includegraphics[width=0.32\linewidth]{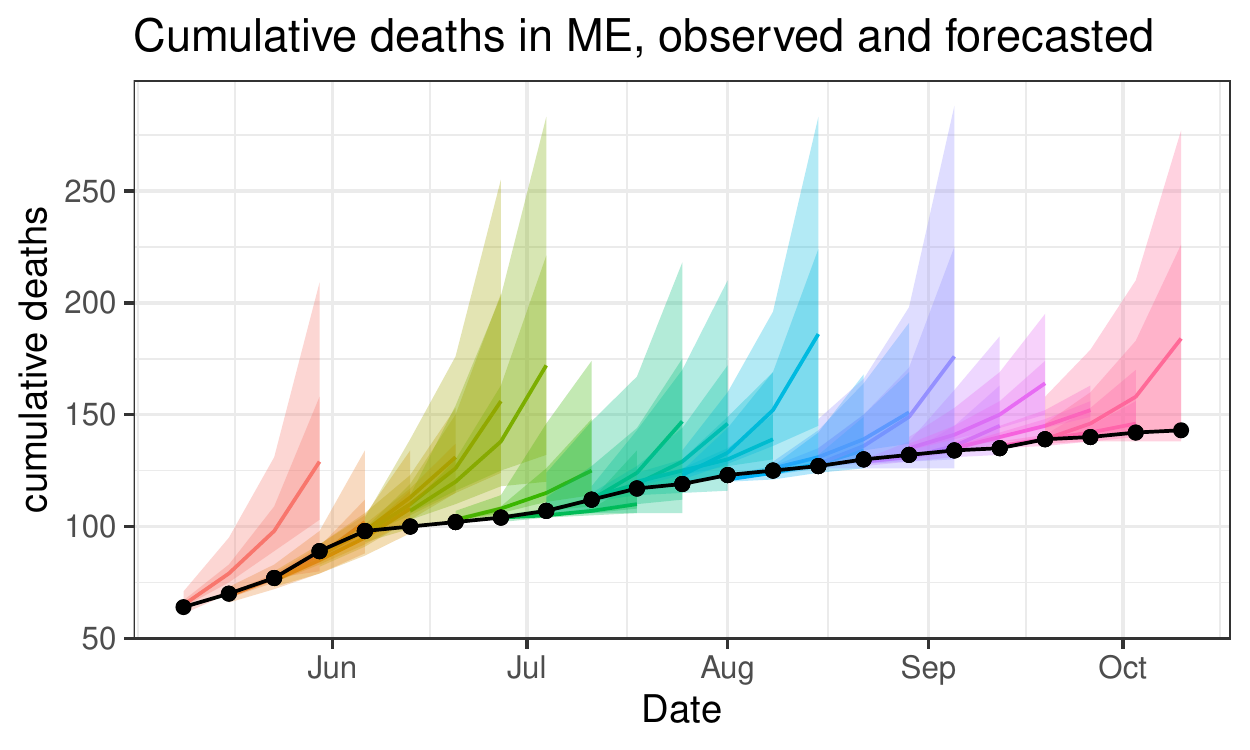}
 \includegraphics[width=0.32\linewidth]{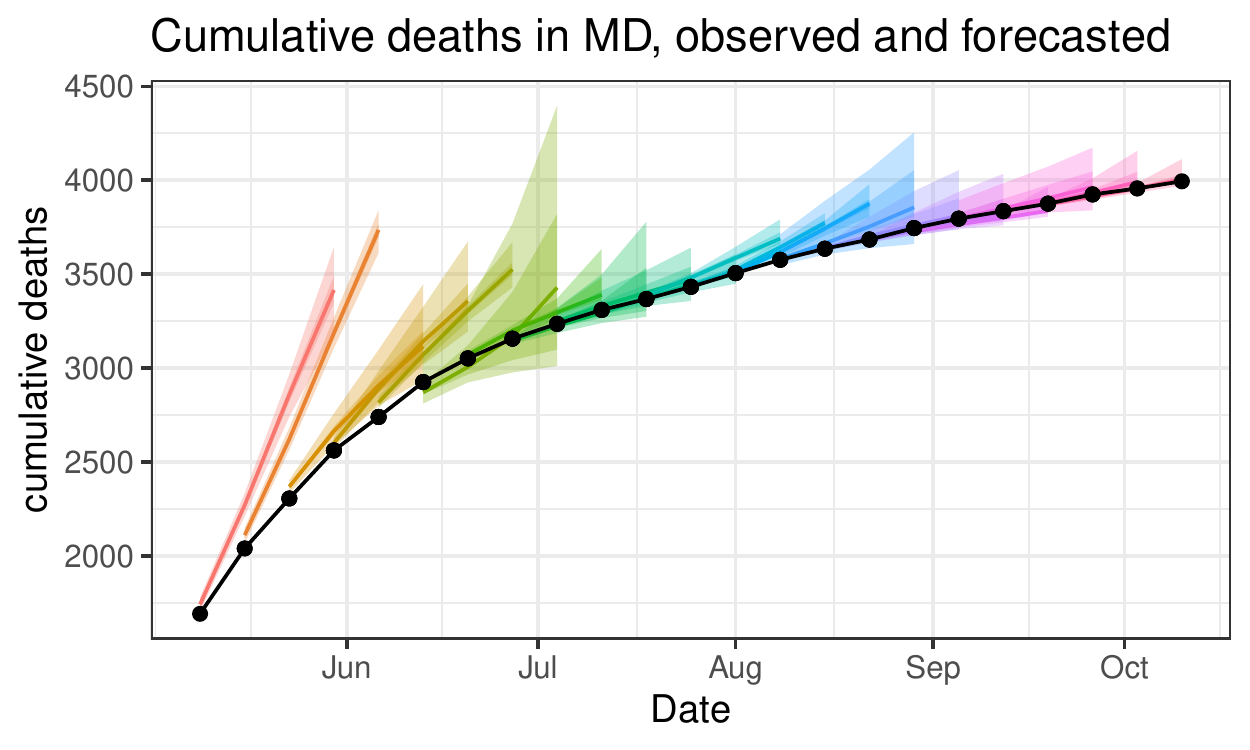}
 \includegraphics[width=0.32\linewidth]{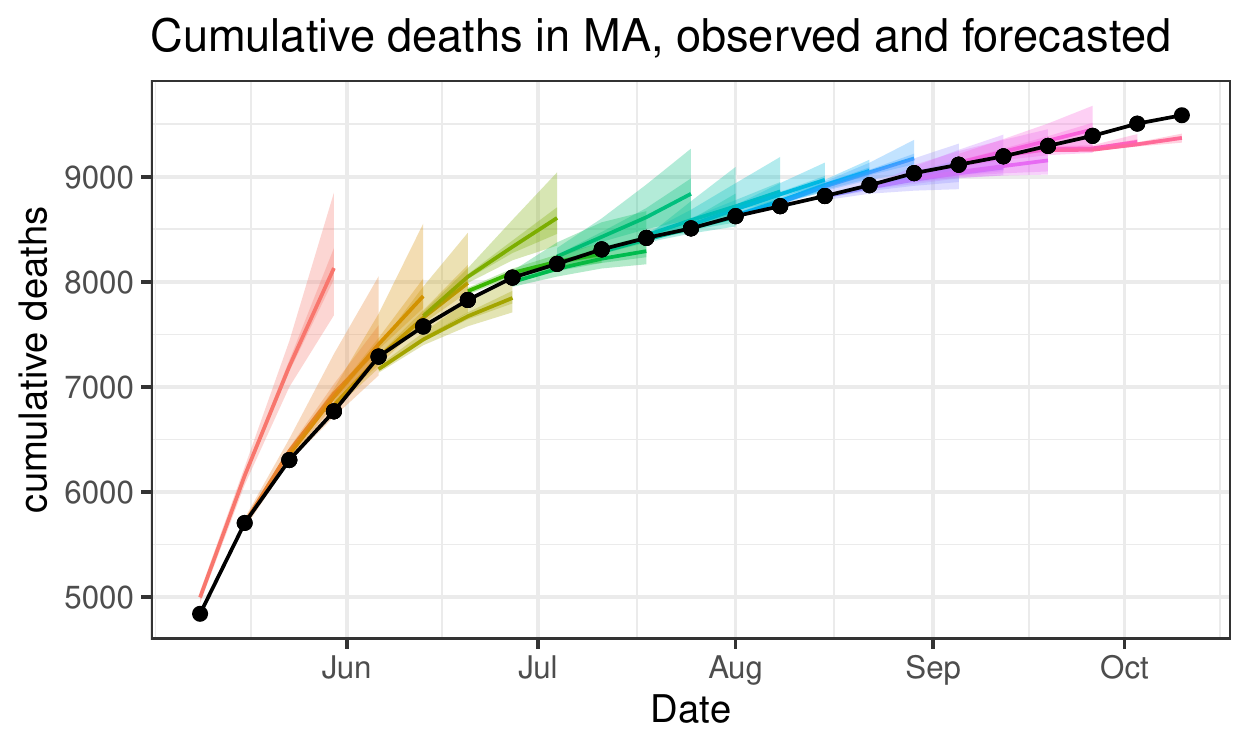}

 \bigskip

 \includegraphics[width=0.32\linewidth]{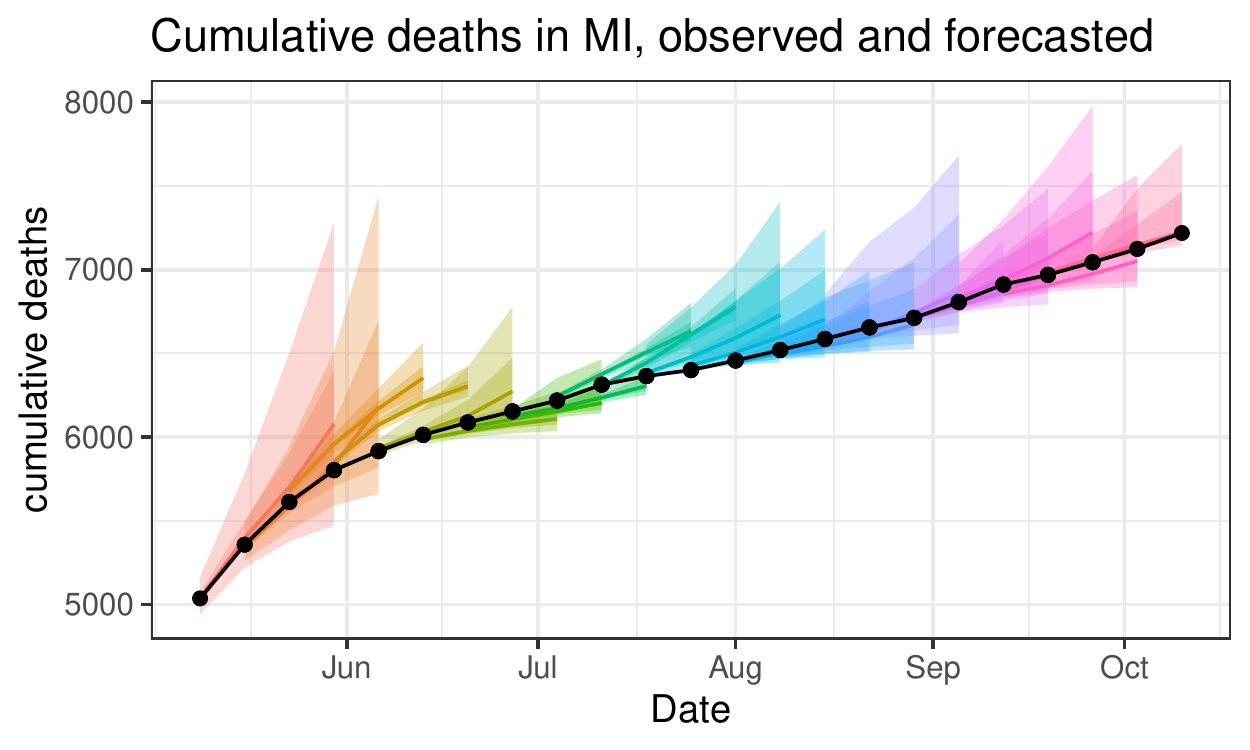}
 \includegraphics[width=0.32\linewidth]{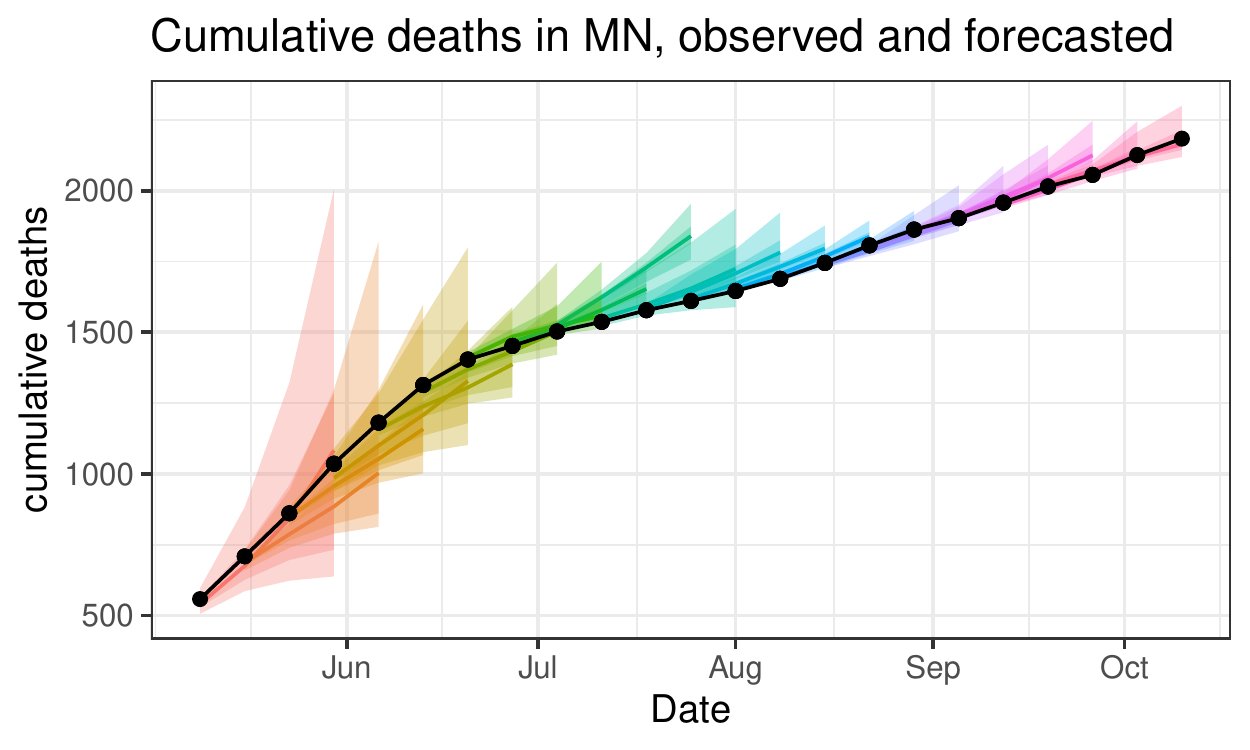}
 \includegraphics[width=0.32\linewidth]{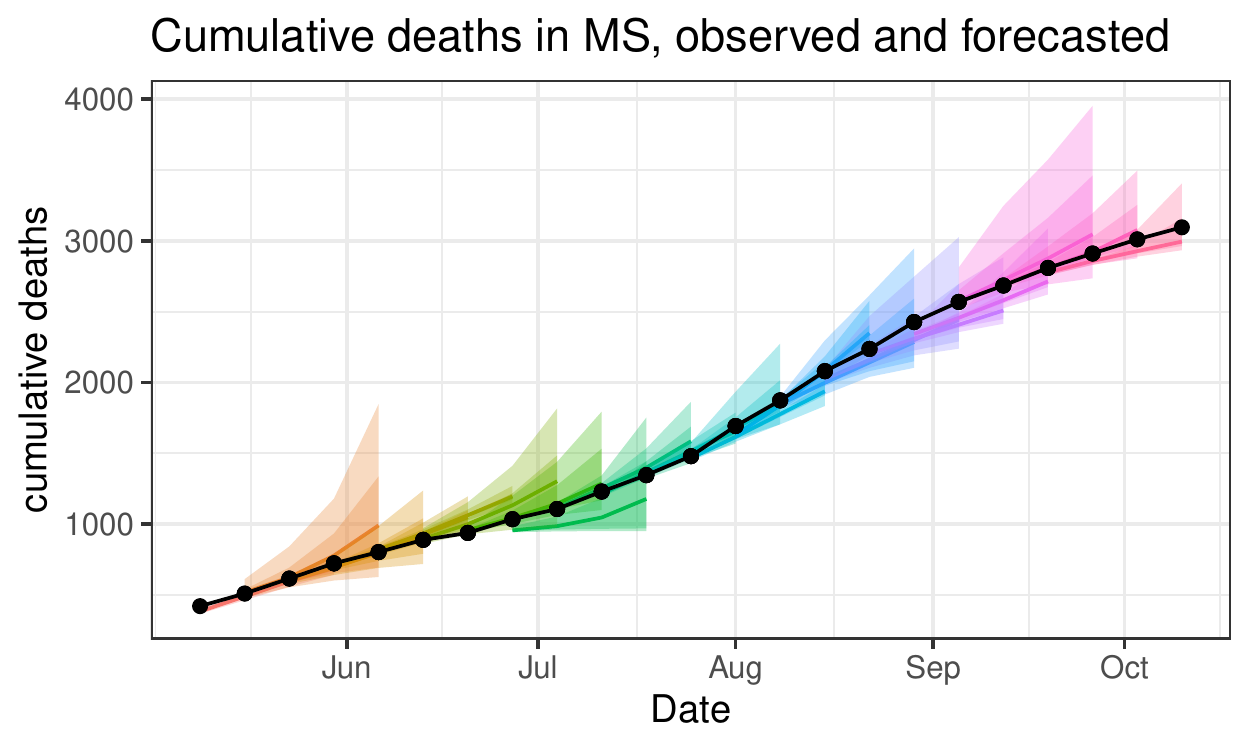}

 \bigskip

 \includegraphics[width=0.32\linewidth]{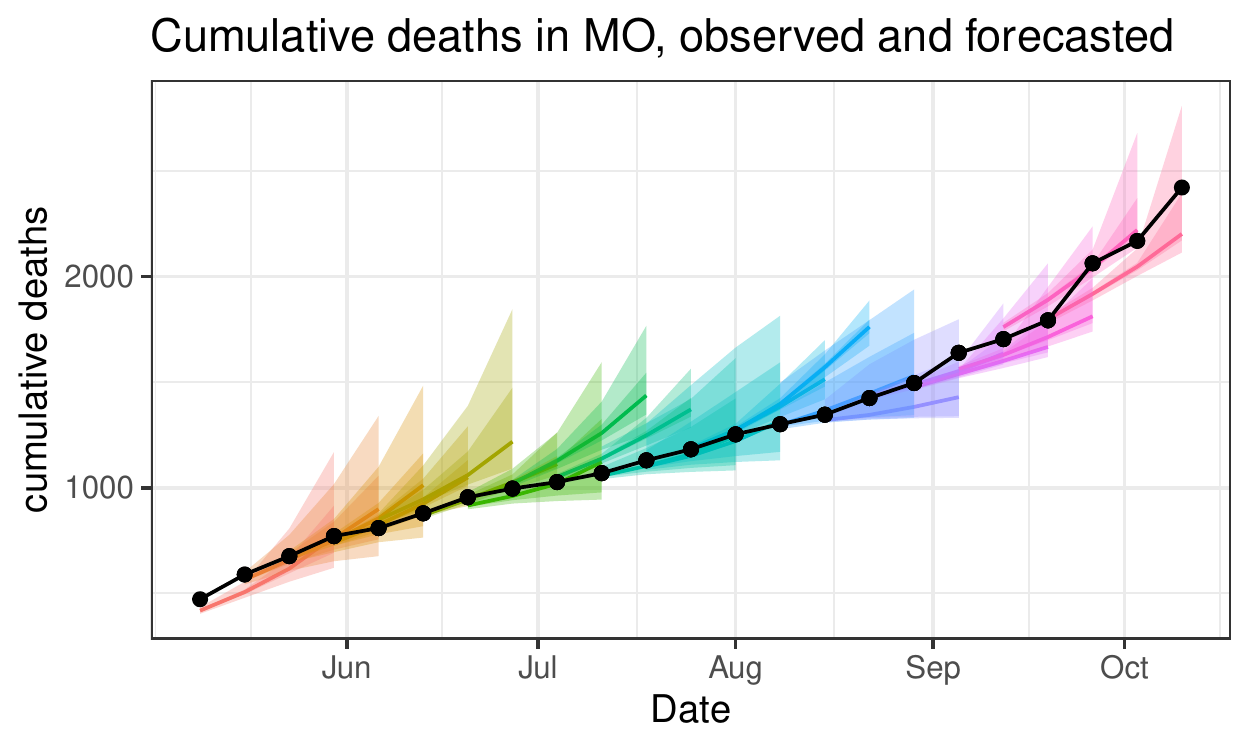}
 \includegraphics[width=0.32\linewidth]{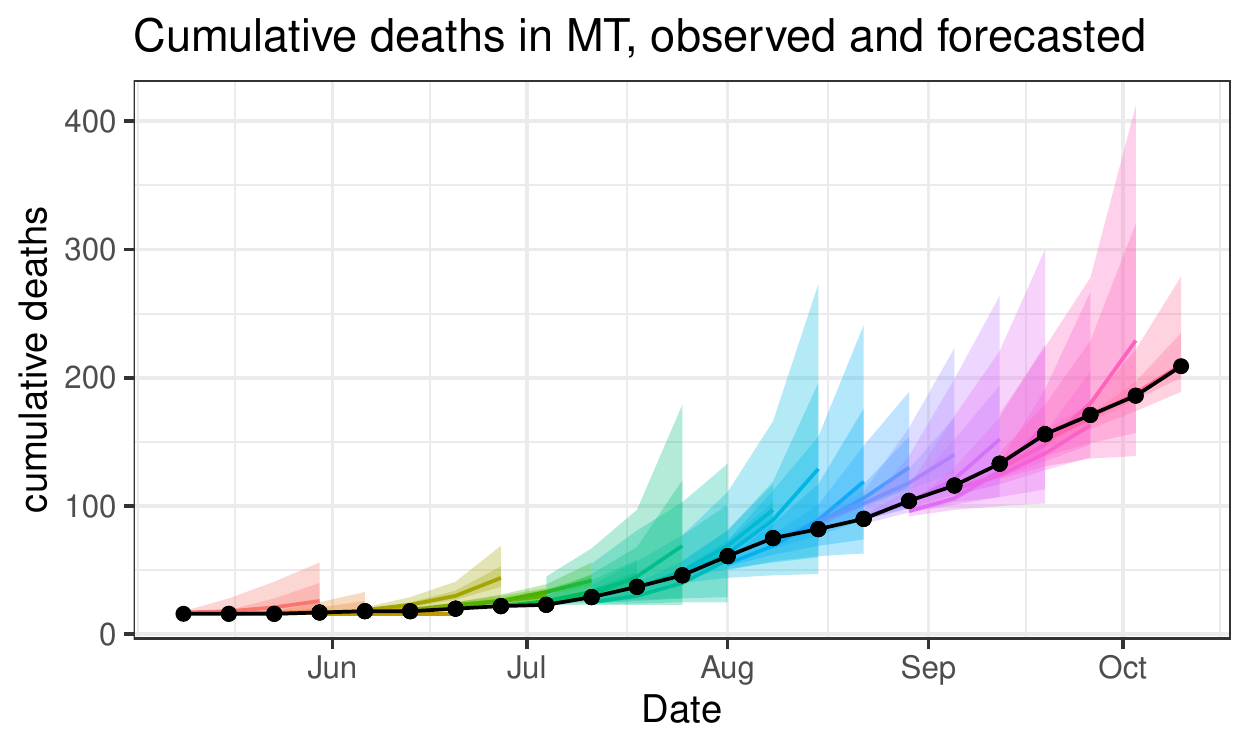}
 \includegraphics[width=0.32\linewidth]{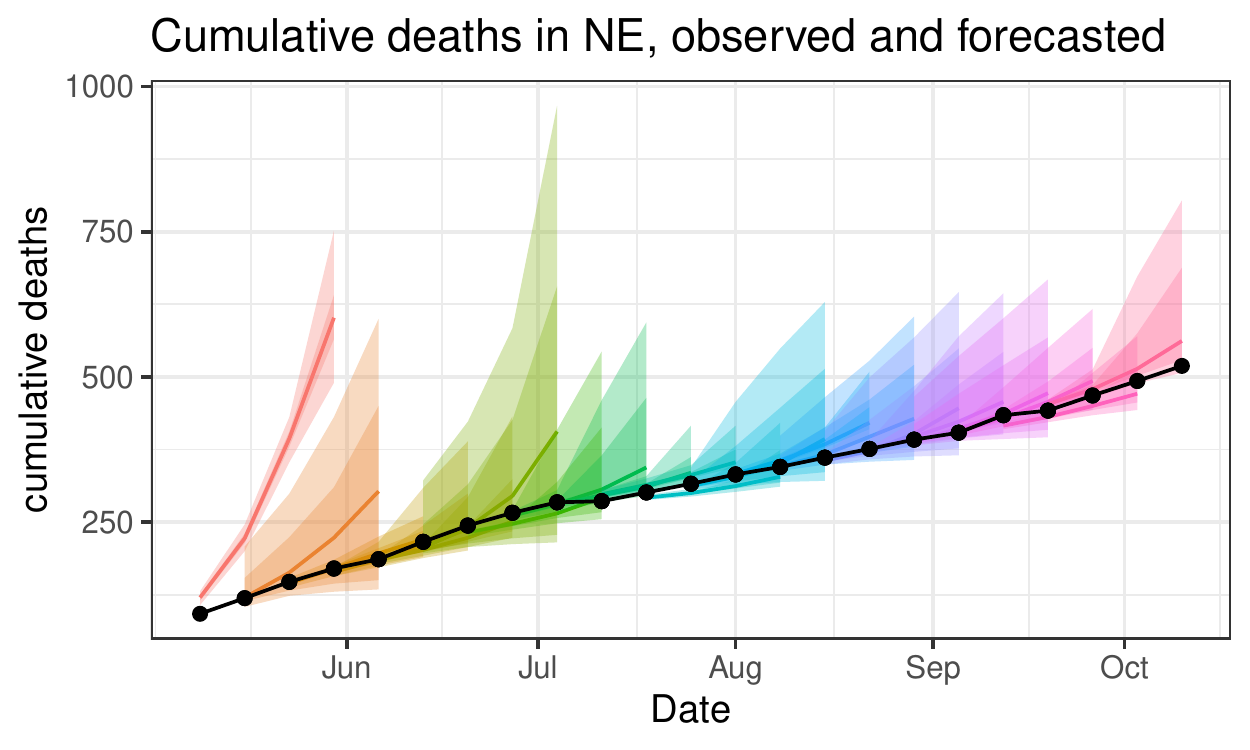}

 \bigskip

 \includegraphics[width=0.32\linewidth]{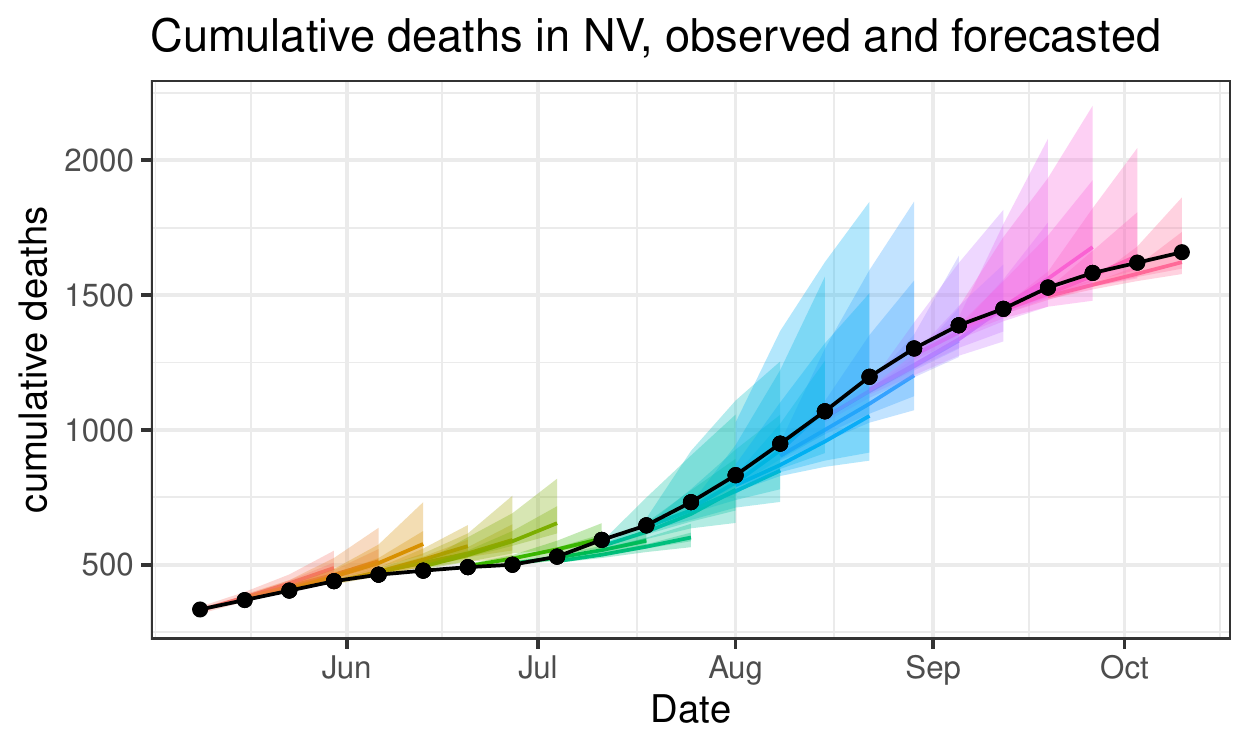}
 \includegraphics[width=0.32\linewidth]{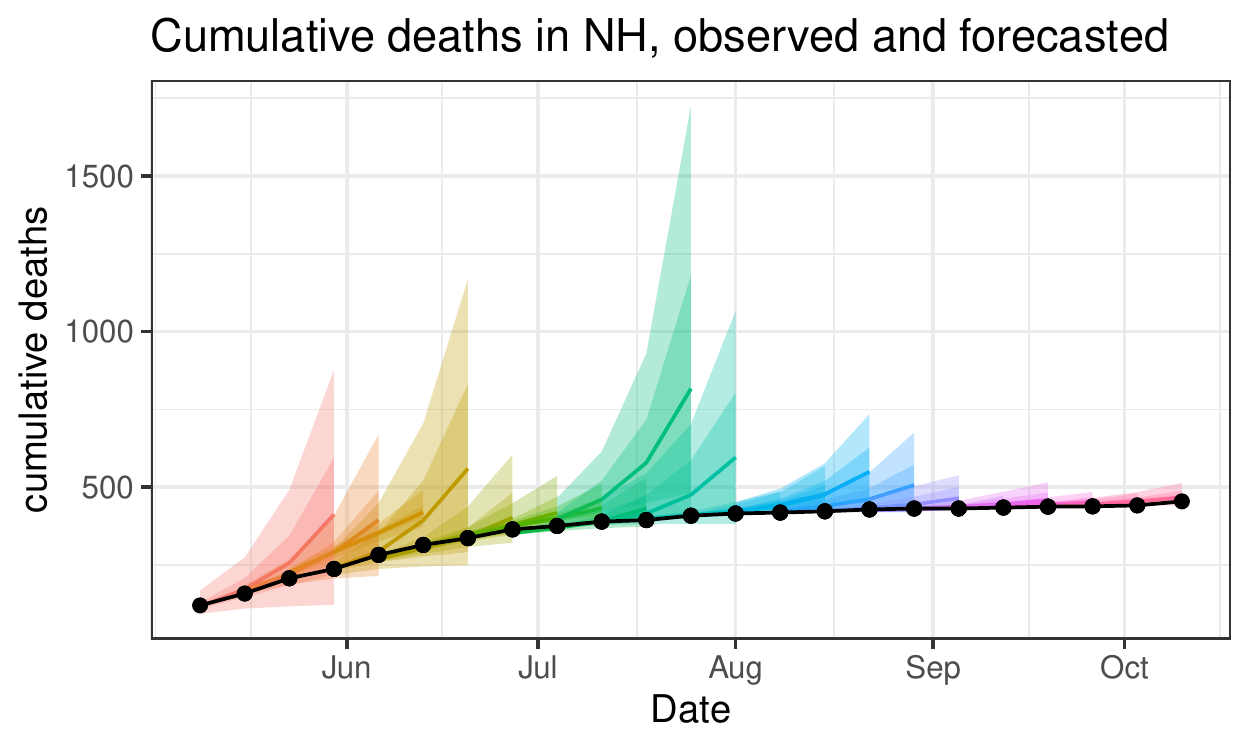}
 \includegraphics[width=0.32\linewidth]{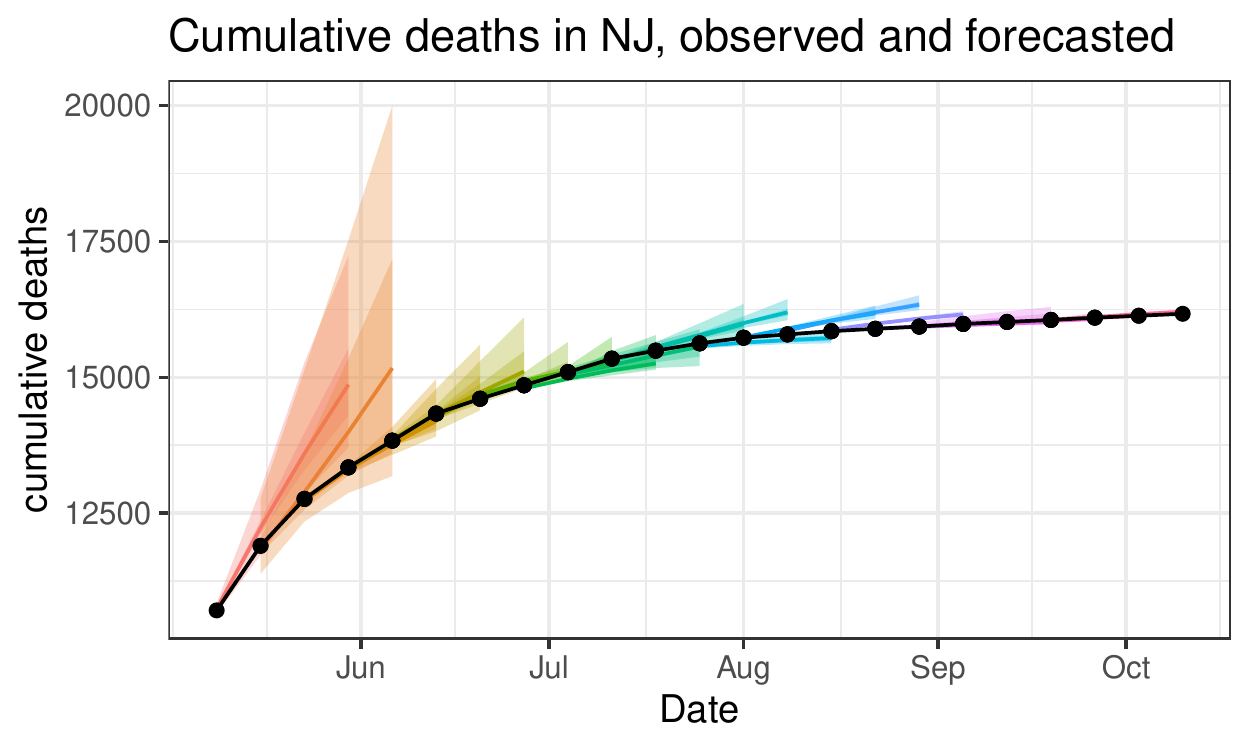}

 \bigskip

 \includegraphics[width=0.32\linewidth]{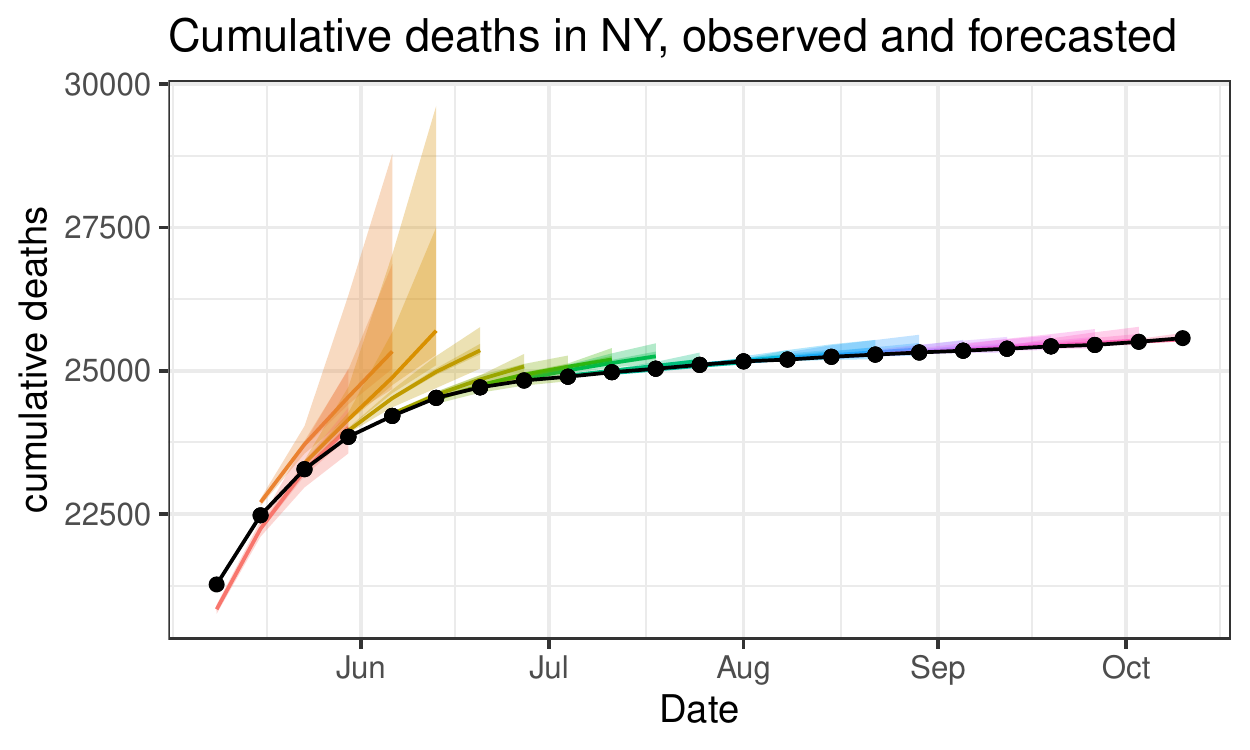}
 \includegraphics[width=0.32\linewidth]{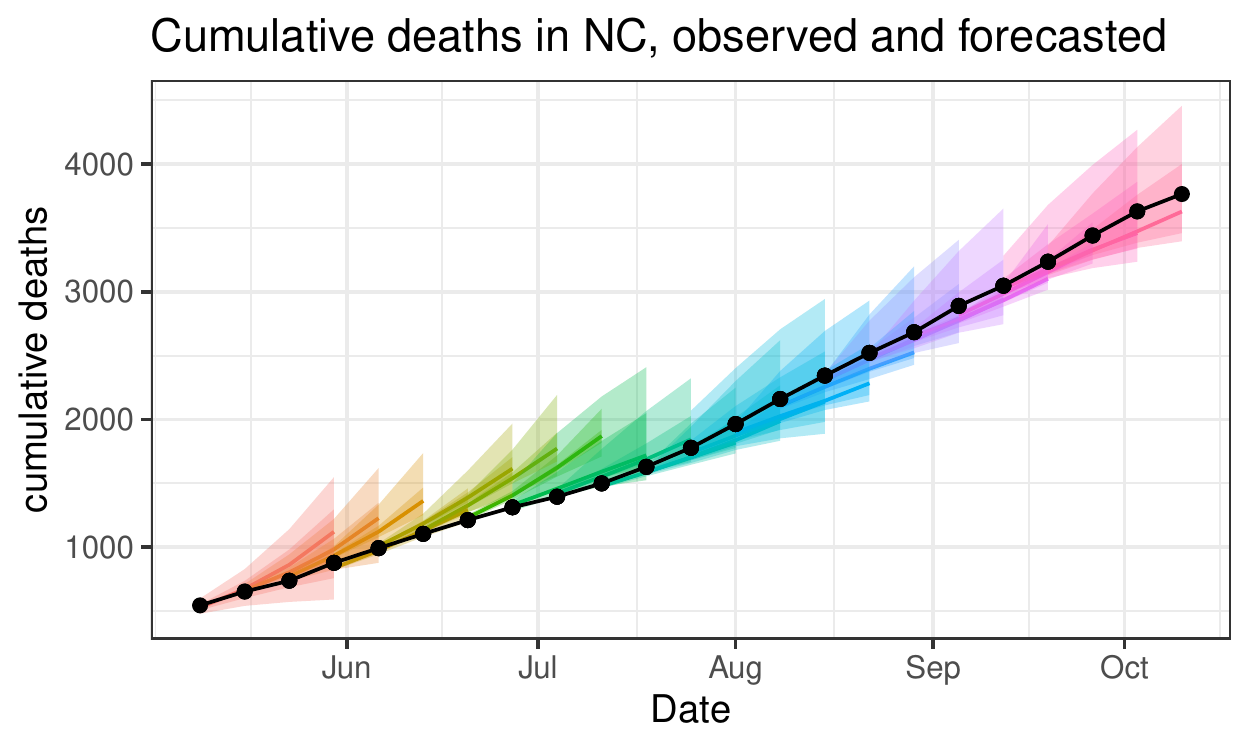}
 \includegraphics[width=0.32\linewidth]{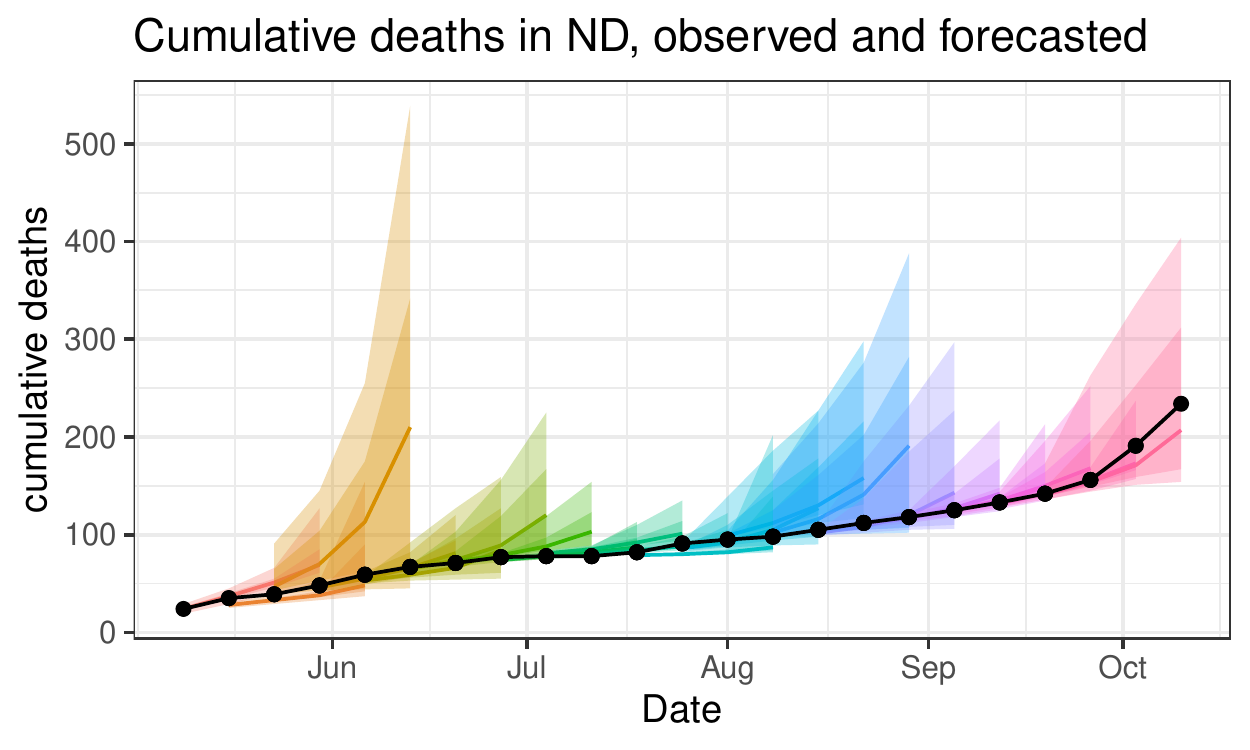}

 \bigskip

 \includegraphics[width=0.32\linewidth]{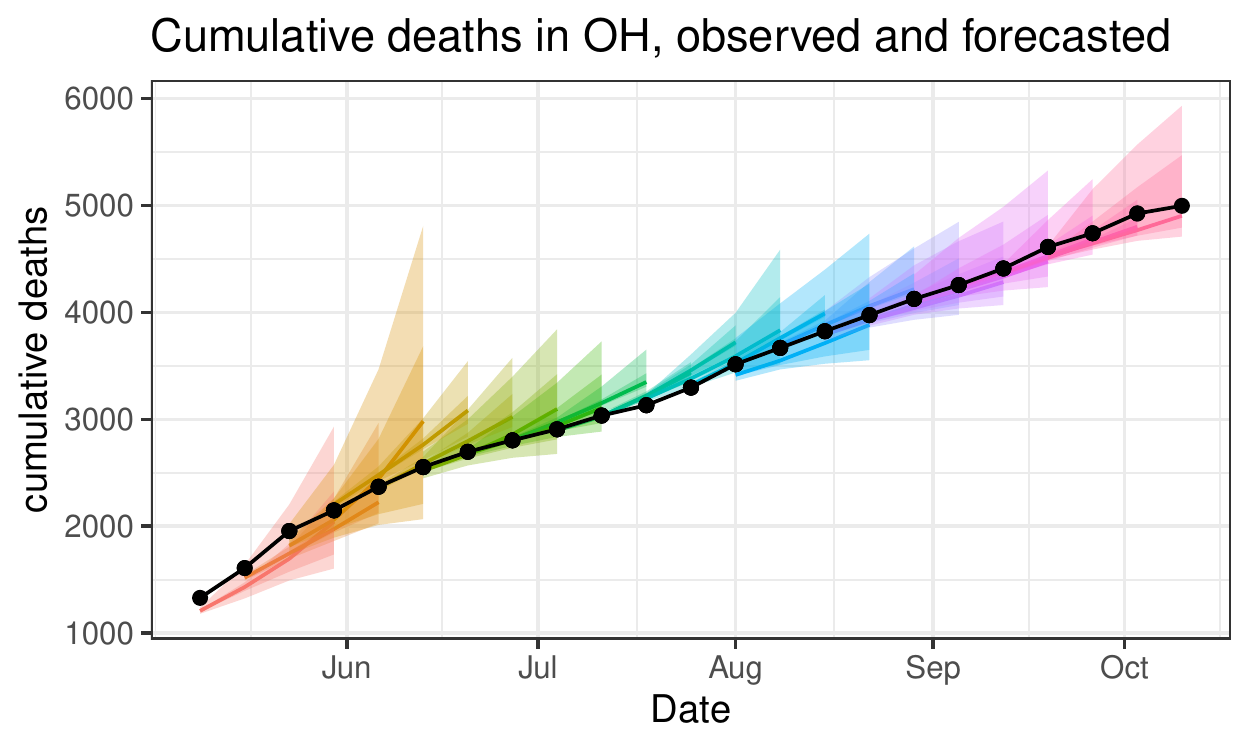}
 \includegraphics[width=0.32\linewidth]{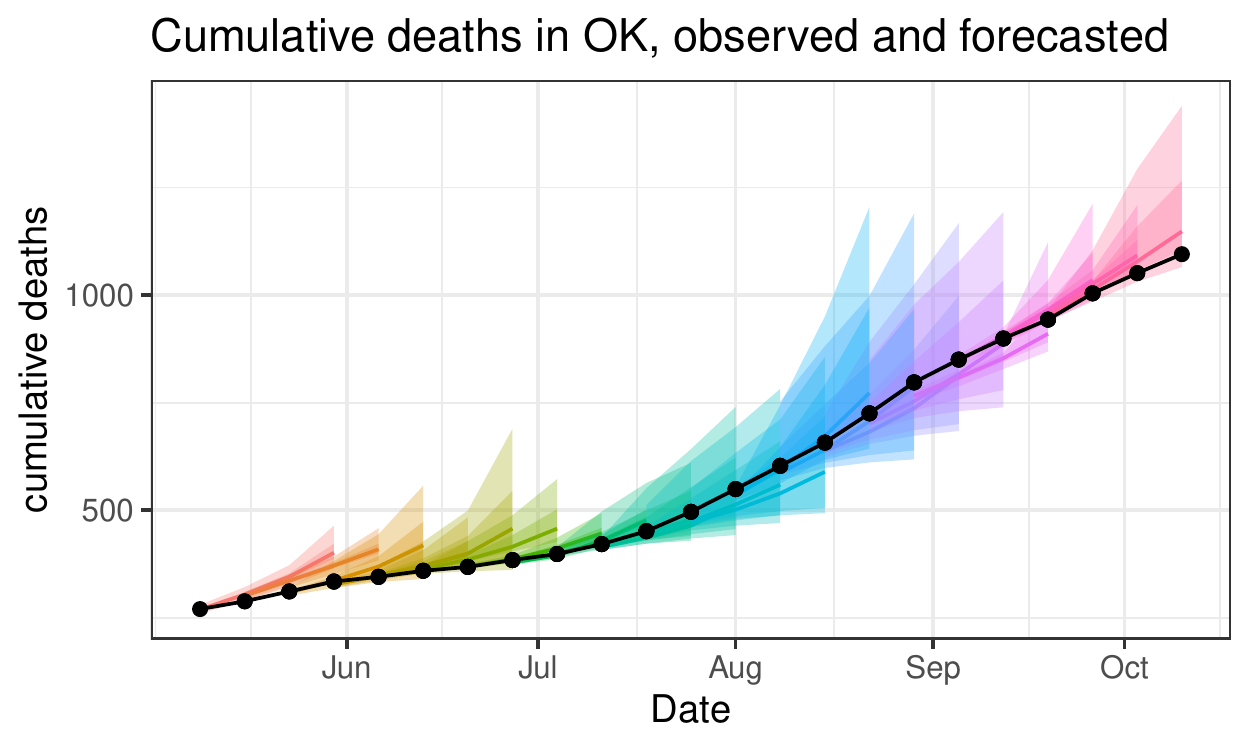}
 \includegraphics[width=0.32\linewidth]{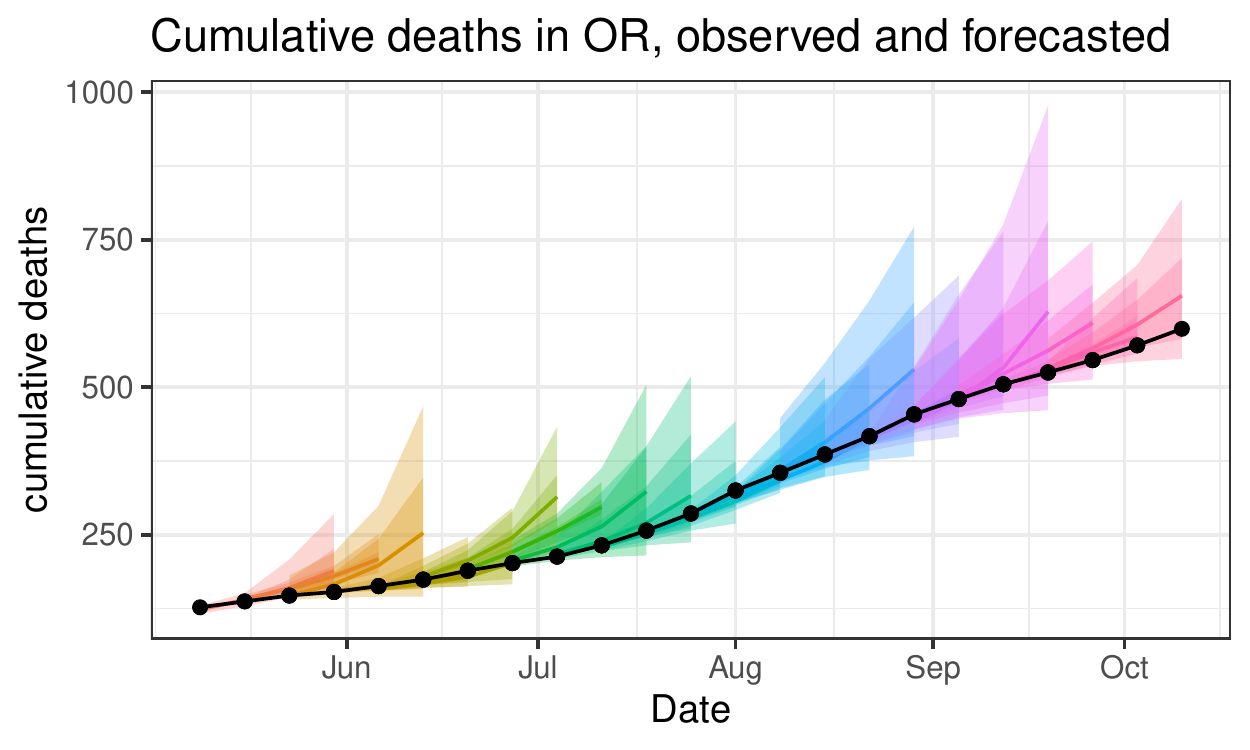}

\bigskip

 \includegraphics[width=0.32\linewidth]{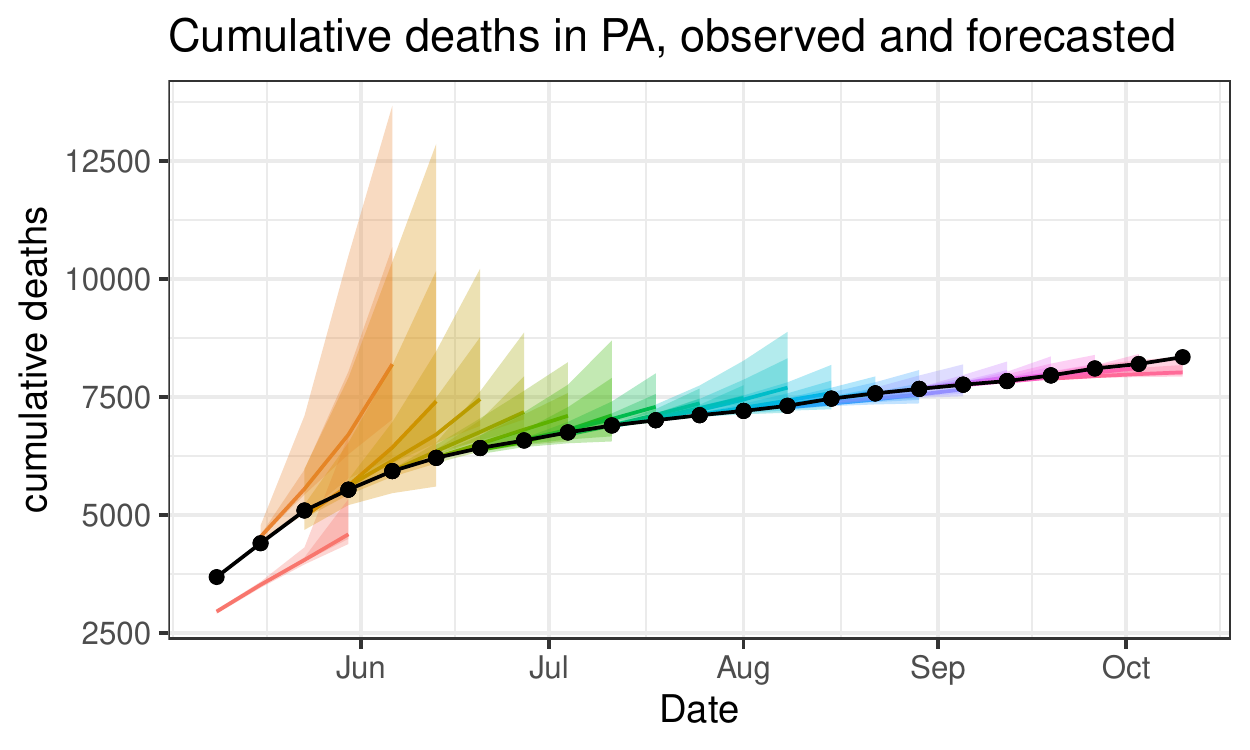}
 \includegraphics[width=0.32\linewidth]{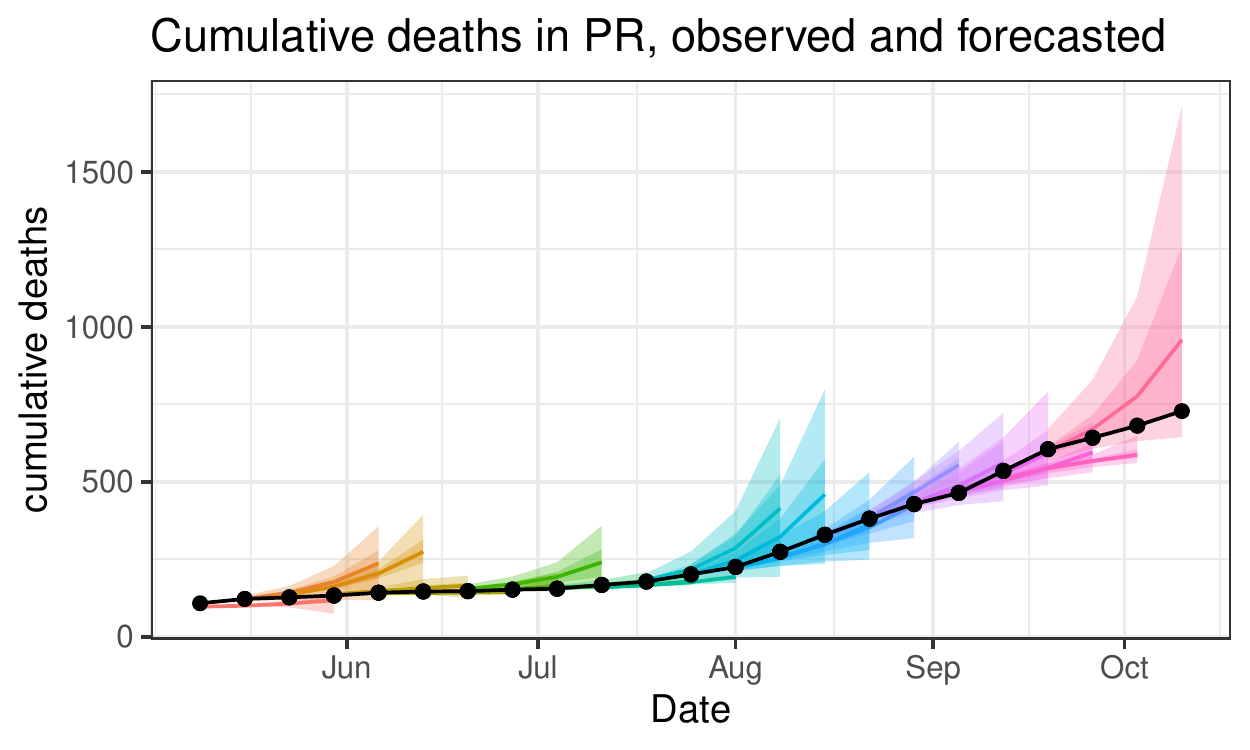}
 \includegraphics[width=0.32\linewidth]{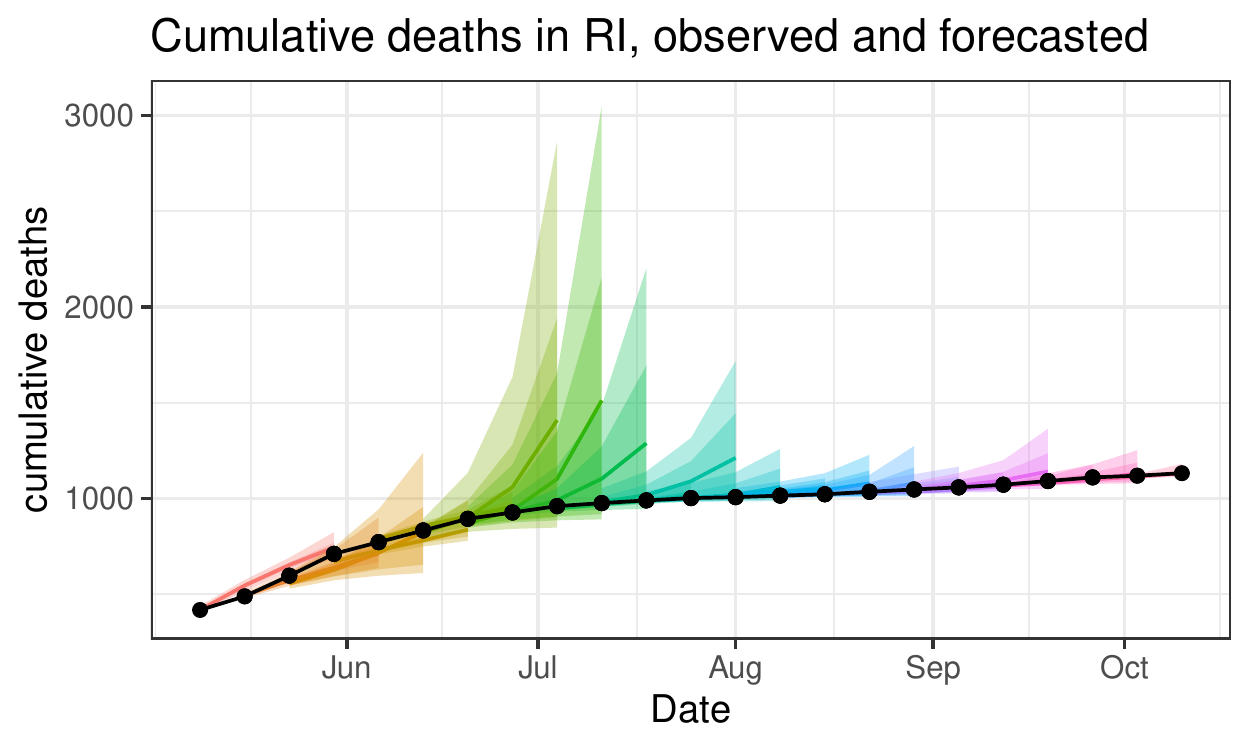}

 \bigskip

 \includegraphics[width=0.32\linewidth]{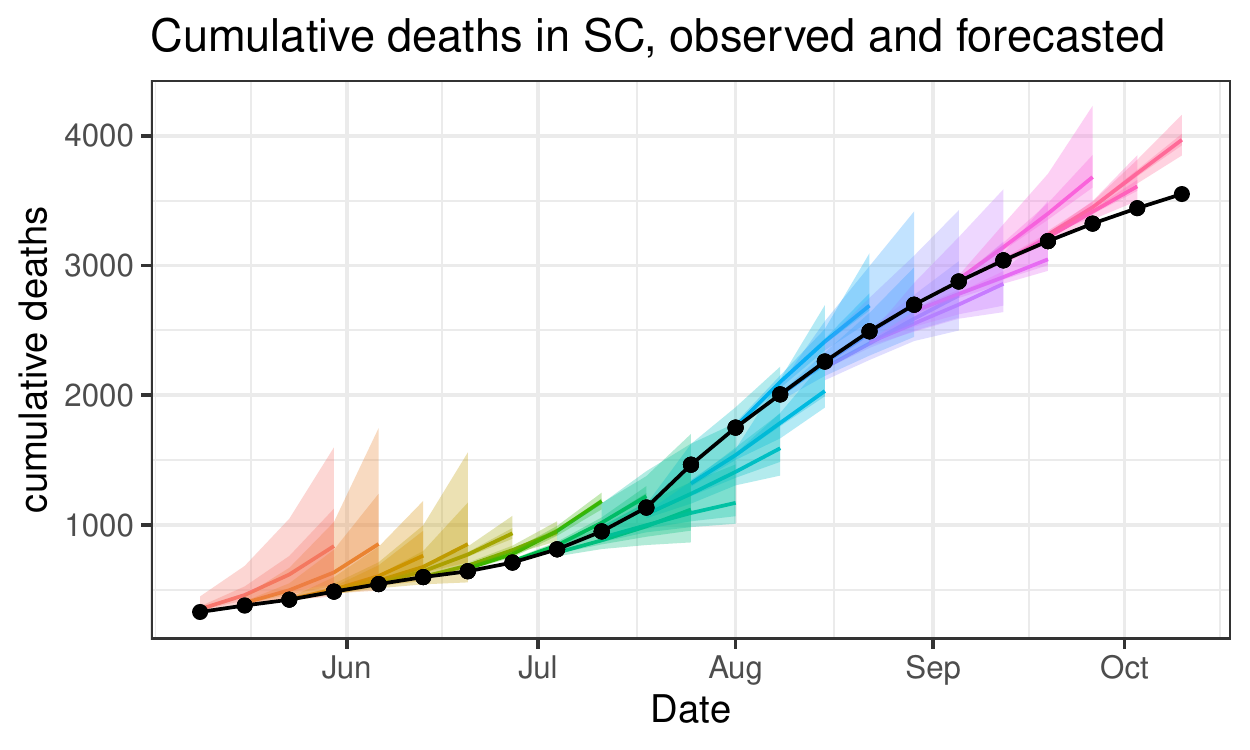}
 \includegraphics[width=0.32\linewidth]{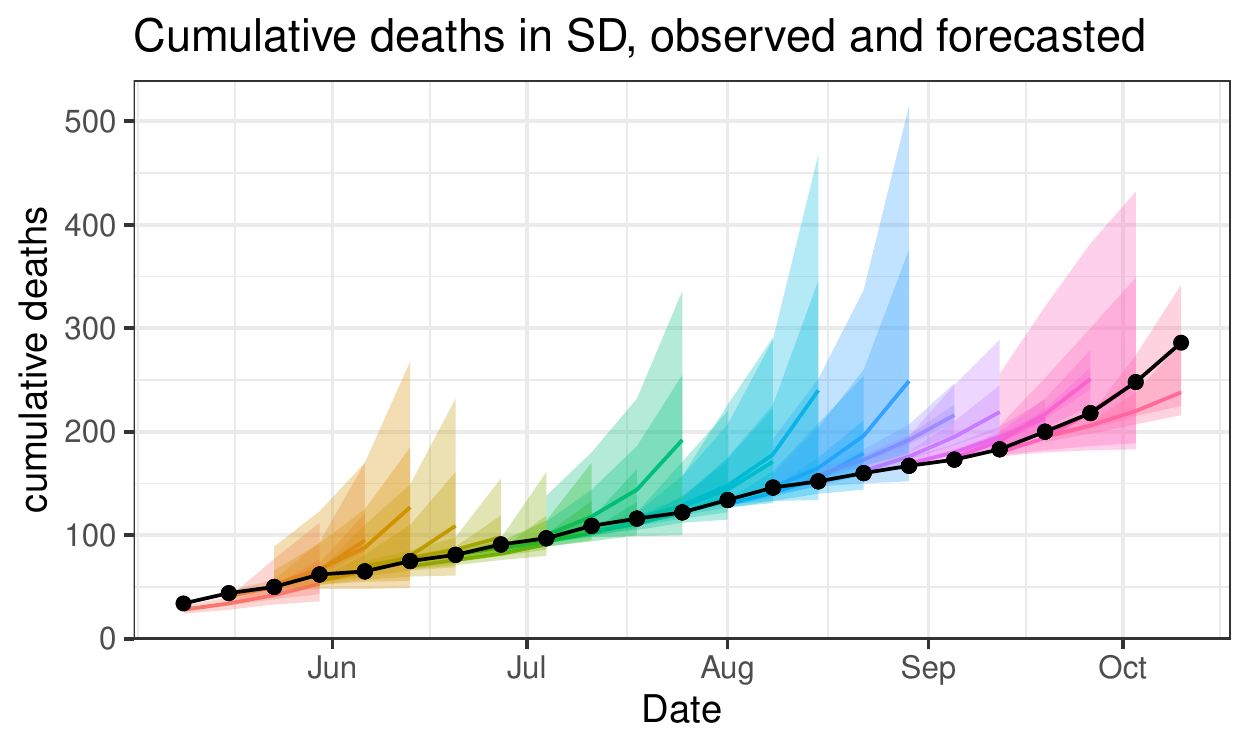}
 \includegraphics[width=0.32\linewidth]{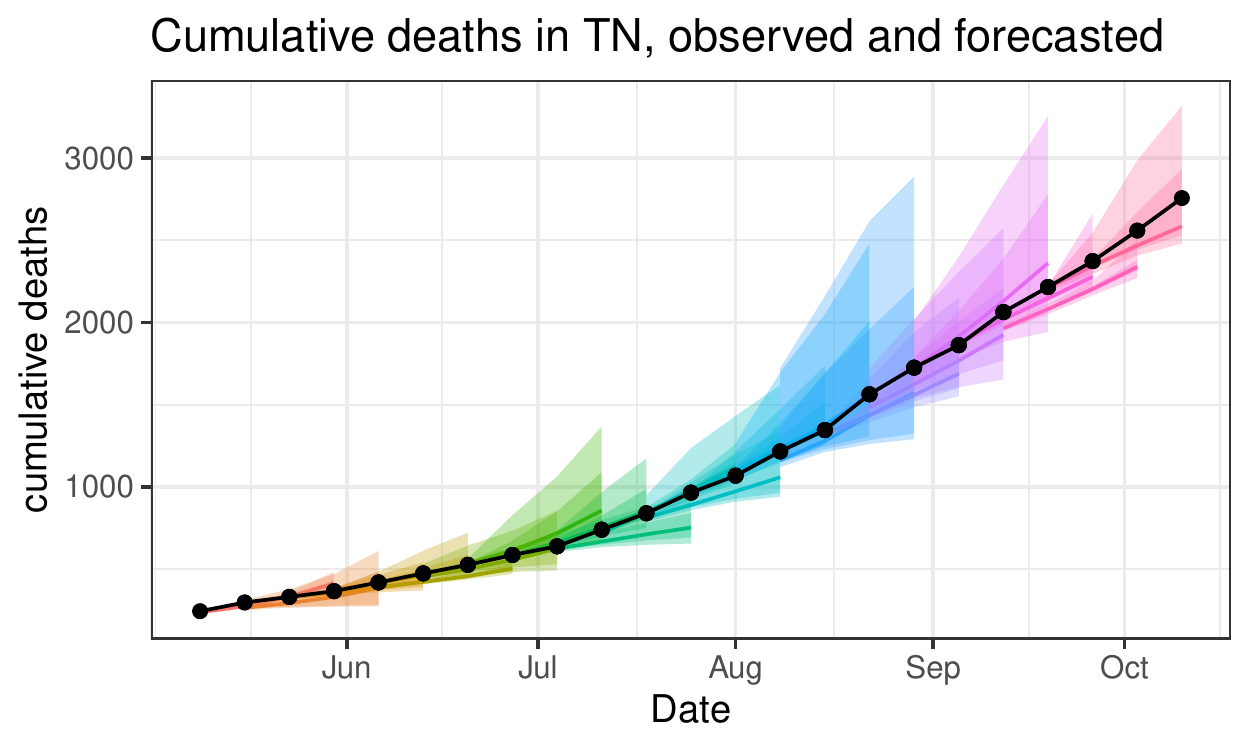}

 \bigskip

 \includegraphics[width=0.32\linewidth]{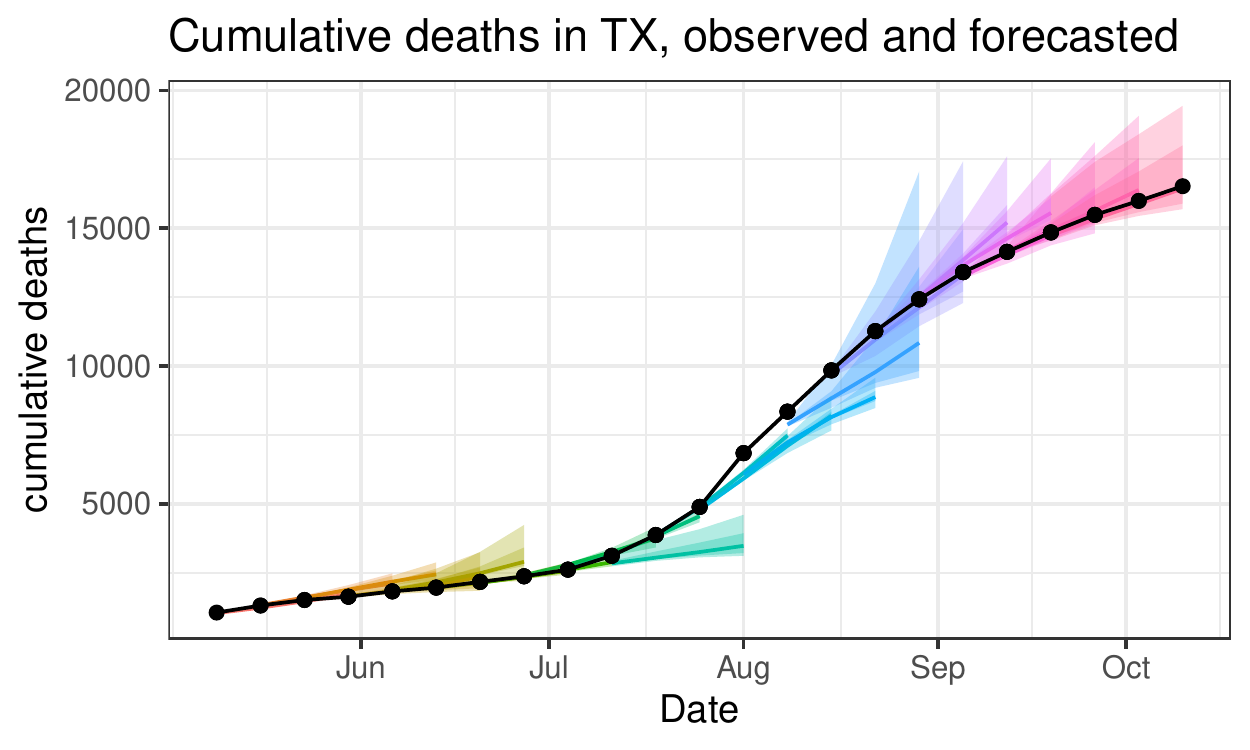}
 \includegraphics[width=0.32\linewidth]{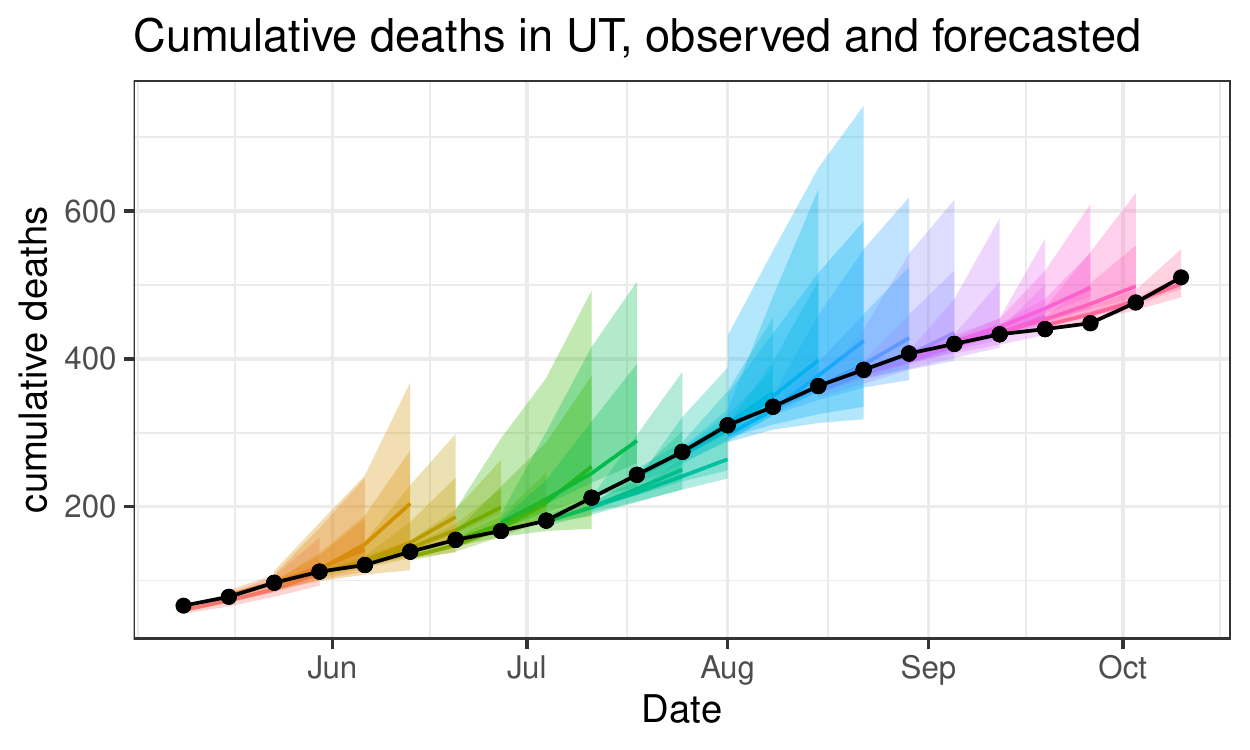}
 \includegraphics[width=0.32\linewidth]{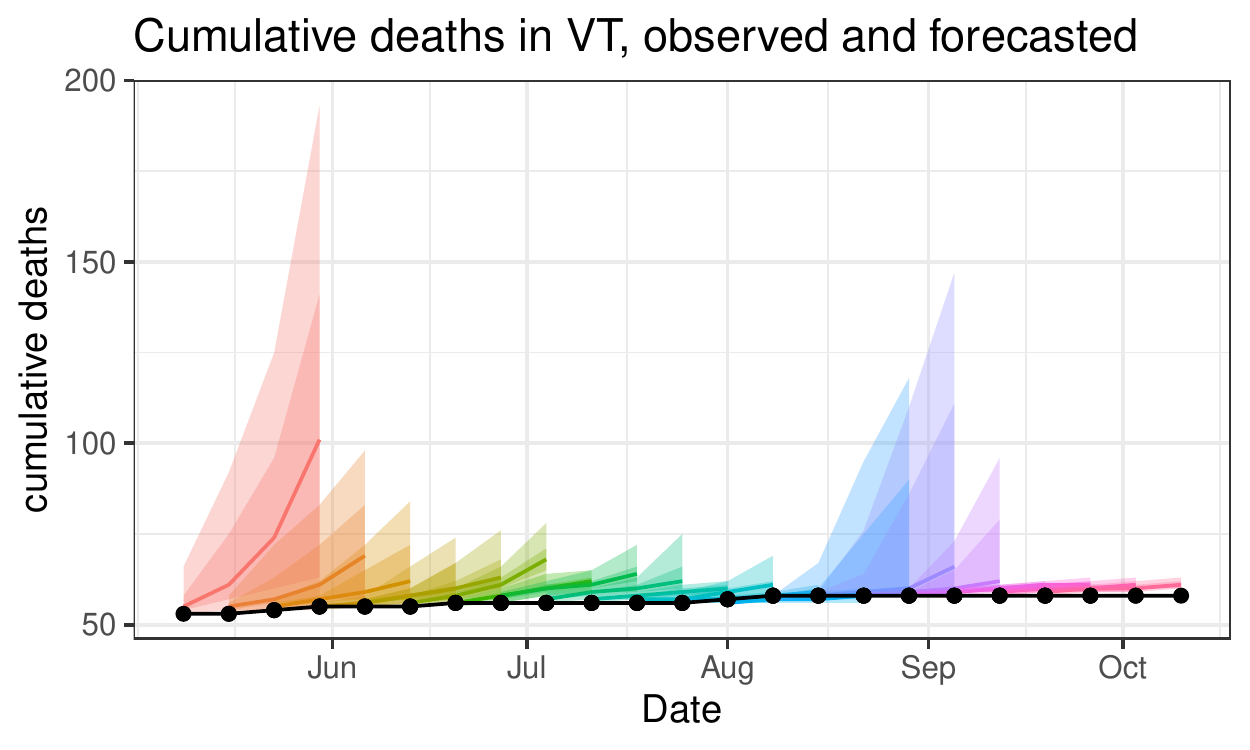}

 \bigskip

 \includegraphics[width=0.32\linewidth]{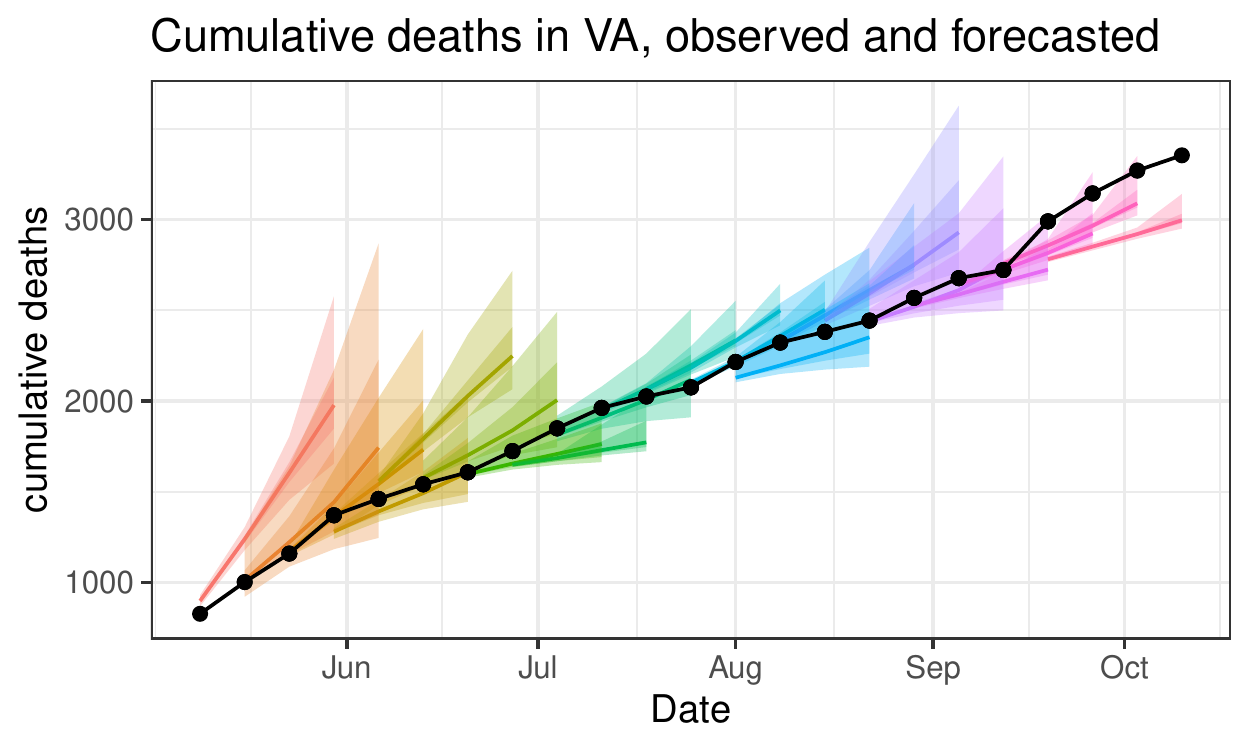}
 \includegraphics[width=0.32\linewidth]{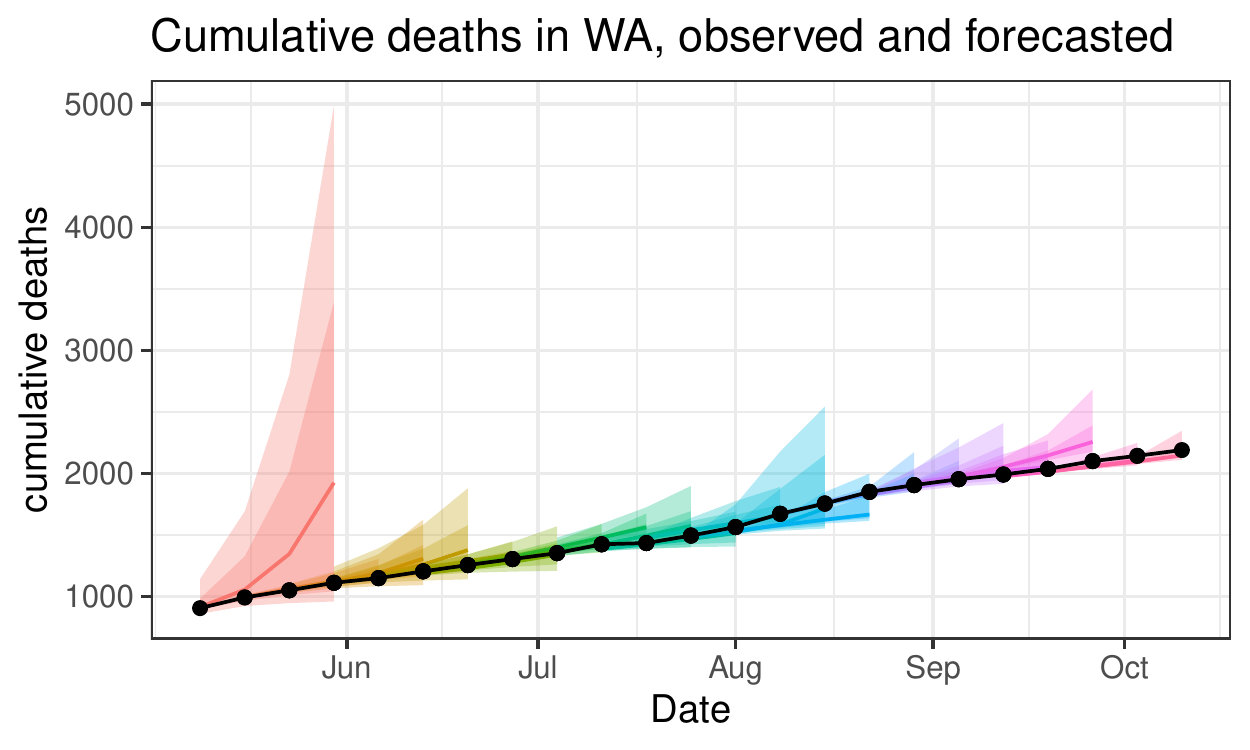}
 \includegraphics[width=0.32\linewidth]{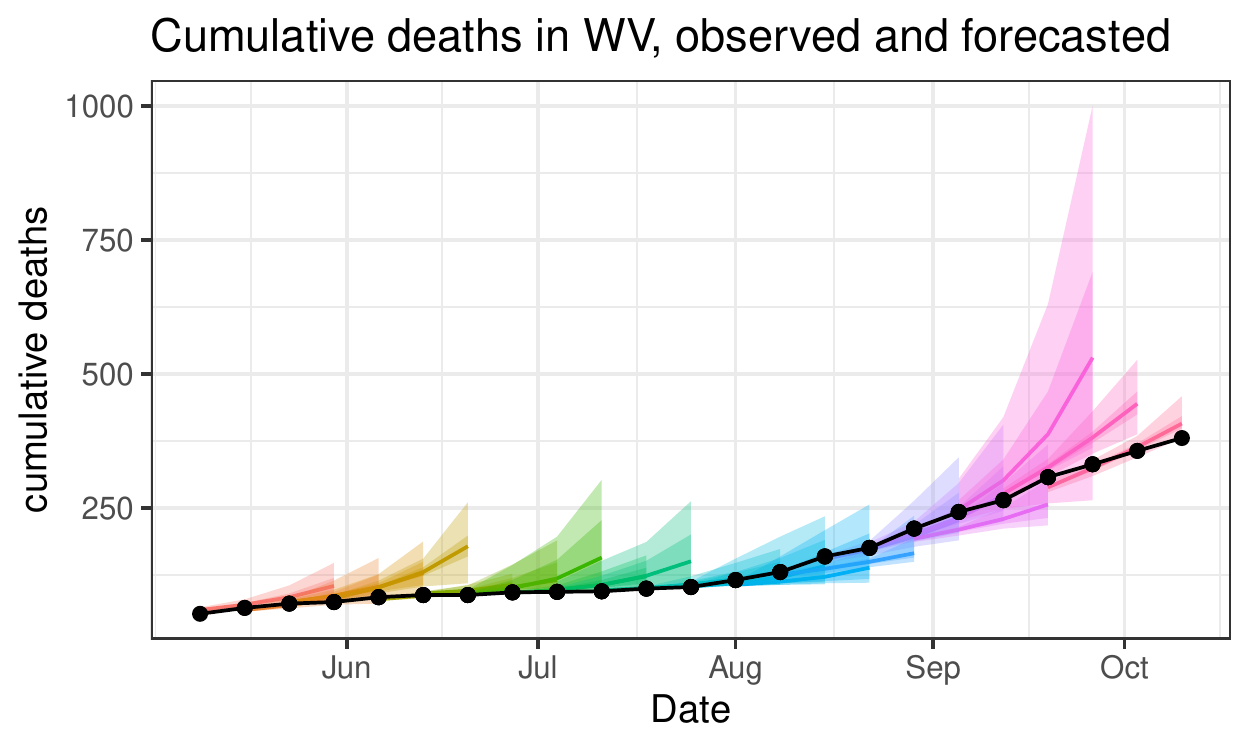}

 \bigskip

 \includegraphics[width=0.32\linewidth]{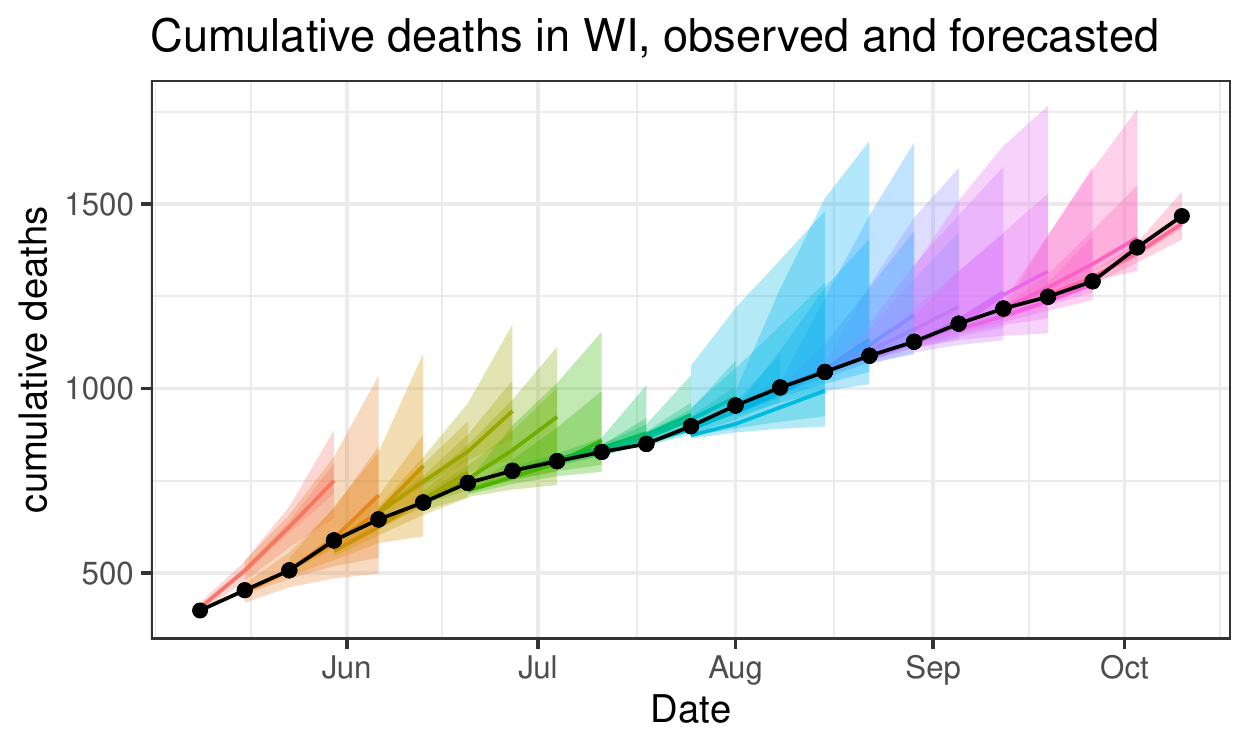}
 \includegraphics[width=0.32\linewidth]{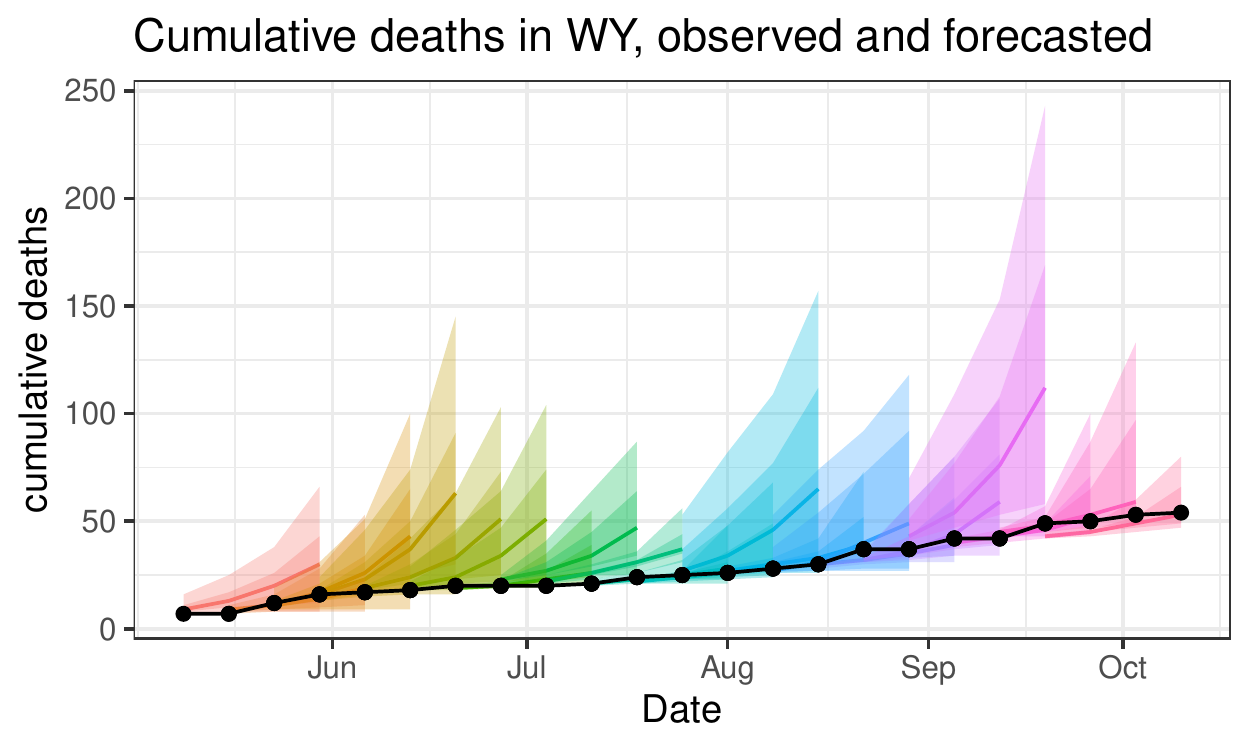}
 \includegraphics[width=0.32\linewidth]{COVIDfigs/US_death_quant_forecasts.pdf}
\end{center}


\subsection{Selected incident case and death forecasts}
We show below the incident case and death forecasts for the overall US and for Arizona. 

\bigskip

\begin{center}

  \includegraphics[width=0.45\linewidth]{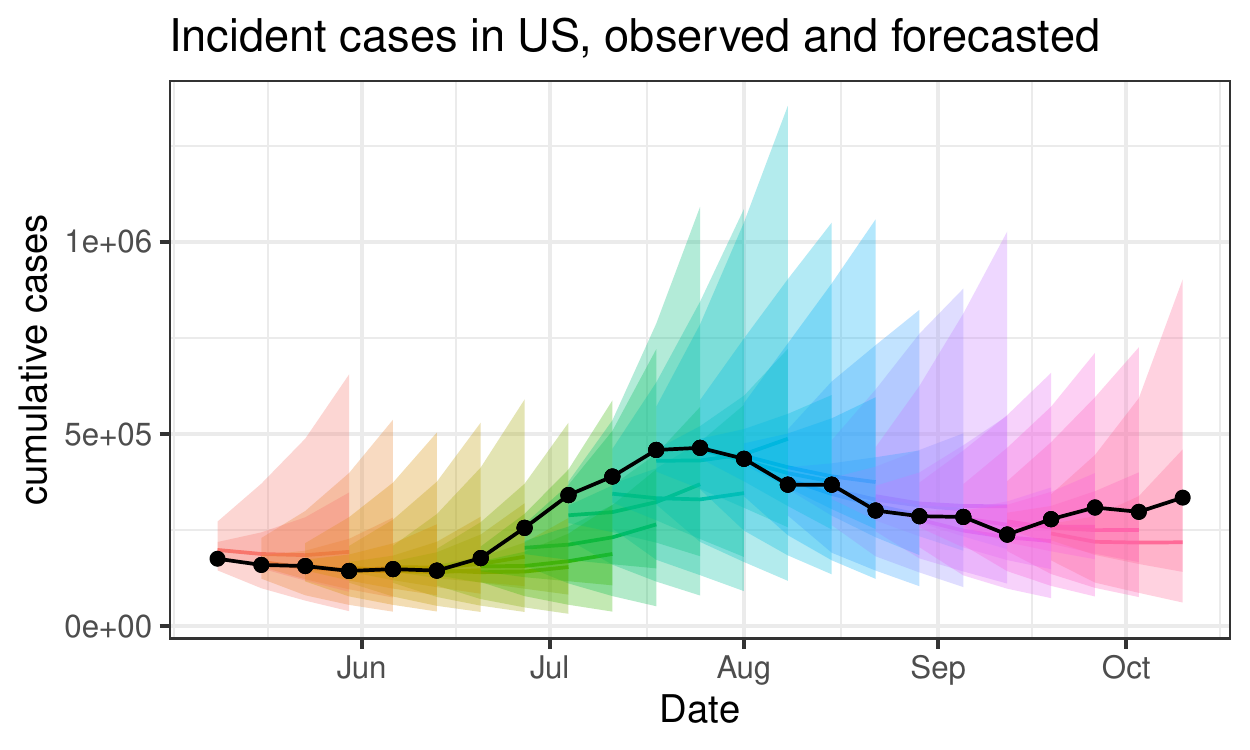}  
 \includegraphics[width=0.45\linewidth]{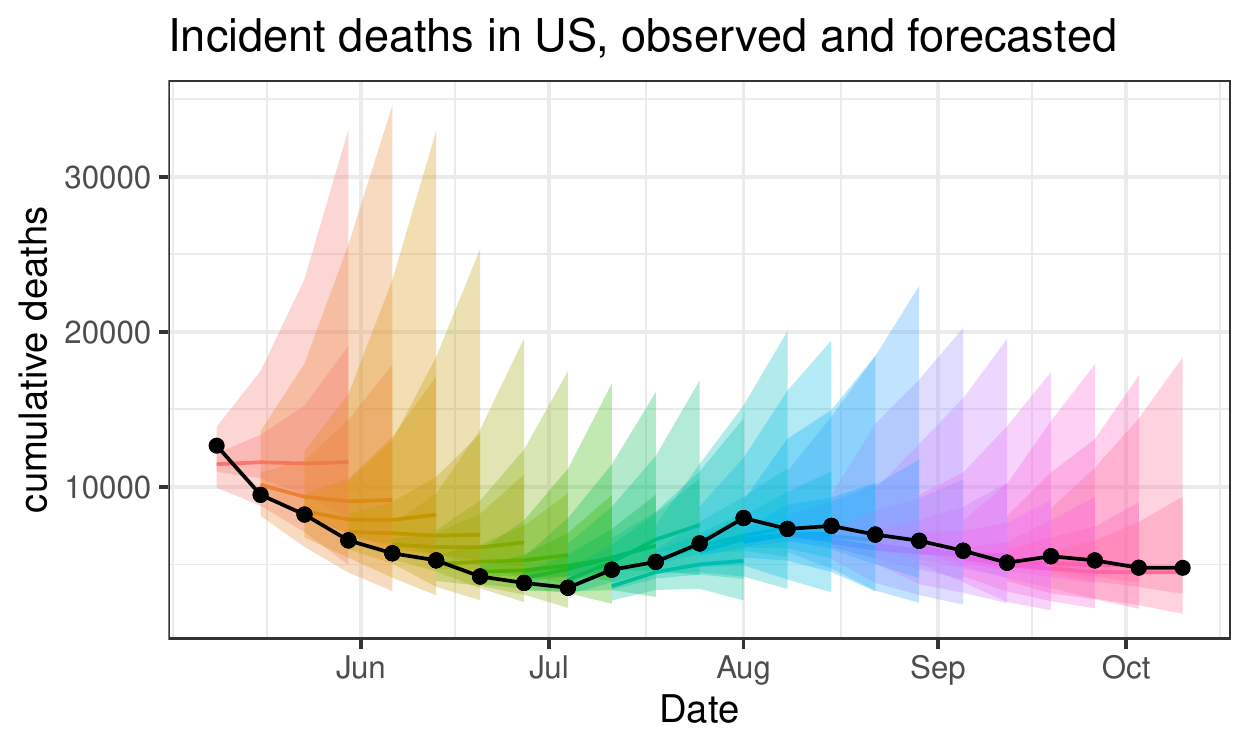}

  \medskip

  \includegraphics[width=0.45\linewidth]{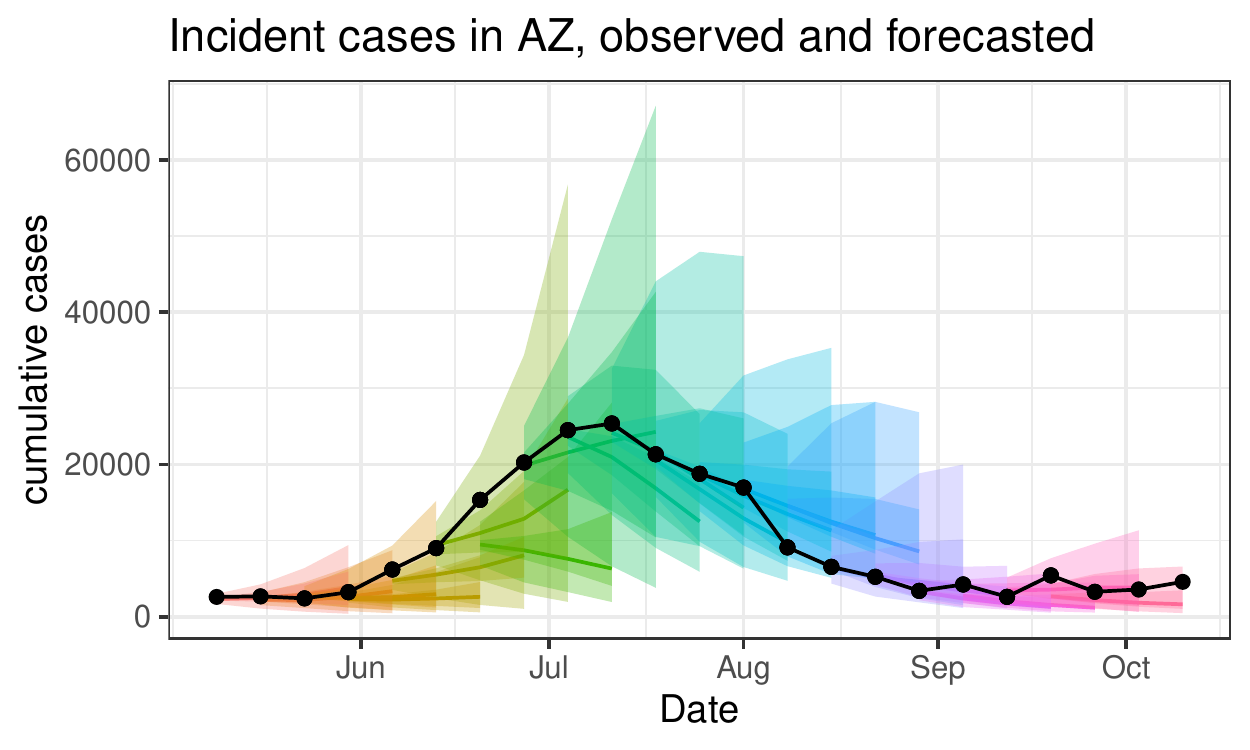}
   \includegraphics[width=0.45\linewidth]{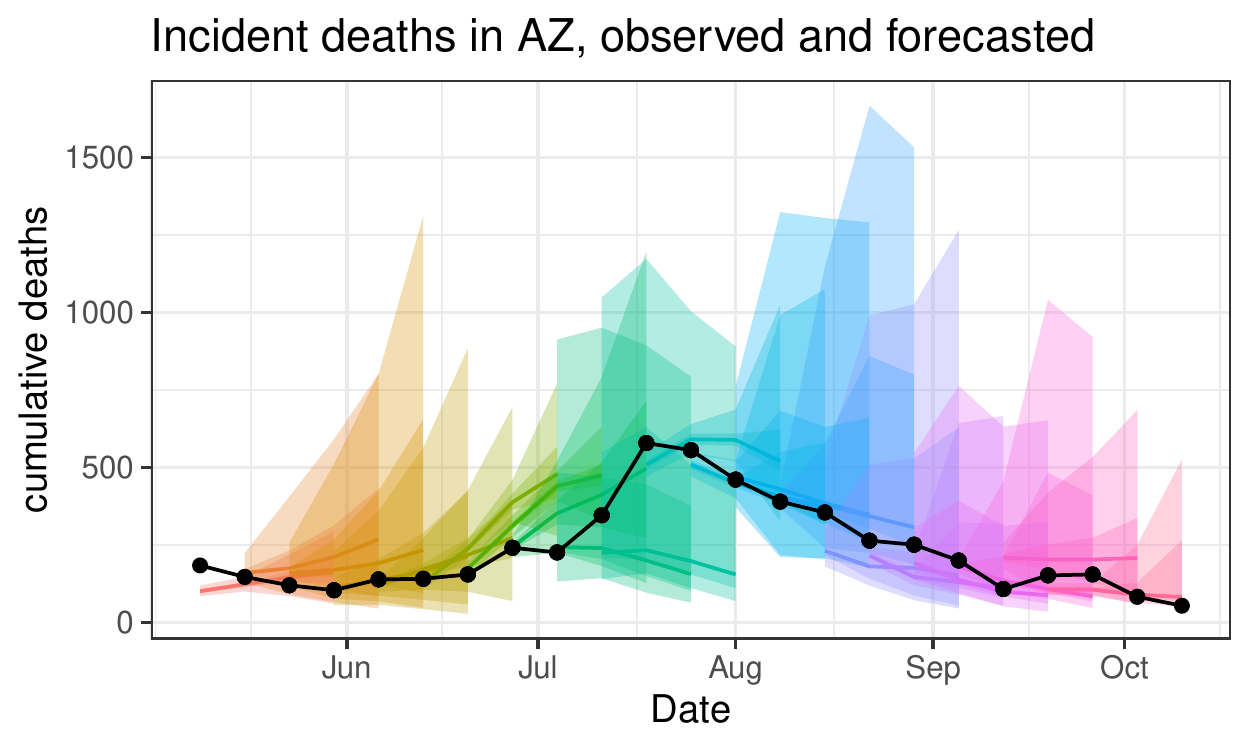}

\end{center}

\noindent The black curves indicate the true values as reported by the COVID Tracking Project \cite{CVDT}. The widest shaded regions correspond to the central 95\% prediction intervals. The smaller shaded regions correspond to the central 50\% prediction intervals. The darker colored curves in shaded regions are the median forecasts. COVID-19 data was provided by The COVID Tracking Project at {\sl The Atlantic} under a CC BY 4.0 license.

\section{Interval Scores}
This section provides heat maps showing interval scores per 100000 population for cases and deaths forecasts, 1 through 4 weeks after the forecast date. We first show the interval score ($\alpha= 0.05$) for case count forecasts. Each location corresponds to a row in the figure below, and each rectangle is a forecast week. The color scale ranges from less than 5 to more than 10,000 cases per 100,000 population.

\bigskip
\begin{center}
\includegraphics[width=\linewidth]{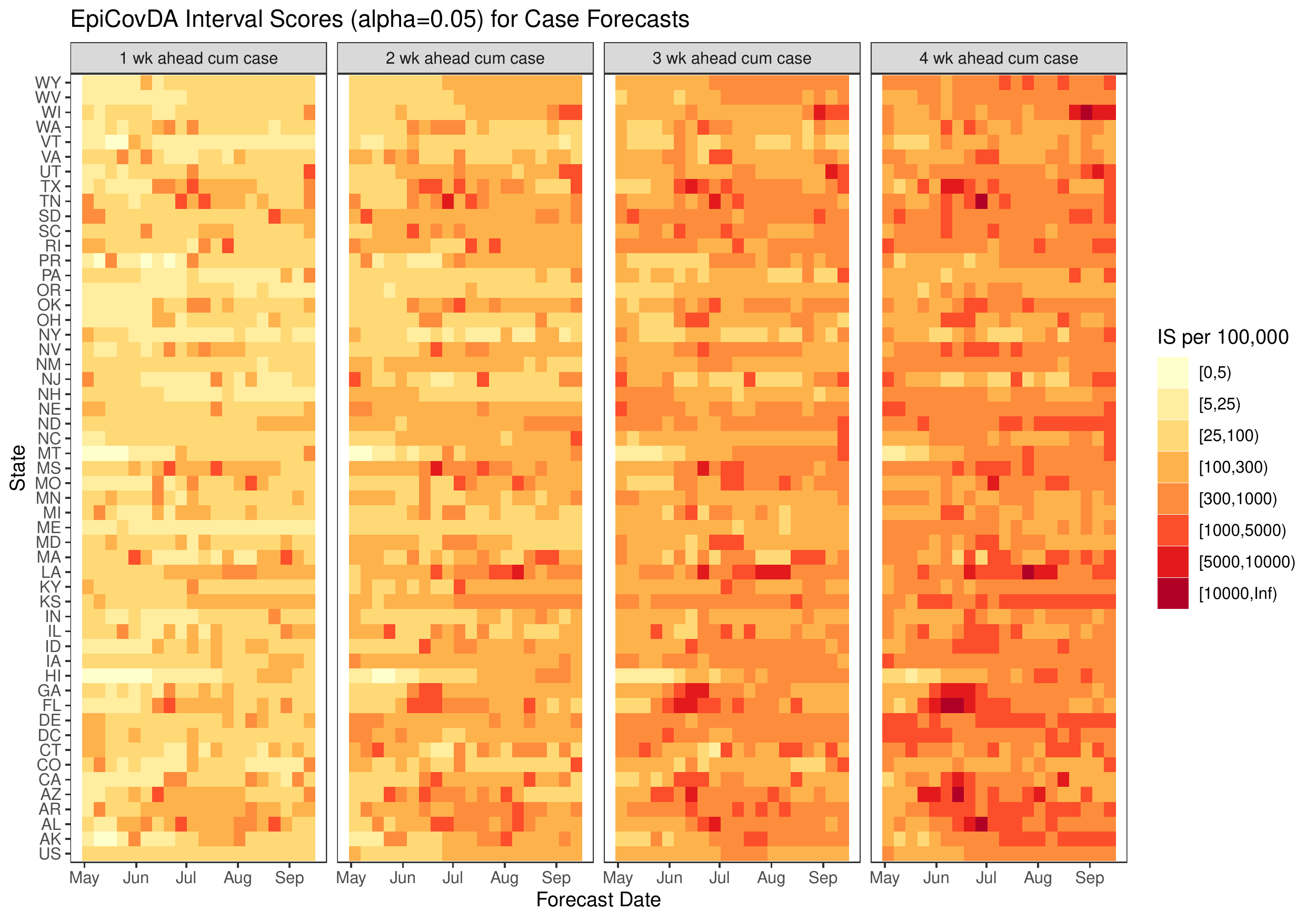}
\end{center}
\newpage
The interval score ($\alpha= 0.05$) for death count forecasts, one through four weeks ahead of the forecast date, is reported below. As before, each location corresponds to a row and each rectangle is a forecast week. The color scale ranges from less than 1 to more than 500 deaths per 100,000 population.

\bigskip
\begin{center}
\includegraphics[width=\linewidth]{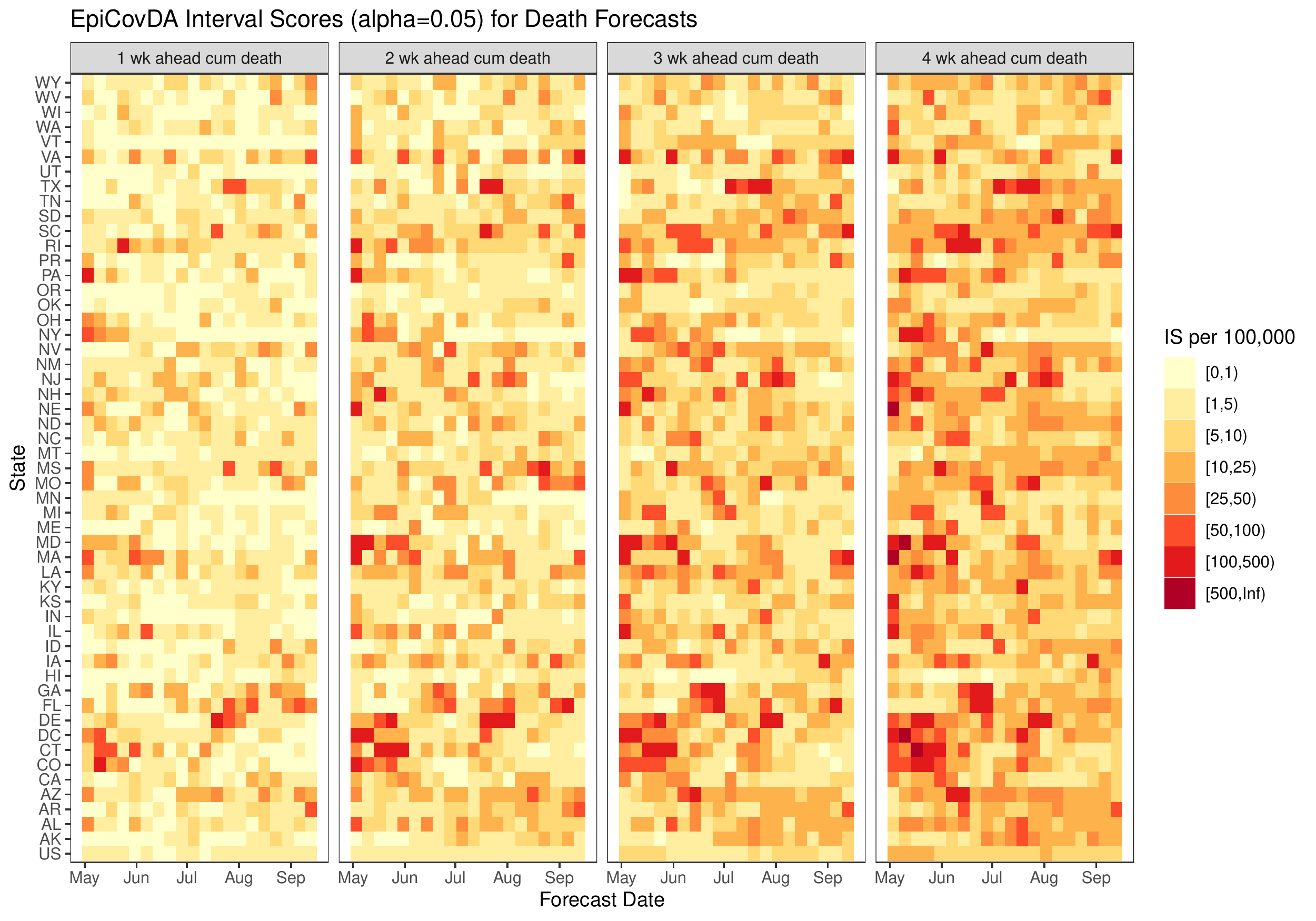}
\end{center}

\clearpage